\documentclass[fleqn,10pt]{JLA_article} 

\usepackage[english]{babel} 
\usepackage[toc,page]{appendix}
\usepackage{xcolor, soul, lipsum} 
\usepackage{multirow}
\usepackage{natbib}
\usepackage{changes}

\usepackage{hyperref} 
\hypersetup{hidelinks,colorlinks,breaklinks=true,urlcolor=color2,citecolor=color1,linkcolor=color1,bookmarksopen=false,pdftitle={Title},pdfauthor={Author}}

\PaperTitle{The FLoRA Engine: Using Analytics to Measure and Facilitate Learners' own Regulation Activities}

\Authors{Xinyu Li\textsuperscript{1}, Yizhou Fan\textsuperscript{2}*, Tongguang Li\textsuperscript{3}, Mladen Raković\textsuperscript{4}, Shaveen Singh\textsuperscript{5}, Joep van der Graaf\textsuperscript{6}, Lyn Lim\textsuperscript{7}, Johanna Moore\textsuperscript{8}, Inge Molenaar\textsuperscript{9}, Maria Bannert\textsuperscript{10}, Dragan Gašević\textsuperscript{11}} 

\affiliation{\textsuperscript{1}\textit{Email: \href{mailto:xinyu.li1@monash.edu}{\color{blue} xinyu.li1@monash.edu} Address: Centre for Learning Analytics, Faculty of Information Technology, Monash University, Australia. ORCID ID: \href{https://orcid.org/0000-0003-2681-4451}{\color{blue} https://orcid.org/0000-0003-2681-4451}}}
\affiliation{Corresponding author \textsuperscript{2}\textit{Email: \href{mailto:fyz@pku.edu.cn}{\color{blue} fyz@pku.edu.cn} Address: Graduate School of Education, Peking University, Beijing, China. ORCID iD: \href{https://orcid.org/0000-0003-2777-1705}{\color{blue} https://orcid.org/0000-0003-2777-1705}}} 
\affiliation{\textsuperscript{3}\textit{Email: \href{mailto:tongguang.li@monash.edu}{\color{blue} tongguang.li@monash.edu} Address: Centre for Learning Analytics, Faculty of Information Technology, Monash University, Australia. ORCID iD: \href{https://orcid.org/0000-0003-4616-5268}{\color{blue} https://orcid.org/0000-0003-4616-5268}}}
\affiliation{\textsuperscript{4}\textit{Email: \href{mailto:mladen.rakovic@monash.edu}{\color{blue} mladen.rakovic@monash.edu} Address: Centre for Learning Analytics, Faculty of Information Technology, Monash University, Australia. ORCID iD: \href{https://orcid.org/0000-0002-1413-1103}{\color{blue} https://orcid.org/0000-0002-1413-1103}}}
\affiliation{\textsuperscript{5}\textit{Email: \href{mailto:shaveen.singh@monash.edu}{\color{blue} shaveen.singh@monash.edu} Address: Centre for Learning Analytics, Faculty of Information Technology, Monash University, Australia. ORCID iD: \href{https://orcid.org/0000-0002-7862-8047}{\color{blue} https://orcid.org/0000-0002-7862-8047}}}
\affiliation{\textsuperscript{6}\textit{Email: \href{mailto:joep.vandergraaf@ru.nl}{\color{blue} joep.vandergraaf@ru.nl} Address: Behavioural Science Institute, Radboud University, Nijmegen 6500 HE, The Netherlands. ORCID iD: \href{https://orcid.org/0000-0003-1205-2193}{\color{blue} https://orcid.org/0000-0003-1205-2193} }} 
\affiliation{\textsuperscript{7}\textit{Email: \href{mailto:lyn.lim@tum.de}{\color{blue} lyn.lim@tum.de} Address: TUM School of Social Sciences and Technology, Technical University of Munich, Munich 80333, Germany. ORCID iD: \href{https://orcid.org/0000-0002-0617-5552}{\color{blue} https://orcid.org/0000-0002-0617-5552}}} 
\affiliation{\textsuperscript{8}\textit{Email: \href{mailto:j.moore@ed.ac.uk}{\color{blue} j.moore@ed.ac.uk} Address: School of Informatics, University of Edinburgh, Edinburgh EH8 9AB, United Kingdom. ORCID iD: \href{https://orcid.org/0000-0001-7247-6823}{\color{blue} https://orcid.org/0000-0001-7247-6823}}} 
\affiliation{\textsuperscript{9}\textit{Email: \href{mailto:inge.molenaar@ru.nl}{\color{blue} inge.molenaar@ru.nl} Address: Behavioural Science Institute, Radboud University, Nijmegen 6500 HE, The Netherlands. ORCID iD: \href{https://orcid.org/0000-0003-4639-2524}{\color{blue} https://orcid.org/0000-0003-4639-2524}}} 
\affiliation{\textsuperscript{10}\textit{Email: \href{mailto:maria.bannert@tum.de}{\color{blue} maria.bannert@tum.de} Address: TUM School of Social Sciences and Technology, Technical University of Munich, Munich 80333, Germany. ORCID iD: \href{https://orcid.org/0000-0001-7045-2764}{\color{blue} https://orcid.org/0000-0001-7045-2764}}} 
\affiliation{\textsuperscript{11}\textit{Email: \href{mailto:dragan.gasevic@monash.edu}{\color{blue} dragan.gasevic@monash.edu} Address: Centre for Learning Analytics, Faculty of Information Technology, Monash University, Australia. ORCID iD: \href{https://orcid.org/0000-0001-9265-1908}{\color{blue} https://orcid.org/0000-0001-9265-1908}}}

\Keywords{Learning Analytics, Self-Regulated Learning, Scaffolding, Learning Tools, Trace data} 

\Submitted{14/01/24}
\Accepted{05/12/24}
\Published{--/--/--}

\Volume{1}
\Number{1}
\Pages{1---10}
\Doi{xxx-xxx-xxx}
\Notesname{Notes for Practice} 

\note{Empirical studies have underscored the positive correlation between self-regulated learning (SRL) proficiency and enhanced learning outcomes. However, the intricacies involved in SRL, spanning cognitive, metacognitive, and motivational processes, present challenges to many learners. The FLoRA engine is designed and developed to detect, measure and facilitate complicated SRL.}
\note{The main contribution of this paper is the description of the design and implementation of an open-source SRL support engine that can unobtrusively collect fine-grained and temporally ordered trace data, map trace data to corresponding SRL processes, conduct analysis to understand learners' learning progress and provide corresponding as-needed personalised scaffolds.}

\Abstract{
The focus of education is increasingly set on learners’ ability to regulate their own learning within technology-enhanced learning environments (TELs). Prior research has shown that self-regulated learning (SRL) leads to better learning performance. However, many learners struggle to self-regulate their learning productively, as they typically need to navigate a myriad of cognitive, metacognitive, and motivational processes that SRL demands. To address these challenges, the FLoRA engine is developed to assist students, workers, and professionals in improving their SRL skills and becoming productive lifelong learners. FLoRA incorporates several learning tools that are grounded in SRL theory and enhanced with learning analytics (LA), aimed at improving learners' mastery of different SRL skills. The engine tracks learners' SRL behaviours during a learning task and provides automated scaffolding to help learners effectively regulate their learning. The main contributions of FLoRA include (1) creating instrumentation tools that unobtrusively collect intensively sampled, fine-grained, and temporally ordered trace data about learners' learning actions, (2) building a trace parser that uses LA and related analytical technique (e.g., process mining) to model and understand learners' SRL processes, and (3) providing a scaffolding module that presents analytics-based adaptive, personalised scaffolds based on students' learning progress. The architecture and implementation of the FLoRA engine are also discussed in this paper.
}

\begin{document}

\flushbottom 

\maketitle 


\thispagestyle{fancy} 


\section{Introduction}

Self-regulated learning (SRL) is a complex ensemble of cognitive, metacognitive, motivational, and emotional learning processes~\citep{panadero2017review}. Productive self-regulated learners are goal-oriented and they actively oversee their learning processes as they advance towards their learning goals. These learners have been found to outperform their less self-regulated peers in different learning tasks and subjects~\citep{uzir2020analytics}. Moreover, given the dynamically changing requirements in a modern job market, SRL skills have been considered critical for future professionals to successfully adapt to those changes, maintain employment relevance, and facilitate future career transfers~\citep{siadaty2016trace}. However, despite the well-documented benefits of SRL and the many opportunities provided to students to acquire, practise, and refine SRL skills in different subjects, students often struggle to productively self-regulate their learning~\citep{jarvela2023human}. Consequently, more research is needed to help students further improve their proficiency in SRL~\citep{lee2019systematic}. Modern TELs can be utilised to advance research and support for SRL, in particular given the fine-grained trace data those environments can gather and the interactive learning support those environments can provide to learners as they work on different learning tasks using the environment. To date, researchers have developed various TELs that support learners' use and development of SRL. For example, Metatutor used four simulated pedagogical agents to nudge learners to adopt effective learning strategies, including summarisation, monitoring, and activation of prior knowledge~\citep{azevedo2011metatutor, azevedo2022lessons}.  


Scaffolding -- i.e., an external guidance provided to students on a skill they could not enact independently and gradually faded out as the students become more proficient in that skill -- has been considered a common approach to supporting SRL in a classroom and digital learning environments~\citep{hmelo2006understanding,sharma2007scaffolding}. Researchers have demonstrated that scaffolding can be utilised to facilitate learners' SRL skills leading to improved learning performance and a better understanding of their SRL~\citep{azevedo2008externally,veenman2007assessment}. Following the promises of scaffolding for supporting SRL, computer-generated SRL scaffolding in digital learning environments has been increasingly investigated over the past decade. Previous research has shown that automated scaffolding of SRL in digital learning environments typically consists of predefined instructions delivered to all students at set time intervals (also known as standardised, fixed, or generalised scaffolds), rather than scaffolds tailored to each student's learning progress and prior SRL activities~\citep{guo2022using}. In addition, most current scaffolding approaches provide clear learning directions or generic instructions, however, without considering students' learning affordances (the affordances of scaffolds)~\citep{devolder2012supporting} and offering continuous feedback after scaffolds throughout learning~\citep{wisniewski2020power}. Thus, tailoring the content and design of scaffolding to provide adaptive support based on students' unique SRL repertoire is considered more effective in prompting effective use of SRL processes \citep{guo2022using}.

To identify students' SRL processes, researchers have analysed educational data using machine learning techniques~\citep{saint2022temporally}. The most commonly used educational data are trace data such as mouse movement and page navigation~\citep{winne2019nstudy}. Numerous innovative data analytic methods~\citep{saint2022temporally}, such as process mining~\citep{ahmad2019discovering,matcha2019detection} and epistemic network analysis (ENA)~\citep{shaffer2016tutorial}, have been used to identify SRL behaviours from the trace data unobtrusively generated by users in TELs. These SRL behaviours have been demonstrated to be predictive of learning outcomes~\citep{fincham2018study,saint2020combining}. Despite progress in the field, the researchers have encountered two major challenges~\citep{matcha2020analytics}. Firstly, students' learning activities are tightly connected to the structure of the learning environment, instead of being anchored in well-theorised SRL processes~\citep{matcha2020analytics}. As a result, changes in the learning platform make it challenging to compare the results of various studies and to consolidate the SRL research knowledge base~\citep{fan2021learning}. Secondly, current SRL models are not sufficiently supported by empirical evidence about contextual factors that may affect the selection and execution of SRL strategies~\citep{winne2018theorizing}. Additionally, there is limited research exploring the relationship between SRL tactics inferred from trace data and other factors such as prior knowledge and metacognitive knowledge~\citep{azevedo2019analyzing}. Finally, most current TEL systems lack the capability to detect students' SRL processes during learning. The presentation of scaffolds at fixed time intervals is insufficient to support students at varying levels of achievement~\citep{molenaar2012dynamic}. While some systems, such as Metatutor~\citep{azevedo2022lessons} and Betty' Brain~\citep{munshi2023analysing}, offer adaptive scaffolds, they are not personalised and do not take into account each student's individual SRL progress.

To address the existing challenges related to SRL measurement and facilitation, we developed FLoRA (short for Facilitating Learners' own Regulation Activities), a web-based, micro-service architectural learning and data analysing engine. The design of the FLoRA engine is grounded in the research on SRL conducted over the past few decades. This includes theoretical models of SRL~\citep{panadero2017review} and empirical studies that examined SRL support using human and digital tutoring~\citep{chi2021translating}, artificial intelligence~\citep{biswas2016design}, and the principles of multimedia learning~\citep{lajoie2021student}. The most pronounced advantage of FLoRA engine is that it offered opportunities in using learning analytics (LA)~\citep{gavsevic2015let} and related data analytic techniques (e.g., process mining)~\citep{matcha2019detection} to model and understand learners' use of SRL processes and based on which informed the design and implementation of adaptive supports (e.g., scaffolding) on SRL. Specifically, the FLoRA engine includes theory-informed (1) instrumentation tools that facilitate work on a task (e.g., tools for text annotation and writing), and unobtrusively collect fine-grained trace data as learners working on a task; (2) a trace parser that automatically analyses learner trace data and, based on the predefined set of rules, detects SRL processes that learners enact during the task, and (3) a scaffolding module that generates adaptive and personalised scaffolds to guide learners towards more productive SRL and improve their performance in the task.



\section{Background}
 
\subsection{Self-Regulated Learning}
Since \citet{zimmerman1986becoming} clarified the potential of SRL, a various of learning models have been developed and evolved rapidly to form a comprehensive research ecosystem, such as COPES \citep{winne1998studying} and Dual Processing model \citep{boekaerts2006far}. While many current SRL models provide unique insights, they all start from the assumption that SRL is best understood as a cycle with distinct phases that each has their own unique set of activities~\citep{panadero2017review}. The comparison of all the extant SRL models reveals that there are, in fact, three distinct phases to the SRL process: preparatory, performance, and appraisal phases~\citep{panadero2017review}. In terms of the preparatory phase, the main activities are analysing the task, planning the learning, and setting goals~\citep{panadero2017review}. Analysing the task is to activate prior knowledge and determine the most effective cognitive methods for completing the task. During this process, the goals can be formulated together with the plan for achieving these goals~\citep{winne2018theorizing}. The actual task is done while monitoring and controlling the progress of performance, which refers to the performance phase. The importance of regulating learning through monitoring and controlling procedures is highlighted by Nelson and Narens~\citep{nelson1994investigate}. They propose dividing cognition into a "meta-level" (the mental representation of a human learner's cognition) and an "object-level" (human learner's cognition), with the interaction between these two levels reflected by the monitoring and controlling of the learning processes. Controlling procedures, such as rereading the material or terminating the current approach, might be modified based on the meta-level representation (e.g., judgements of their learning) gleaned via monitoring~\citep{bjork2013self}. In the last phase -- appraisal, students reflect and regulate the current learning, such as contrasting current progress with the goals set at the first phase~\citep{zimmerman2000attaining}, leading to adaptation for better performance of the next iteration.

\subsection{Detecting and Measuring SRL}
According to previous studies~\citep{panadero2017review}, SRL encompasses a spectrum of cognitive and metacognitive activities that learners engage in to control their learning processes. These activities include setting goals, monitoring progress, and adjusting strategies as needed. To comprehensively understand these SRL processes in the computer-based learning environment, log data and learning analytics has been utilised~\citep{siadaty2016trace}. For example, a learner's decision to attempt a quiz before moving on to reading materials can be used as a criterion for evaluation~\citep{saint2020trace}. The identification of a specific SRL process requires locating the corresponding log entries that represent SRL actions. However, linking specific SRL actions to log data traces is challenging, as not all SRL actions are easily recognised in logs~\citep{papamitsiou2014learning, bannert2014process}. Furthermore, the methods for recognising metacognitive level SRL actions are limited~\citep{clarebout2013metacognition}. To address this issue, instrumentation tools embedded within learning environments have been proposed to capture SRL processes that otherwise remain hidden, such as organisation and orientation~\citep{marzouk2016if}. As learners interact with these instrumentation tools, trace logs that indicate SRL processes are generated. For instance, when learners highlight texts, it may indicate that they are engaging in the process of organising information. An example of a learning technology with such instrumentation tools is nStudy, proposed by Winne and colleagues~\citep{winne2017learning, winne2019nstudy}. The term “instrumentation tool” is used because the designed tools (e.g., writing or timer tools) not only provide convenience for the reading and writing process but also enable the measurement and collection of learning activities that reflect students' SRL processes. From students' perspective, these tools offer them the opportunity to initiate and reflect on their cognitive and metacognitive activities. Meanwhile, from the perspective of educators and researchers, these tools provide a means to collect and analyse trace data, which are indicative of students' cognitive and metacognitive engagement. \citet{winne2018theorizing} emphasised the importance of such traces, arguing that they should be a standard feature to understand learners' SRL.

SRL can be conceptualised as a series of events, which highlights three successively more complex levels -- occurrence, contingency, and patterned contingency \citep{winne2019nstudy}. The first level -- occurrence -- describes the appearance or existence of one specific learning action (e.g. taking notes). This level captures the frequency of learning actions, but it fails to account for the context of the enacted learning action. The second level -- contingency -- captures a short sequence of learning actions, which adds the context of a single learning action (e.g. taking notes followed by reading or taking notes followed by writing). The third level -- patterned contingency -- is a more complex form of contingency, which adds extra dimensions about the learning pattern (normally involving more than two learning actions) and frequency of the pattern. For example, \textit{it is revealed that a learner takes notes immediately followed by highlighting, and then paraphrases the highlighted information} is a typical example of capturing SRL at the level of patterned contingency. The patterned contingency level is useful in understanding not only the occurrence of learning actions and the flow of learning actions from one to another, but also explaining learners' SRL patterns. As suggested by~\citet{reimann2009time} and \citet{winne2018theorizing}, the analysis of SRL within computer-based learning environments can benefit from a trace parser that offers both event-centred and process-based perspectives. As such, there is a need for a learning environment and learning trace data that can unpack the temporal patterns of engagement on learning tasks~\citep{winne2000measuring, winne2019nstudy}. Many recent studies have started using the temporal data to recognise learners' SRL processes~\citep{paans2019temporal,cerezo2020process,huang2021process,saint2022temporally}. Recent literature has concluded five major strands using temporally-focused analytical methods to reveal learners' dynamic SRL processes in various learning platforms~\citep{saint2022temporally}. For instance, one of the strands emphasised the usage of trace data collected from Learning Management Systems (LMSs) and Massive Open Online Courses (MOOCs) and utilised, for example, First-Order Markov Model (FOMM) and process mining to model learners' tactic and strategy use~\citep{matcha2020analytics,saint2020trace}. 

To further enhance the utility of trace data, a trace parser is developed within the FLoRA engine. This trace parser is designed to map raw trace data (e.g., navigational logs, mouse and keyboard traces) generated by learners into distinct SRL processes~\citep{fan2022towards}. In this way, we ensure a real-time measurement of SRL processes. The development of the SRL trace parser in the FLoRA engine was informed by the theoretical framework of SRL processes proposed by \citet{bannert2007metakognition}. The Bannert framework originated from Zimmerman's model. It is an operationalisation of existing theoretical work on SRL in the form of a coding scheme for analysis of  SRL data. Bannert’s team utilised this SRL framework and think-aloud coding protocol to research higher education learning tasks~\citep{bannert2007metakognition, bannert2009promoting, bannert2013scaffolding, bannert2014process}. According to this framework, SRL processes can be categorised into three broad categories: metacognitive, cognitive, and other processes. The metacognitive processes include \textit{task orientation} (i.e., surveying task requirements and constraints), \textit{planning} (i.e., arranging learning activities for the task), \textit{monitoring} (i.e., overseeing the effectiveness of learning throughout the task), and \textit{evaluation of learning}. For example, a learner orients to understand task instructions and resources available for the task; plans to come up with learning strategies for the task; monitors their progress relative to task requirements and a scoring rubric; and evaluates whether the previously used learning strategies worked well, given the progress achieved. The cognitive category spans several low- and high-cognition processes. As articulated in the \citet{bannert2007metakognition} theoretical framework, the low-cognition processes are \textit{first time reading} and \textit{re-reading}, whereas high-cognition processes are \textit{elaboration} and \textit{organisation}, e.g., summarising or synthesising the information from source texts in a writing task. Other processes, although not included in the real-time analysis, contain learner motivation (e.g., a learner's excitement for the topic) and processes enacted due to procedural issues (e.g., a learner asking questions about using the writing tool). Each component of Bannert's framework can be mapped to preparation, performance, and appraisal stages, specifically: (1) preparation—orientation and planning; (2) performance—low/high cognition, monitoring, and evaluation; and (3) appraisal—evaluation. It is noteworthy that evaluation can occur in both the performance stage and the appraisal stage. More details about the SRL trace parser are provided by the~\citep{fan2022improving}.

\subsection{Examining Analogous Systems}

Numerous studies have led to the design and development of various tools aimed at supporting SRL from diverse perspectives. However, the majority of these tools predominantly focus on one or two specific phases of SRL~\citep{alvarez2022tools}. For instance, most tools are oriented towards activities such as planning~\citep{nussbaumer2014framework}, goal-setting~\citep{alexiou2015managing, davis2018srlx}, time management~\citep{yau2009mobile}, annotation~\citep{yousef2015video}, and Learning Analytics dashboards~\citep{matcha2019systematic}. Given that SRL involves cyclical processes where different phases can recur multiple times throughout the learning period~\citep{zimmerman2013cognitive}, tools that concentrate solely on a single phase are insufficient for fully fostering and promoting SRL. Therefore, a comprehensive system encompassing all SRL theory-based instrumentation tools is essential to detect and measure all phases of SRL effectively.

Currently, there are only a few comprehensive systems designed to support SRL, including nStudy~\citep{winne2019nstudy}, MetaTutor~\citep{azevedo2022lessons}, and Betty's Brain~\citep{munshi2023analysing}. nStudy is a comprehensive web-based learning tool that supports SRL by providing students with a variety of tools to enhance study strategies, metacognition, and learning progress tracking. It features functionalities such as searching, note-taking, writing, and information organisation through concept maps~\citep{winne2019nstudy}. MetaTutor is an intelligent tutoring system that supports SRL by guiding learners through complex scientific content using adaptive feedback and scaffolding. It integrates cognitive, metacognitive, and motivational strategies to help learners develop effective learning habits, offering real-time prompts, hints, and progress tracking to facilitate learning through reflection and self-monitoring~\citep{azevedo2011metatutor}. Betty’s Brain is an educational software tool designed to teach students about complex systems and scientific concepts through a learning-by-teaching approach. In this interactive environment, students assist a virtual character named Betty by creating and modifying concept maps that visually represent relationships between different concepts, thereby engaging in SRL activities such as planning, monitoring, and evaluating their understanding~\citep{munshi2023analysing}.


Despite the careful design of these systems to support SRL and promote learning, they exhibit limitations in processing trace data and supporting SRL. Although these systems can capture a range of fine-grained digital raw log data (e.g., clicks, highlights), they do not have well-developed and validated functionalities for mapping these raw log data onto SRL processes. As a result, while they can perform data analysis after the learning task is completed, the scaffolding they provide is not grounded in learners’ real-time SRL processes at different stages of learning. Furthermore, these systems are primarily designed for research purposes and have been predominantly utilized in laboratory settings. This focus raises concerns about their applicability and effectiveness in authentic learning environments, such as traditional classrooms and online learning platforms, where variables are more dynamic and less controlled. The complexities and variabilities inherent in real-world educational settings may significantly impact the systems’ ability to support SRL effectively, indicating a need for additional investigation and adaptation. To address these limitations, the FLoRA engine has been specifically designed and developed.

\section{FLoRA Architecture}

\begin{figure}[ht]\centering
\includegraphics[width=0.9\linewidth]{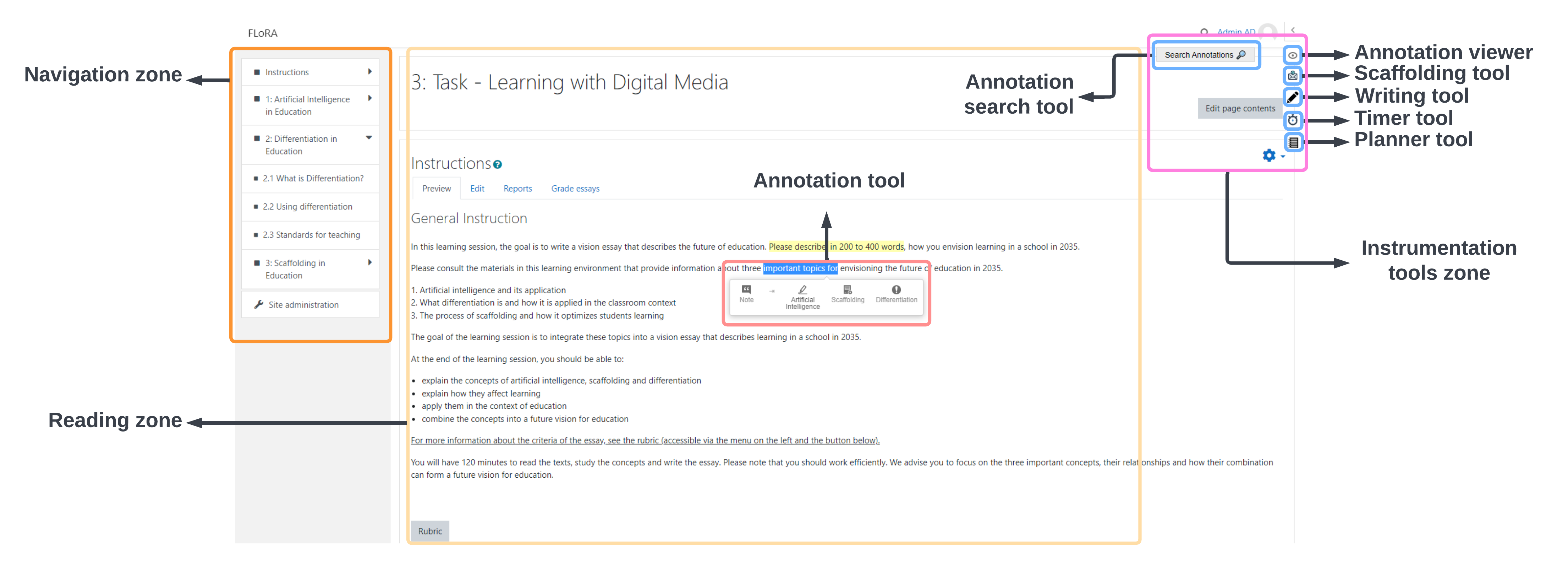}
\caption{The user interface of the FLoRA integrated into Moodle LMS}
\label{fig:overview_user_interface}
\end{figure}

FLoRA is an interactive, AI-powered engine, incorporating an instrumentation tools module, a trace parser module, and a scaffolding module, designed to support research and improve SRL skills in complex learning tasks that involve searching, navigating, reading, and interacting with textual or video sources, and developing written responses grounded in those sources. Not like other learning systems or platforms which can provide learning materials and conduct learning activities. The FLoRA is independent from the LMSs and all three modules are used together with the LMSs. Hence, the FLoRA can be seamlessly integrated into various web-based learning platforms, such as Moodle. Figure~\ref{fig:overview_user_interface} illustrates the integration of the FLoRA engine within the Moodle platform, providing an overview of its appearance. Three main working zones are pointed out in the picture, which are the Navigation zone, Reading zone, and the Instrumentation tools zone. Only the Instrumentation tools zone is provided by the FLoRA engine. The other parts of the system are original Moodle components. Because the FLoRA driven instrumentation tools are injected into the LMS, all the learners' actions happen on the web pages can be tracked and processed in the backend of the FLoRA engine.

\begin{figure}[ht]\centering 
\includegraphics[width=0.70\linewidth]{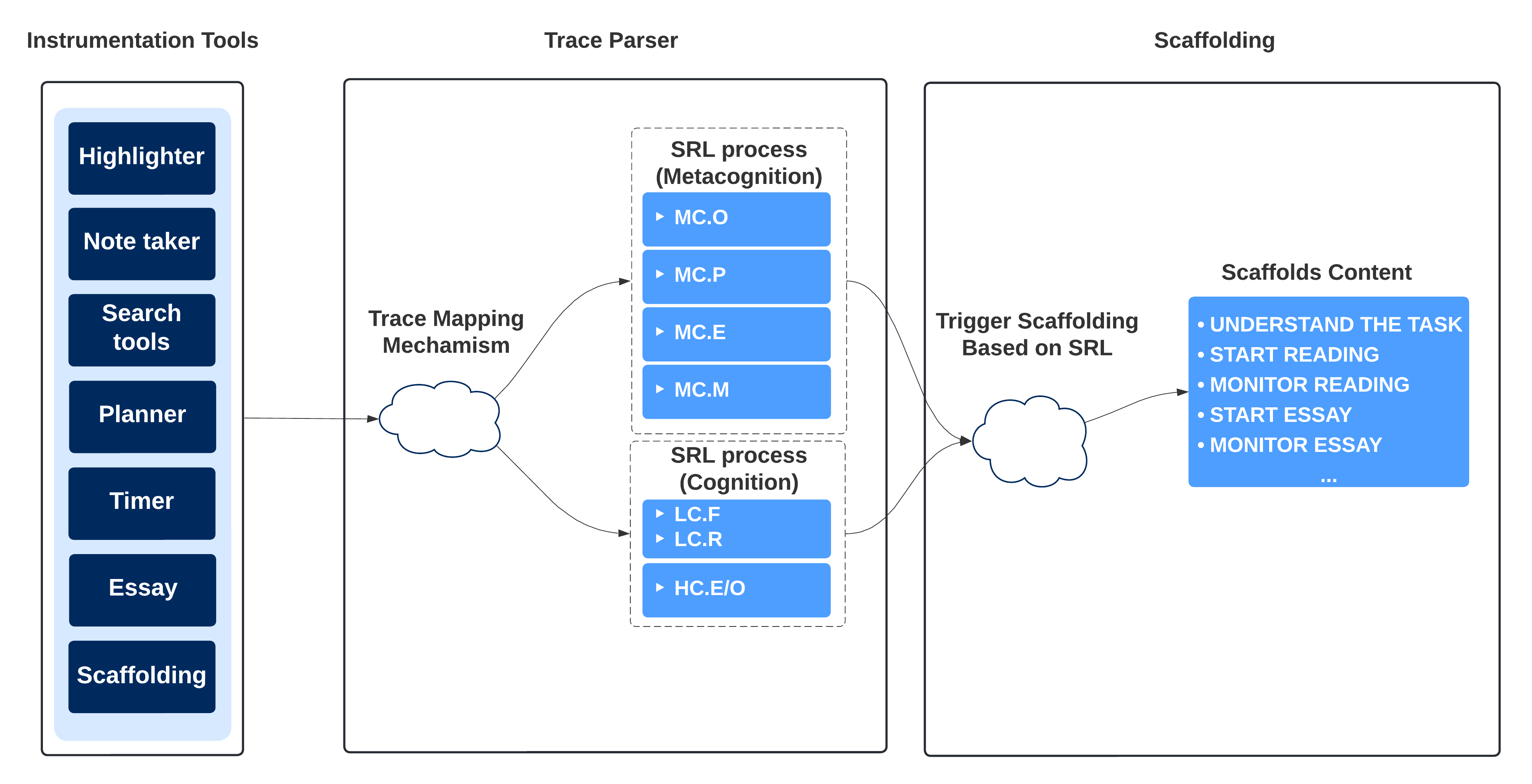}
\caption{The architecture of the FLoRA engine}
\label{fig:flora_architecture}
\end{figure}

A high level architecture of the FLoRA engine can be found in Figure~\ref{fig:flora_architecture}. The instrumentation tools are designed to be easily integrated into the different LMSs, ensuring a smooth and uninterrupted experience for learners. While students use the learning system and the instrumentation tools, trace data are collected by the engine and systematically analysed by the trace parser to identify SRL events. As shown in Figure~\ref{fig:flora_architecture}, the SRL processes are labelled into meta-cognition level and cognition level. The details about each code are further explained in Section~\ref{subsec:trace_parser}. Subsequently, the scaffolding module is employed to generate personalised scaffolds. The scaffolds will be displayed via the scaffolding displaying tool, which are instrumental in augmenting students' learning performance and fostering productive SRL behaviours, as discussed by \citet{lim2023effects}. This section aims to describe the three primary modules constituting the FLoRA engine, elucidating their individual and collective contributions to the facilitation of advanced learning processes. 

\subsection{Module I -- Instrumentation Tools}
As described in the background, there remains a significant gap in understanding the specific SRL processes from trace logs. This gap is particularly noteworthy given that instrumentation tools are specifically designed to capture these elusive SRL processes. The FLoRA engine exemplifies this by offering a suite of instrumentation tools designed to facilitate the detection and measurement of SRL processes. These tools, which include annotation tool (highlighter and notetaker), annotation search tool, timer, planner, writing tool, and scaffolding tool, capture high-quality, granular, and temporally ordered log data, thereby enabling a more nuanced analysis of SRL dynamics.

\subsubsection{Annotation Tool}
\begin{figure}[ht]\centering
\includegraphics[width=0.75\linewidth]{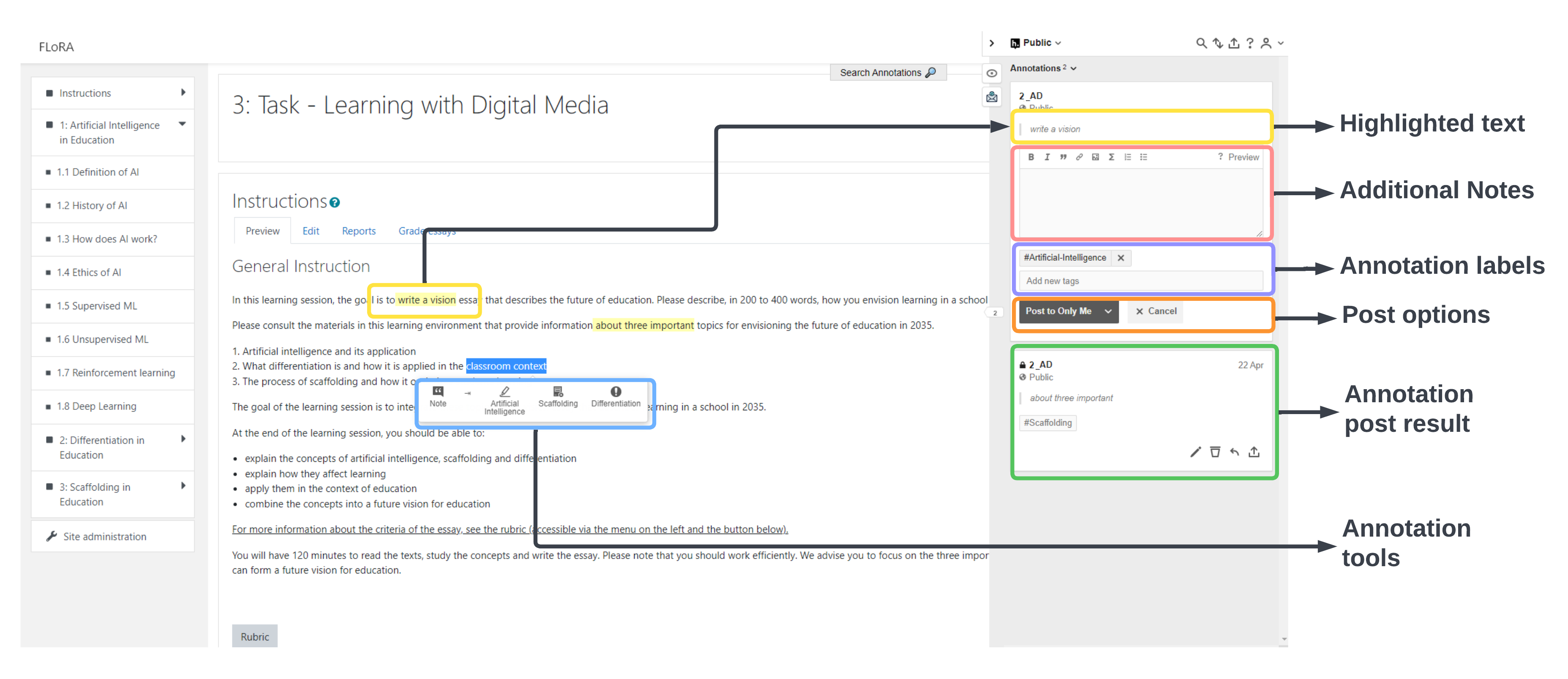}
\caption{Annotation tool}
\label{fig:annotation_tools}
\end{figure}

The instrumentation tools feature an annotation tool including a highlighter and a side panel for note taking, which is shown in Figure~\ref{fig:annotation_tools}. When text is selected on a web page, a pane emerges, presenting an array of labelling options for the user's selection. Post-selection, the annotation is generated in the right-side panel, where users can append notes or assign customised labels. The default label mirrors the choice made during the highlighting phase. Once annotations are completed, users have the option to publish them and, if desired, locate them via the search tools. Additionally, the tool offers the functionality of hiding or displaying all highlights, thereby granting users agency over the annotations' influence on their reading experience. This tool aims to track students' highlighting and note-taking behaviours. Previous research has established a positive relationship between the highlighting action and learning performance~\citep{van2021instrumentation}, and thus the FLoRA leverages this approach to shed light on students' SRL strategies. In addition to the highlighter tool, a note-taking tool is also provided to consist of the annotation tool. While learners use the highlighter, the highlighted texts are linked with notes, allowing learners to write down extra information. During this process, learners will generate long-term memory for the encoded information from the annotation, and save additional information for future check~\citep{azevedo2004does,di1973listening}. This indicates that notes can be used to memorise information (low cognition), organise or elaborate information (high cognition), and monitor the relevance of information (metacognition)~\citep{lim2021temporal}. Beyond the conventional note-taking process, FLoRA also records all keystroke events during note-writing and note-browsing events, providing a more comprehensive trace data that helps researchers to gain a deeper understanding of the SRL processes.

\subsubsection{Annotation Search Tool}
\begin{figure}[ht]\centering
\includegraphics[width=0.75\linewidth]{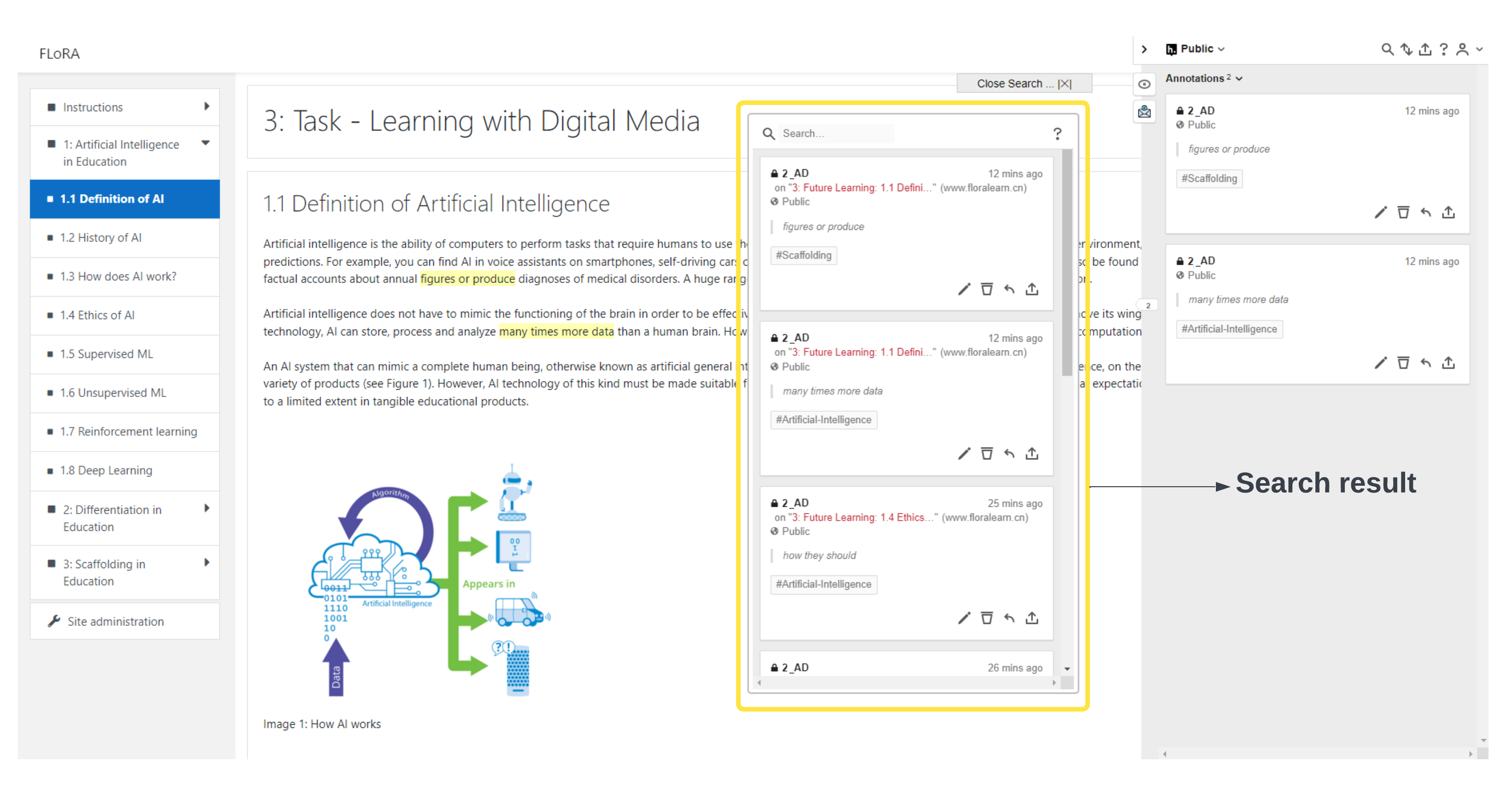}
\caption{Annotation search tool}
\label{fig:search_tools}
\end{figure}

Figure~\ref{fig:search_tools} displays the annotation search tool. The interface displays a list of annotations and a text-input box. Users can view public annotations in the search results area or employ the text-input box for fuzzy searches, with filtered outcomes presented. Furthermore, the tool enables users to locate highlighted texts tagged with specific labels, offering a streamlined method to track annotations across various web pages in the learning environment. The use of a search tool is indicative of planning behaviour, as the initiation of a search suggests that the learner has a specific goal in mind and has chosen keywords to direct their search~\citep{greene2009macro}. Furthermore, the process of comparing search results aligns with the monitoring phase of SRL, wherein learners critically evaluate the pertinence of the information they have retrieved~\citep{greene2009macro}. While the connections between search actions and SRL processes remains an area not yet exhaustively investigated, the FLoRA implements a sophisticated annotation search tool. The FLoRA engine tracks a comprehensive array of events associated with the usage of the search tool. This includes but is not limited to, the opening and closing of the tool, and the specific content of these keywords. Such detailed tracking is instrumental in providing insights into the learners' search strategies and behaviours. In our previous studies, the use of annotation search tool connect with the SRL process - planning and monitoring. Besides, the frequencies of the timer, search and planner usage was positively correlated, as they share the SRL processes they captured. But the search function is used least often compare with other tools. An explanation for the low usage of search tool is that the learning materials in our studies were limited to three topics, which might have reduced the need for goal-directed search behaviour~\citep{van2021instrumentation}.

\subsubsection{Timer Tool}
\begin{figure}[ht]\centering
\includegraphics[width=0.75\linewidth]{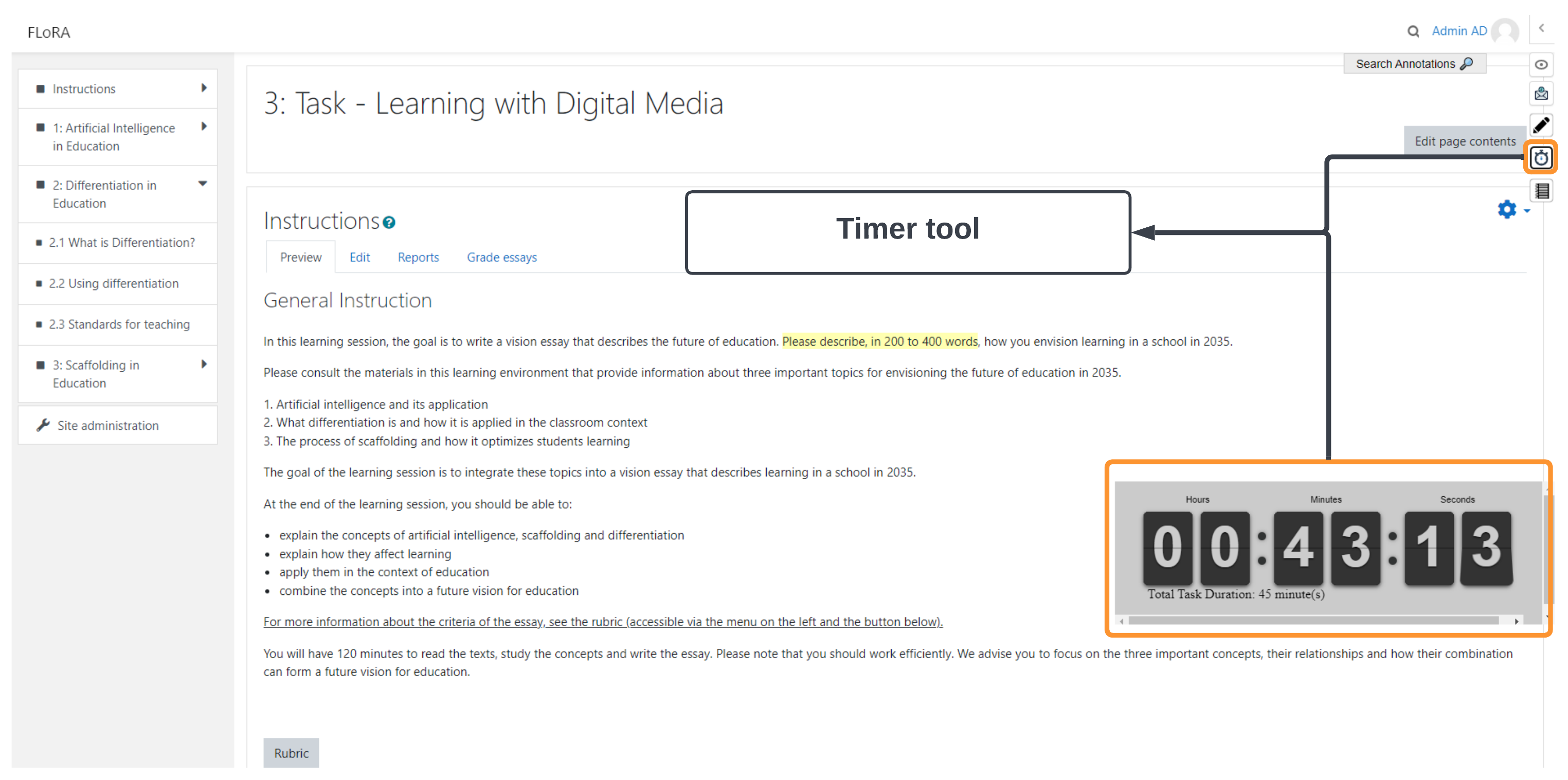}
\caption{Timer tool}
\label{fig:timer_tools}
\end{figure}

As shown in Figure~\ref{fig:timer_tools}, the timer tool, triggered by an alarm clock icon, displays a countdown timer with adjustable duration via the configuration file. To mitigate the issue of constant timer monitoring by students, the tool automatically closes after a two-second display. This feature is crucial in facilitating the detection of student interaction with the timer. The tool remains accessible throughout the task, enabling students to monitor the time remaining for the completion of a given task.

Base on the theoretical model, the checking timer is identified as a meta-cognitive process - Monitoring~\citep{bannert2007metakognition}. This feature is instrumental in enabling students to track their learning progress in alignment with their predefined objectives~\citep{winne2017learning}. Moreover, the systematic tracking of all actions associated with the use of the timer, including but not limited to the opening and closing of the tool, is essential in capturing a comprehensive and accurate representation of the students' SRL behaviours. This level of detailed monitoring is crucial for providing insights into the temporal aspects of students' learning strategies, thereby contributing to the understanding and enhancement of SRL processes in educational settings. In our previous studies, the timer is used frequently in a task and it was used by all students. Since timer can be used to assess monitoring of time left, it can help to evaluate progress in relation to the set goals~\citep{van2021instrumentation}.

\subsubsection{Planner Tool}

\begin{figure}[ht]\centering
\includegraphics[width=0.75\linewidth]{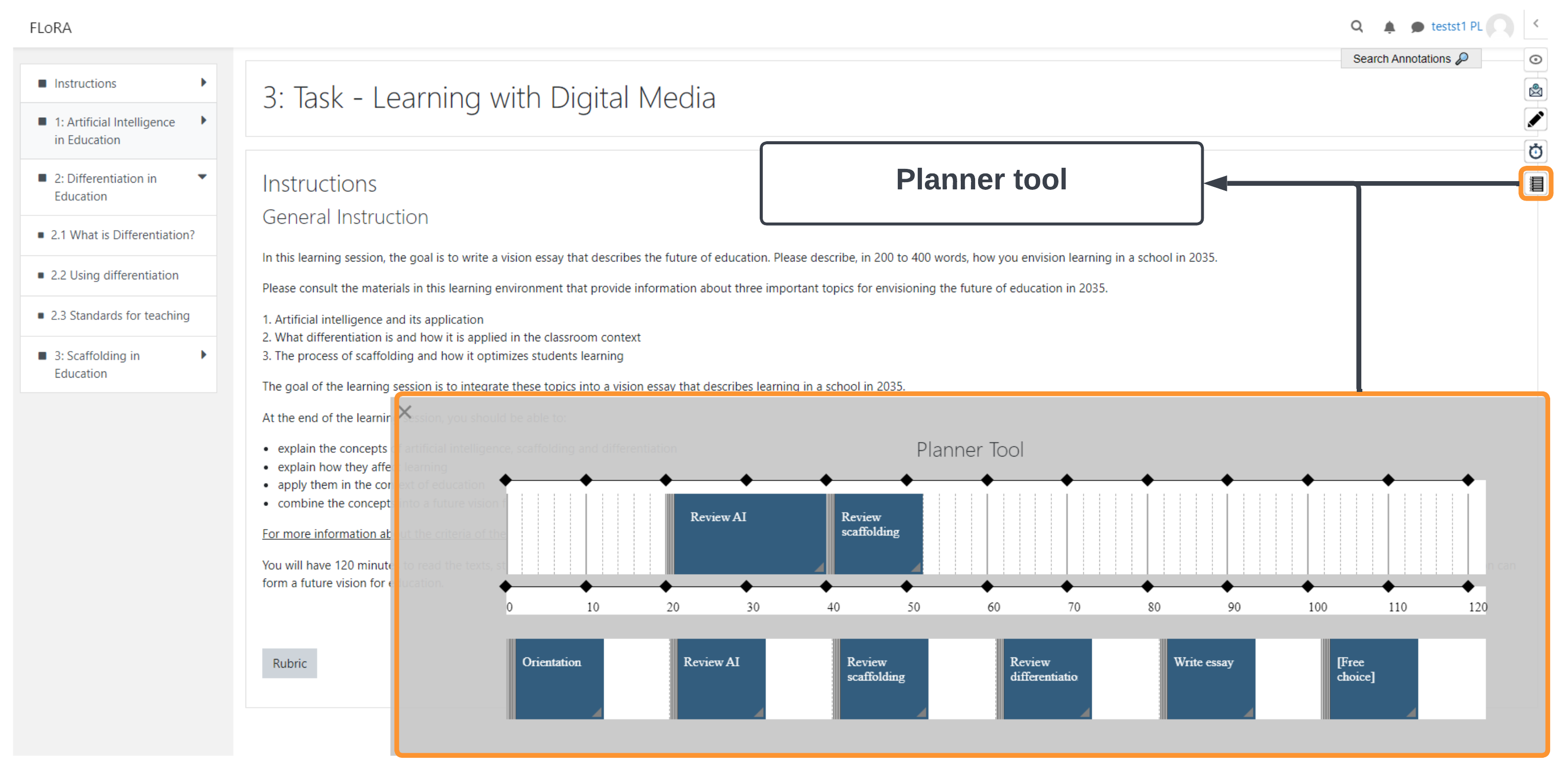}
\caption{Planner tool}
\label{fig:planner_tools}
\end{figure}

Given that planning constitutes a principal facet of metacognition~\citep{winne2019nstudy}, the FLoRA engine incorporates a planner tool (as shown in Figure~\ref{fig:planner_tools}) designed to foster engagement in planning and goal-setting activities among learners. The significance of such activities is extensively documented in the literature, highlighting their positive impact on motivation and learning performance~\citep{bowman2020impact}. This tool enables learners to methodically organise their tasks in a timeline format, offering the flexibility to adjust the allocated time for each specific task.

This tool is engineered to offer a drag-and-drop functionality, thereby significantly reducing the need for manual text or time entry for scheduling tasks. Users can effortlessly relocate predefined plan items to specific time slots, with the added functionality of adjusting the duration through the resizing of the item block. This intuitive design enhances user experience by simplifying the task planning process. Furthermore, FLoRA ensures a comprehensive record of all interaction events associated with the planner tool. This includes, but is not limited to, the opening and closing of the planner, as well as detailed tracking of mouse clicks and movements. Such extensive data collection is crucial for analysing learners' planning behaviours, providing valuable insights into their metacognitive processes. This, in turn, contributes to a deeper understanding of the role of planning in enhancing learning outcomes. Planner is also hardly used in our study. An explanation for the low usage of the planner is that the learning task in our study were 45 minutes, which would reduce the need for a planner~\citep{van2021instrumentation}. In another study we conducted in non-English speaking country which has a 2 hours task time, the planner usage increased a lot~\citep{fan2023towards}.

\subsubsection{Writing Tool}
\begin{figure}[ht]\centering
\includegraphics[width=0.75\linewidth]{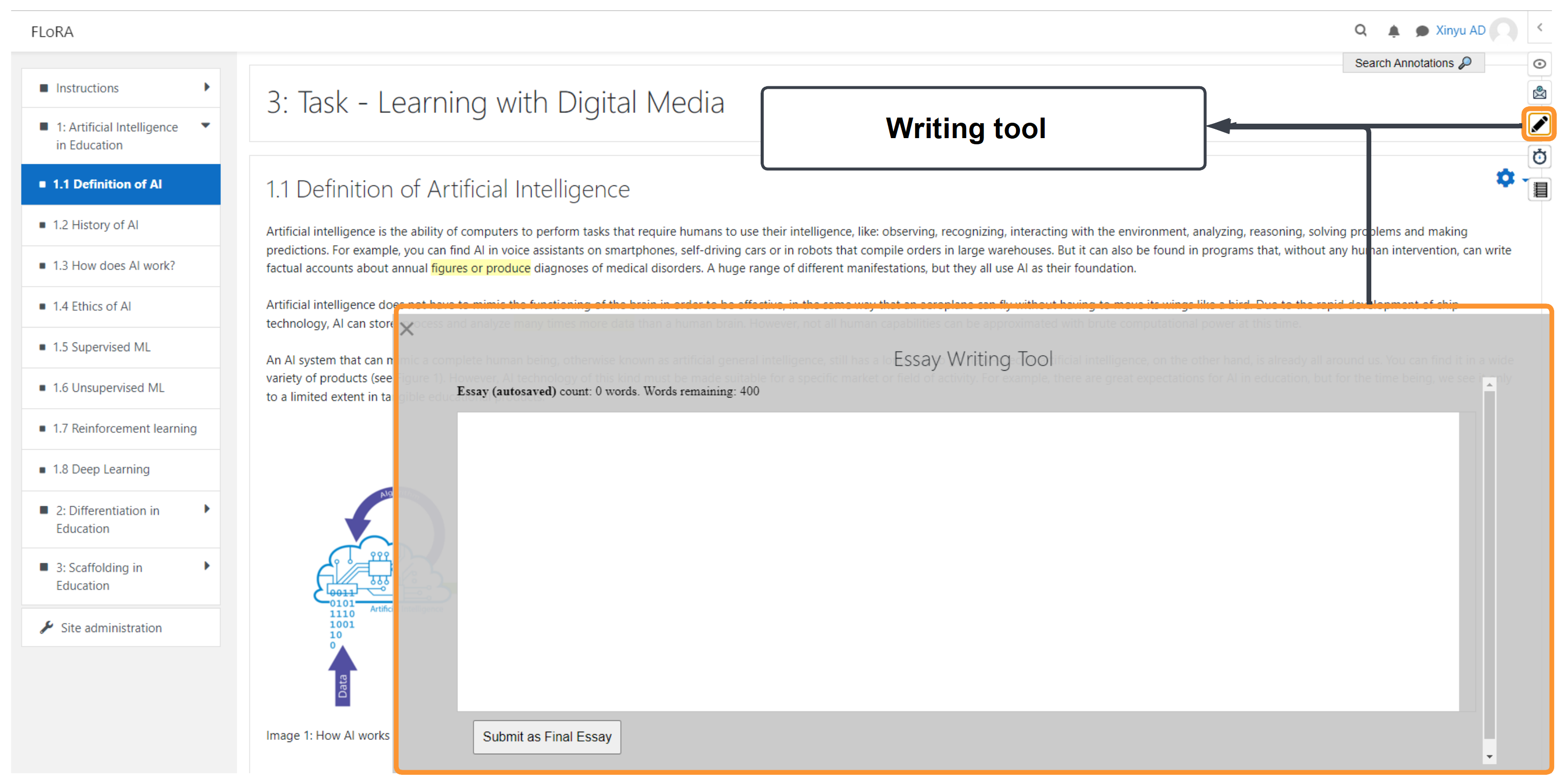}
\caption{Writing tool}
\label{fig:essay_writing_tools}
\end{figure}

The FLoRA engine is designed to facilitate a diverse array of learning activities, utilising textual or video-based materials. In prior research endeavours~\citep{fan2022improving}, a heightened focus has been placed on the reading and writing task. According to the theoretical framework of SRL processes proposed by~\citet{bannert2007metakognition}, the actions related to the writing tool, such as writing text and open tool, can be identified as meta-cognitive process - evaluation and the cognition process - elaboration/organisation. To capture learners' more SRL processes during a task, a writing tool has been developed and integrated into the engine, as displayed in Figure~\ref{fig:essay_writing_tools}. This tool features an automatic recording mechanism that captures and stores keystroke events and the corresponding essay content in a database, with each keystroke being automatically logged. This feature is particularly beneficial as it ensures the preservation of the essay text, safeguarding against potential losses due to unforeseen closures of the writing tool. Additionally, the opening and closing of the writing tool events are tracked to further contribute to the comprehensive data collection.

\subsubsection{Scaffolding Display Tool}
\begin{figure}[ht]\centering
\includegraphics[width=0.75\linewidth]{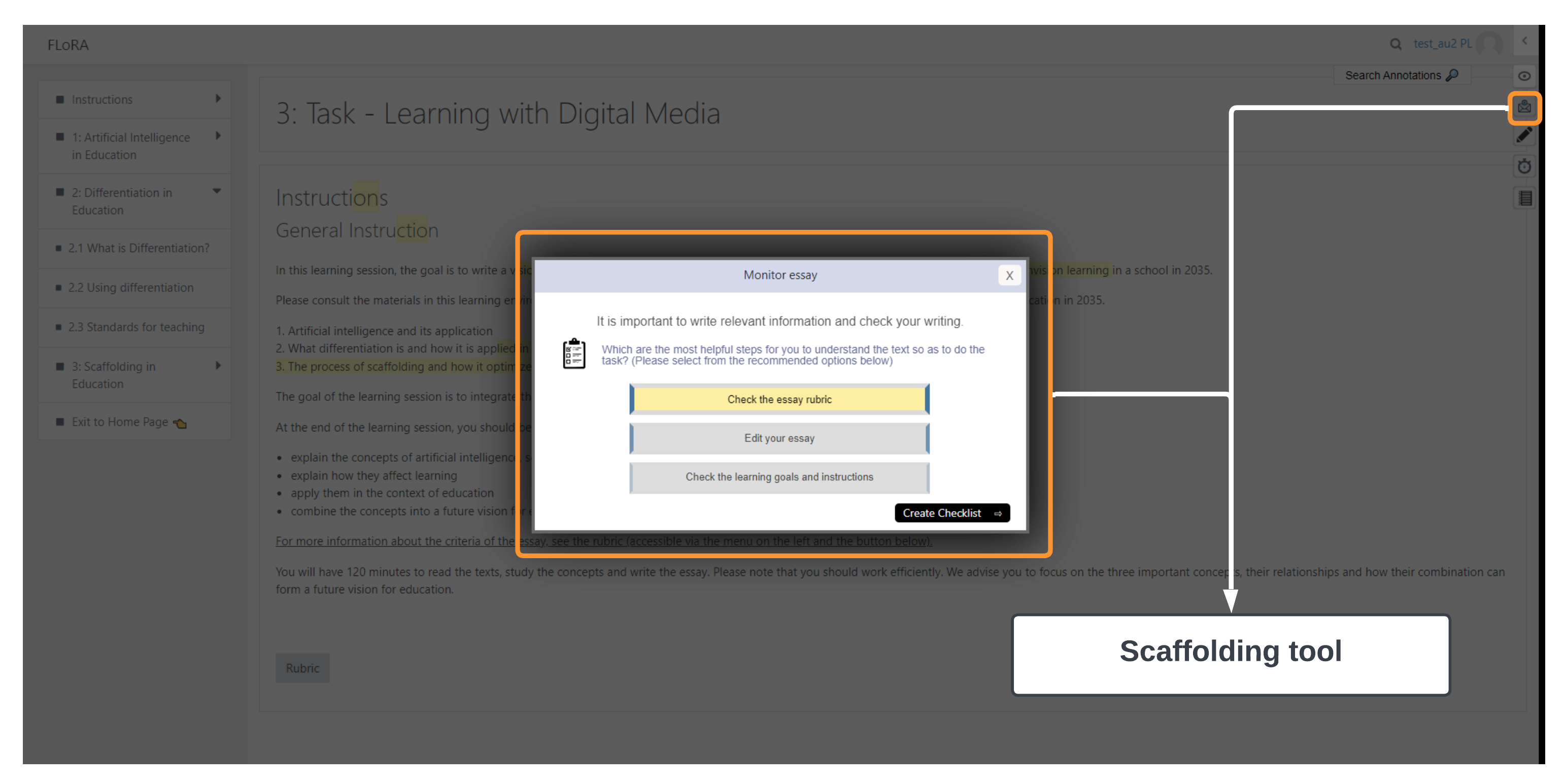}
\caption{Scaffolding display tool}
\label{fig:scaffolding_tools}
\end{figure}

The scaffolding display tool, as depicted in Figure~\ref{fig:scaffolding_tools}, is designed to activate upon the fulfilment of predefined conditions, detailed in Table~\ref{tab:content_of_scaffolding_example} and section~\ref{subsec:scaffolding}. This tool is presented as a pop-up window set against a shadowed backdrop, providing learners with the opportunity to select various options for the creation of a to-do list. Significantly, the tool's design allows researchers to customise both the timing for the scaffold's activation and its content through modifications in the backend configuration file. This adaptability ensures that the tool can be tailored to meet specific research requirements and learning contexts.

The scaffolding tool is intricately connected to the scaffolding module. The scaffolding module determines the appropriate scaffold content to present to students based on the detection of various SRL processes. Moreover, all interactions with this tool are captured and recorded. Given the recognised deficiency in SRL proficiency among many learners, this scaffolding display tool, in conjunction with the scaffolding module, has been specifically developed to provide more effective and targeted assistance, tailored to the needs of individual learners. A comprehensive elaboration of the backend scaffolding module, including its design logic and applications, is presented in the subsequent Section~\ref{subsec:scaffolding}.

\subsection{Module II -- Trace parser}\label{subsec:trace_parser}


Within the FLoRA trace parser, a sophisticated protocol for the measurement of SRL processes has been established, utilising trace data. This protocol is underpinned by prior scholarly research~\citep{fan2022improving}, and a theoretical framework~\citep{bannert2007metakognition} has been rigorously developed to delineate specific SRL processes. To be more specific, the learning process is categorised into three major domains: metacognition (MC), low cognition (LC), and high cognition (HC). Each cognitive level encompasses distinct SRL processes, such as \textit{MC.Orientation}, \textit{MC.Planning}, \textit{MC.Monitoring}, \textit{MC.Evaluation}, \textit{LC.First-reading}, \textit{LC.Re-reading}, and \textit{HC.Elaboration/Organisation}. These processes are elaborated with detailed definitions in Appendix~\ref{sec:appendixB}. Besides, the protocol encompasses an action library and a process library, which are instrumental in translating raw trace data into meaningful learning actions and their corresponding SRL processes (a comprehensive library is available in Appendix~\ref{sec:appendixB}). 

The action library defines a total of 17 learning actions, which correlate with specific trace data, such as RELEVANT\_READING, WRITE\_ESSAY, and EDIT\_ANNOTATION. For example, the OPEN\_ESSAY label denotes the action of opening the writing tool. When students commence typing in the writing tool, all keystroke events are captured and mapped to the WRITE\_ESSAY action. Further information on the complete action definitions and the labelling procedure is provided in the section under Appendix~\ref{sec:appendixB}. 

\begin{figure}[ht]\centering 
\includegraphics[width=0.9\linewidth]{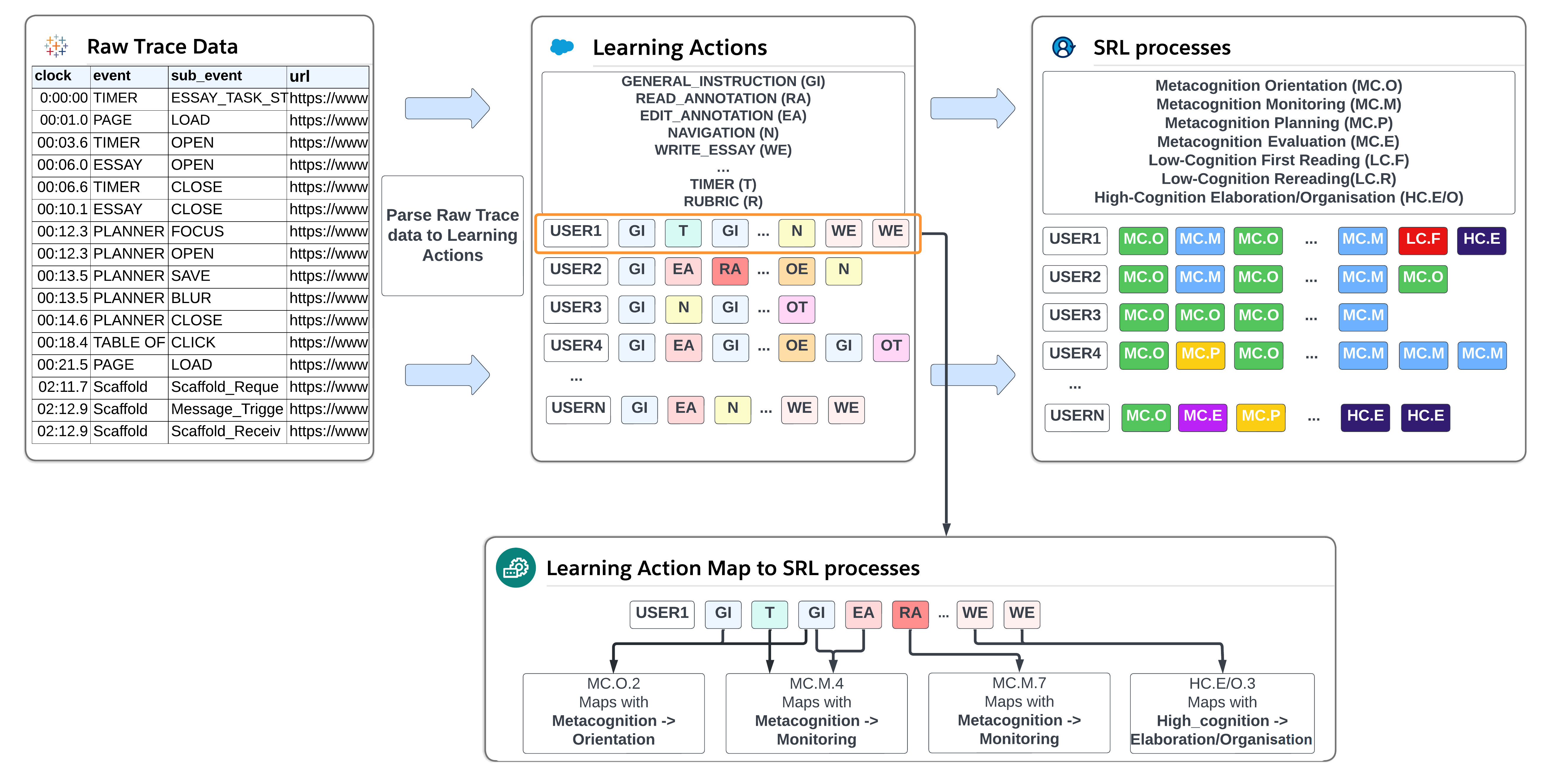}
\caption{Parsing trace data to SRL process}
\label{fig:trace_data_to_SRL}
\end{figure}

The process library employs the SRL process definitions from Appendix~\ref{sec:appendixB} as a framework to translate various actions (as described in the paragraph above) into SRL events (as depicted in Figure~\ref{fig:trace_data_to_SRL}). For example, when a learner highlights text or edits a note while surfing the general instruction page of the task, the actions are labelled as GENERAL\_INSTRUCTION and EDIT\_ANNOTATION, respectively. The combination of these actions forms a pattern "$GENERAL\_INSTRUCTION <-> EDIT\_ANNOTATION$", which, according to the process library, is mapped to \textit{MC.Orientation}. The symbol "$<->$" indicates that there is no sequential dependency between the actions GENERAL\_INSTRUCTION and EDIT\_ANNOTATION, allowing for flexibility in the order of these actions. In another scenario, when a learner engages with the timer to check the time, this action is labelled as TIMER and categorised under \textit{MC.Monitoring}, signifying the learner's effort to monitor the remaining time for the task. In the analysis of trace data, all items labelled as NO\_PROCESS are excluded, as they do not correspond to any meaningful sequences in the process library. Finally, if a learner does not interact with the FLoRA tools or learning systems for a certain period of time (5 minutes defined in FLoRA. This time is decided based on our previous studies~\citep{fan2022towards} and can be changed in the config file), an action named OFF\_TASK is recorded. In summary, through the integration of the action and process libraries, all trace logs are systematically transformed into labels indicative of the SRL processes.

To validate the protocol, both trace data and think-aloud data are collected and used. Two previous studies have been conducted to collect the think-aloud data. The first one is from 44 participants ($M_{age}=26.56years, SD_{age}=4.18years$) across various universities in Germany. The second one was conducted at a university in the Netherlands and involved 44 learners ($M_{age} = 21years, SD_{age} = 3years$). Participants joined the onsite session individually. The learning environment used is an older version of the FLoRA which also contains annotation tool, timer tool, planner tool, and writing tool. Throughout the task, participants are required to read and think aloud. Whenever they were silent for more than 5 seconds or spoke quietly, they were prompted verbally by the experimenter. More details can be found in our prior work in~\citep{lim2021temporal}. The trace data was collected via the FLoRA engine. 

After collecting all the data, the validation process is achieved through an innovative methodology that effectively merges both theory-driven and data-driven approaches, as proposed in prior research~\citep{fan2022towards}. The theory-driven approach starts with the development of a comprehensive coding scheme based on an established theoretical framework of SRL. This involves labelling raw trace data into meaningful learning actions and categorising these actions into SRL processes according to the Bannert's framework~\citep{bannert2007metakognition}. Multiple rounds of expert discussions are conducted to refine the process library and ensure its alignment with theoretical constructs. On the other hand, the data-driven approach leverages empirical evidence by aligning coded think-aloud data with trace data. This involves segmenting trace data based on think-aloud codes and using process mining techniques to identify dominant action patterns within each segment. These patterns are then interpreted and validated against the think-aloud data to construct a data-driven process library. The final step involves integrating empirical evidence with theoretical rationale to resolve any ambiguities and finalise the interpretation of action patterns, resulting in an improved process library. This integrated methodology ensures a robust and validated measurement of SRL processes by combining the strengths of both theoretical and empirical approaches.

This methodology establishes three pivotal alignments between trace data and think-aloud data. The first alignment contrasted the theory-driven process library with coded think aloud data, highlighting matches, mismatches, and unique aspects of each data type, thereby validating the SRL processes extracted from trace data. The second alignment, adopting a data-driven approach, refined the process library by directly aligning learning actions with think aloud data, utilising process mining techniques for more accurate mapping between SRL processes and think aloud codes. Finally, the third alignment involved an empirical evaluation of this improved process library against think aloud data from both training and test sets. This comprehensive approach ensured a deeper and more nuanced understanding of the SRL processes, aligning them more closely with the learners' cognitive and metacognitive strategies as expressed in their think aloud data, thereby significantly enhancing the validity of the measurement. In addition, the protocol employs a quantitative indicator, termed the match rate, to assess the degree of correspondence between SRL processes ascertained from trace data and those derived from think-aloud data. This match rate is further augmented by three supplementary metrics: sensitivity, specificity, and trace coverage. Each of these metrics plays a vital role in providing a comprehensive and robust evaluation of the protocol’s overall validity.

\subsection{Module III -- Scaffolding}\label{subsec:scaffolding}

The scaffolding module constitutes a pivotal component in the FLoRA engine. The overall design of scaffolding was informed by SRL theory~\citep{winne1998studying} and supplemented with Multimedia Learning theory~\citep{clark2023learning}. Within this module, scaffolds are crafted and implemented to provide as-needed-based assistance. With the advanced instrumentation tools and trace parser, log data are collected and processed to detect learners' SRL processes in real-time. The triggering mechanism of these scaffolds is complexly designed to be contingent upon the learner's real-time detected SRL processes \citep{maier2022personalized}.

Informed by previous empirical studies, the scaffolding module incorporates five distinct scaffolds at the moment. This decision is supported by findings from our previous lab studies which indicate that, during a two-hour reflective writing task, the use of five distinct scaffolds — each serving a specific purpose and introduced at designated intervals — maximises learners' overall performance~\citep{van2023design}. However, it was also noted that an excessive number of scaffolds could potentially impede students' learning progress during their learning activities. Consequently, these findings advocate for a moderated use of scaffolds, specifically the incorporation of five, to optimise educational outcomes~\citep{van2022dynamics}.


In terms of the content, each scaffold has a generalised and personalised version. The generalised scaffolds uniformly present identical messages and options to all students, whereas the personalised scaffolds are tailored to individual learners’ needs, based on real-time detection of their SRL processes. Each scaffold has a specific purpose. As shown in Table~\ref{tab:content_of_scaffolding_example}, the first (Orientation) scaffold is aligned with the initial stages of SRL where learners analyse the task and set their learning goals. This scaffold can help learners understand the task requirements, clarify objectives, and establish clear and achievable goals. The second (Start Reading) scaffold corresponds to the phase where learners select and implement strategies for understanding and assimilating information. This scaffold can provide guidance on effective reading techniques, how to identify key information, and methods for active engagement with the text. The third scaffold (Monitor Reading) aligns with the metacognitive aspect of SRL, specifically monitoring one's understanding and effectiveness during the reading process. It can help learners assess their comprehension, identify areas of confusion, and decide whether to continue with the current reading strategy or modify it. The fourth scaffold (Start Writing) maps to the phase where learners implement strategies to perform the task, in this case, writing. This scaffold can assist in organising thoughts, structuring the writing, and applying effective writing strategies. The fifth scaffold (Monitor Writing) is aligned with the self-reflection phase of SRL. It supports learners in evaluating their written work, reflecting on the writing process, and making necessary adjustments. It can guide learners in revising their work, ensuring it meets the objectives, and adjusting strategies for future tasks. 

These scaffolds are aligned with the SRL cycle definition~\citep{winne1998studying, zimmerman2000attaining}. Each scaffold provides four options from which learners can choose to create a to-do list that can help them plan, take, and monitor further learning actions. In our design, we recommend learners to read before writing. This is because our previous field study has confirmed that learners with different adoption of learning strategies demonstrated significant differences in medium effect size regarding the transfer test, with significant differences between ‘read first then write’ and ‘write intensively, read selectively’ strategies~\citep{srivastava2022effects}. As such, the current FLoRA design used in reflective writing tasks, both our lab study and field study confirmed that ‘writing after reading’ should be promoted. This is not to say that FLoRA can only be used for reflective writing tasks which promote ‘writing after reading’ strategy, but rather emphasising that scaffolding design should be contextualised and empirically evidenced so that to support effective SRL processes.

\begin{table}[]
\centering
\caption{Example of five scaffolds that are presently implemented in FLoRA: content of suggested actions and examples of times when scaffolds are triggered in some of the previous studies where tasks were 45min long and tasks focused on essay writing based on a given set of readings}
\label{tab:content_of_scaffolding_example}
\resizebox{\textwidth}{!}{%
\begin{tabular}{p{4.8cm} p{5.1cm} p{5.4cm}}
\hline
\textbf{Targeted perspective and Timing} & \multicolumn{1}{c}{\textbf{Prompt message}} & \multicolumn{1}{c}{\textbf{Learning suggestions}} \\ \addlinespace
\midrule
\textbf{\begin{tabular}[c]{@{}l@{}}Scaffold 1 - Orientation \\ (2nd minute)\end{tabular}} & Accurate understanding of the content and requirements of literacy task is critical. Based on your learning behaviour so far, we recommend the following steps: & (a) Use table of contents to get an overview and skim text; (b) Check the essay rubric carefully; (c) Make sure you understand the learning goals and instructions; (d) Process information by taking notes. \\ \addlinespace
\midrule
\textbf{\begin{tabular}[c]{@{}l@{}}Scaffold 2 - Start reading \\ (7th minute)\end{tabular}} & Efficient and high-quality reading of information on different topics in the material is essential. Based on your learning behaviour so far, we recommend the following steps: & (a) Note down important information; (b) Select what to read next using the table of contents; (c) Check time and monitor your reading progress; (d) Search for (specific) information. \\ \addlinespace
\midrule
\textbf{\begin{tabular}[c]{@{}l@{}}Scaffold 3 - Monitor reading \\ (16th minute)\end{tabular}} & Make sure you only read task-related pages and think about the relationship between reading and writing is the key to learning success. Based on your learning behaviour so far, we recommend the following steps: & (a) Review annotations and check what you have learned so far; (b) Review the learning goals and instructions to focus on relevant content; (c) Check your essay structure to determine what to read next; (d) Check the essay rubric. \\ \addlinespace
\midrule
\textbf{\begin{tabular}[c]{@{}l@{}}Scaffold 4 - Start writing \\ (21st minute)\end{tabular}} & Starting writing early and conscientiously writing a high-quality essay is central to the success of this task. Based on your learning behaviour so far, we recommend the following steps: & (a) Check the remaining time for writing; (b) Check and re-read the essay rubric; (c) Draft essay by using your own language and transferring learning to main points; (d) Write the essay with help from notes. \\ \addlinespace
\midrule
\textbf{\begin{tabular}[c]{@{}l@{}}Scaffold 5 - Monitor writing \\ (35th minute)\end{tabular}} & At the end of the task, complete the essay according to the task instructions and rubric to get a high score. Based on your learning behaviour so far, we recommend the following steps: & (a) Check the essay rubric to revise your essay; (b) Edit your essay to make sure it's complete; (c) Check the learning goals and instructions to avoid digress; (d) Check the timer to manage your time. \\ \addlinespace
\midrule
\end{tabular}%
}

\end{table}

As for triggering of scaffolding, the FLoRA engine employs a request/response communication model~\citep{oluwatosin2014client}, where the client dispatches a scaffold request encapsulating client parameters (user identification, scaffolding condition, elapsed time) at predetermined scaffold intervals. Subsequently, the scaffolding module conducts an analysis of trace actions, responding with a scaffold imbued with pertinent message text and options. In existing research, FLoRA has been mostly used in studies that support tasks requiring students to produce essays based on a given set of readings. In that context, the triggering of scaffolding was informed by the work of \citet{fan2022towards}. Their study utilised learners' think-aloud data to identify when a specific SRL process should be executed, with learners categorised into three groups based on their essay score (poor/average/good). All the essays are marked by two expert researchers. The scoring passed the reliability test -- inter-annotator agreement, measured using Cohen's kappa coefficient~\citep{cohen1960coefficient}. For example, compared to the average and poor essay groups, the high-performing essay group tended to begin the orientation process earlier and completed it sooner than the other two groups. This finding suggests the importance of prompting orientation processes at the early stages of the task. Subsequently, based on how the good essay group learned compared to the others, in a 45 minutes task, the scaffolds trigger timing is determined: orientation at 2 minutes since the start of the task; reading at 7 minutes; monitoring of reading at 16 minutes; writing at 21 minutes; and monitoring of writing at 35 minutes. For learners for whom the language they are working on the task is not their first language, the scaffold triggering time can be adjusted accordingly. Table \ref{tab:content_of_scaffolding_example} lists the contents of the five scaffolds that are have been used in the FLoRA engine for essay writing tasks with a given set of reading materials. As the FLoRA engine is designed to be easily reusable, the scaffold triggering time and the content of each scaffold is adjustable for different studies. Furthermore, we repeatedly collect learners' opinions (through post-task interviews) regarding the perceived usefulness and adaptivity of scaffolding to understand how the scaffolding content can be further improved~\citet{li2023learners}.

\begin{figure}[ht]\centering 
\includegraphics[width=0.88\linewidth]{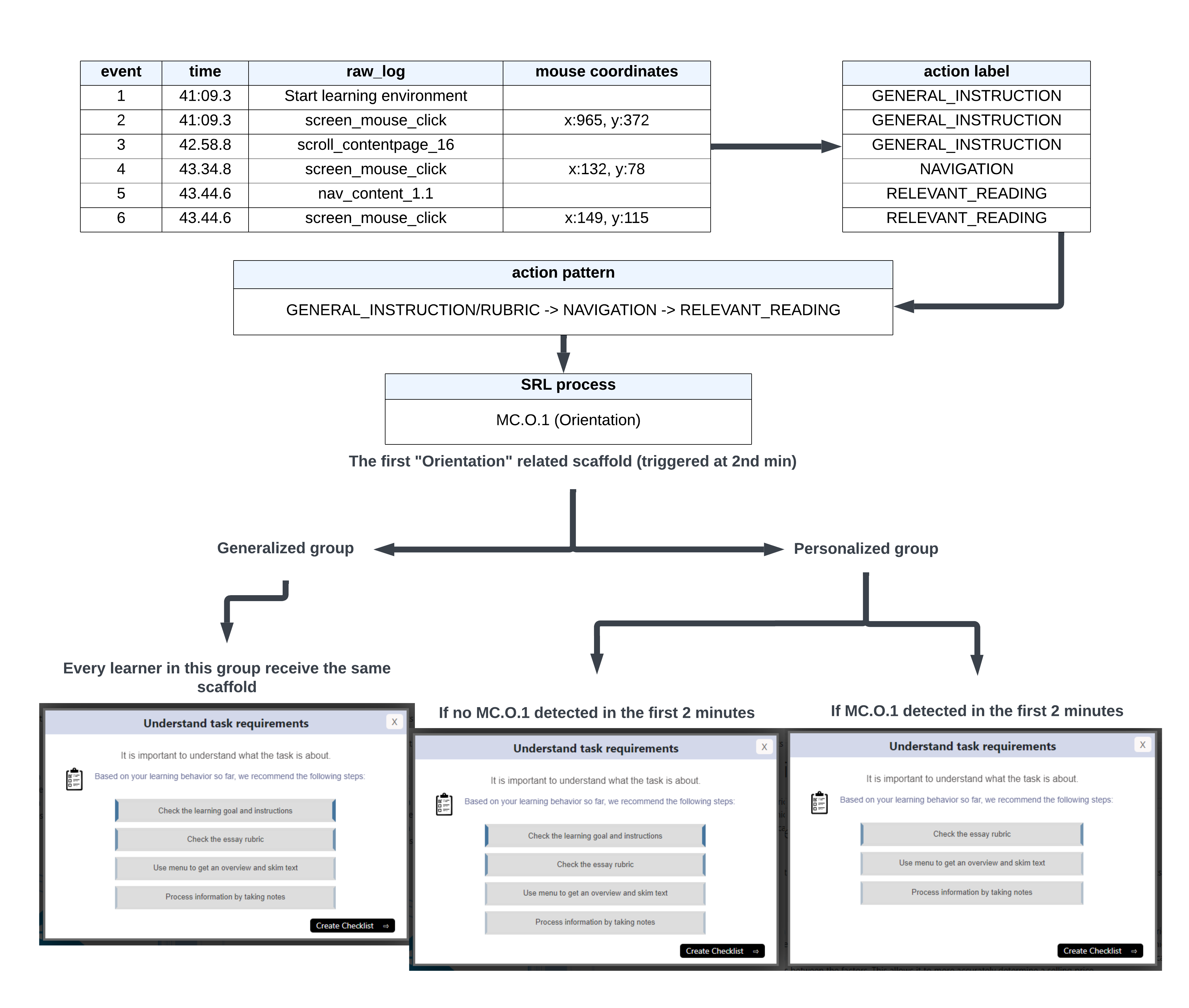}
\caption{Data parsing example and one scaffold rule}
\label{fig:trigger_scaffolding}
\end{figure}

An example of the scaffolding triggering mechanism for both generalised and personalised groups at the two-minute time point is illustrated in Figure~\ref{fig:trigger_scaffolding}. This representation underscores the dependence of the prompting process on three critical factors: the student's progression (time elapsed during the task), study condition (generalised or personalised), and, in the case of personalised scaffolding, the student's identity to authenticate individual action labels. According to our scaffolding framework, Orientation-related scaffolds are activated upon the student reaching the two-minute threshold in the writing activity. For students in the generalised condition, the scaffold content for Orientation (including message text and all four options) is sourced from the database, formatted as a JSON object, and transmitted to the client browser for display. In the context of personalised scaffolds, the module assesses the learner's SRL process concerning Scaffold 1 in Table~\ref{tab:content_of_scaffolding_example} before generating the corresponding message and options. For instance, in Figure~\ref{fig:trigger_scaffolding}, at the two-minute mark, a learner exhibiting adequate engagement (exceeding 15 seconds) with the general instruction page and demonstrating reading behaviours (a sequence categorised as \textit{ORIENTATION - MC.O.1} in the process library, involving "$GENERAL\_INSTRUCTION -> NAVIGATION -> RELEVANT\_READING$"), which causes the deactivation of the option to reorient (i.e., "Check the learning goal and instruction") in the first Personalised scaffold. This leaves the learner with only three options. Should all SRL processes be detected, all options within the scaffold are deactivated, leading to the omission of the entire scaffold. This contrasts with generalised scaffolds, where users receive all scaffolds complete with all options, regardless of their divergent learning behaviours.

\section{Implementation}
\label{sec:implementation}
This section explores the practical implementation of the FLoRA engine across diverse learning activities, utilising advanced and scalable technologies. Developed in accordance with the architecture outlined in the previous section, the FLoRA engine is available for download on GitHub (https://github.com/Xinyu-Li/flora-lms) under the BSD-2-Clause license. As an open-source software, FLoRA aims to significantly contribute to the advancement of SRL research. The subsequent sections clarify several key aspects of FLoRA's deployment and technological choices, specifically: 1) the FLoRA Engine, 2) the Log Server, 3) the Infrastructure, and 4) Case Studies.



\subsection{FLoRA Engine}
The FLoRA engine is designed for integration into various LMSs, such as Moodle. In alignment with this integration, its interface has been crafted to resonate with the thematic elements of the LMSs and can be easily adjusted for different LMSs. The primary interface elements presented to students are the instrumentation tools as well as the LMS interface (i.e. web pages). These tools are predominantly developed using JavaScript, supplemented with PHP programming, ensuring a seamless and efficient user experience. In this integrated environment, both learners and educators engage not only with the instrumentation tools but also with the broader LMSs' interface. This strategic design decision significantly reduces the learning curve for users, as LMSs such as Moodle are extensively utilised across educational institutions globally. Within this ecosystem, the LMSs serve as repositories and management systems for educational content created by teachers. As learners engage with these materials, the FLoRA instrumentation tools unobtrusively record all interaction logs to a back-end database.


The technological choices made in the development of FLoRA underscore its commitment to leveraging robust, widely recognised open-source tools to enhance functionality and scalability. By integrating the Hypothesis\footnote{https://github.com/hypothesis} JavaScript library, FLoRA benefits from a well-established platform for collaborative annotation and interaction within learning materials. The annotation search tool employs ElasticSearch~\citep{gormley2015elasticsearch} as its back-end service provider, offering powerful, real-time search and analytics capabilities that are essential for handling large datasets and ensuring quick, accurate information retrieval. The timer tool is built using the open-source FlipDown.js\footnote{https://github.com/PButcher/flipdown} library, which provides a visually appealing and highly customisable countdown timer. Additionally, the planner tool is developed with the JQuery-UI library, known for its rich set of user interface interactions, effects, widgets, and themes built on top of the jQuery JavaScript library. These technological choices not only enhance the performance and user experience of FLoRA but also ensure that it remains adaptable and scalable, facilitating continuous improvements and contributions from the open-source community.

\subsection{Log Server}
\begin{figure}[ht]\centering 
\includegraphics[width=0.75\linewidth]{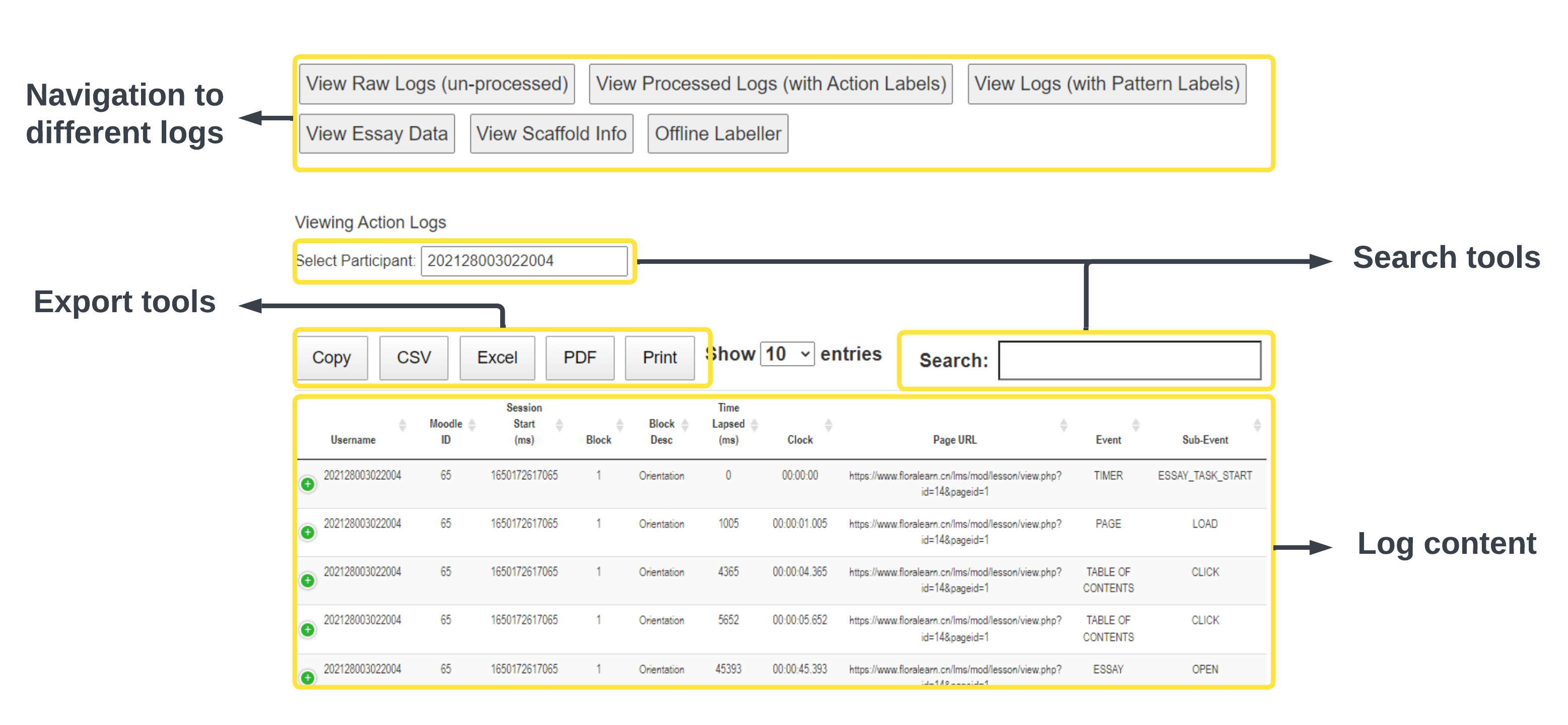}
\caption{Log server interface}
\label{fig:view_logs}
\end{figure}

An advanced log portal has been established to facilitate researchers' access to and retrieval of student trace data. This log interface, as illustrated in Figure~\ref{fig:view_logs}, comprises a well-structured layout. The header section of the interface serves as a navigation area, offering links to various categories of logs, encompassing both processed and raw (unprocessed) data. The processed logs are presented with action labels and process labels, as delineated in the section~\ref{subsec:trace_parser} and Appendix~\ref{sec:appendixB}, offering a clear and systematic view of the data. A search tool embedded within the interface enables the efficient location of logs through participant ID or specific keywords. This enhances the accessibility and usability of the log data for research purposes. The lower segment of the page is dedicated to displaying the logs, providing a comprehensive view of the data collected. Additionally, the portal is equipped with an export tool, allowing for the straightforward extraction of all log data. While it is feasible to retrieve student logs directly from the backend database, this approach necessitates advanced technical skills and poses a potential risk to the integrity of the original data. Therefore, the implementation of the log portal serves as a vital measure to safeguard the raw data against inadvertent modification or damage, ensuring the preservation of data integrity for research purposes.

\subsection{Infrastructure}
\begin{figure}[ht]\centering 
\includegraphics[width=0.75\linewidth]{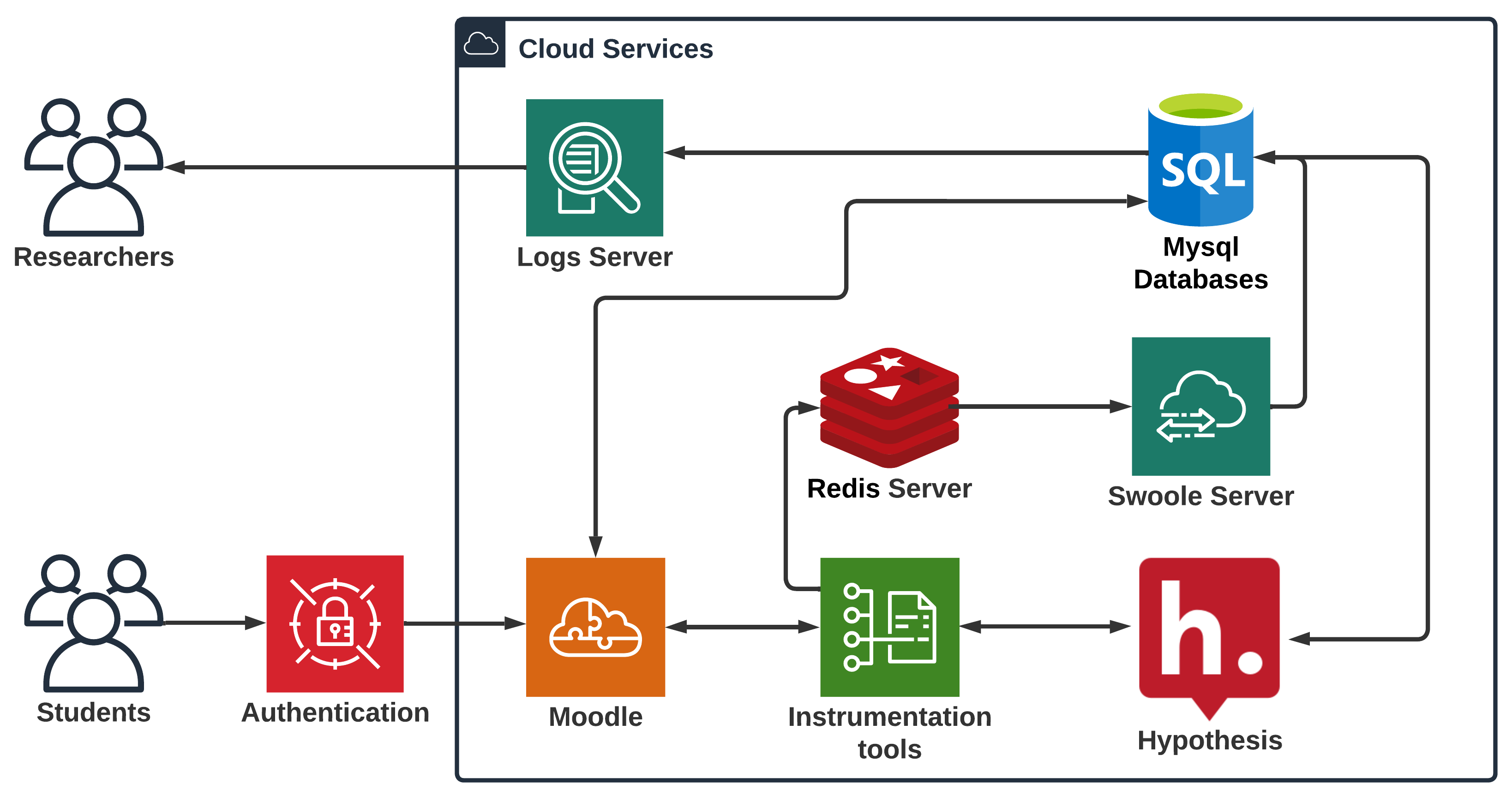}
\caption{FLoRA engine infrastructure}
\label{fig:flora_software_infrastructure}
\end{figure}

The FLoRA engine, now implemented in over ten universities worldwide and serving a significant student demographic, requires a robust, scalable, and reliable infrastructure (as displayed in Figure~\ref{fig:flora_software_infrastructure}). As previous described, the architecture comprises three core modules: instrumentation tools, trace parser, and scaffolding. Instrumentation tools are developed using JavaScript, HTML, and CSS, enhanced by key JavaScript libraries such as JQuery and Hypothesis. Conversely, the trace parser and scaffolding module are crafted with PHP, aligning with modern programming norms to ensure reliability and robustness. The trace parser and scaffolding module are deployed on an Nginx server, known for its excellent performance, security, and scalability. This infrastructure choice highlights a strategic commitment to optimising the engine's functionality and user experience, ensuring FLoRA's effective and stable operation in educational settings. 

In addressing the challenges posed by the substantial volume of trace data generated during user interaction with FLoRA –- including keystrokes, mouse movements, clicks, and URL visits –- it becomes evident that a mere fast database is insufficient. To effectively manage this influx of data, the engine employs two additional services, Redis and Swoole, as middleware systems to facilitate the communication between the front-end and the database. 

Redis serves as an open-source in-memory data store, functioning as a data message queue. Apart from Redis, other common caching middleware includes Memcached, Ehcache, Apache Ignite, and Hazelcast. Each has its own strengths and weaknesses. Memcached is simple and high-performing but lacks data persistence and supports only basic data types. Ehcache integrates well with Java and supports data persistence, but its distributed caching features are less mature. Apache Ignite offers in-memory computing and SQL support but is complex to configure. Hazelcast provides strong distributed caching and rich data structures but requires significant memory and can be complex to set up. Redis stands out due to its support for diverse data structures, high performance, data persistence, robust distributed capabilities, and a rich feature set, making it a preferred choice for many applications. Its large community and extensive ecosystem further enhance its flexibility and usability in various scenarios. This configuration allows for the efficient caching of data prior to its insertion into MySQL database.

Meanwhile, the Swoole server, functioning as a HTTP server, is utilised for the asynchronous processing of database insertions. There are some other similar including ReactPHP, Workerman, and Amp. ReactPHP is known for its event-driven architecture and simplicity but lacks built-in coroutine support, leading to potential performance bottlenecks. Workerman offers high performance and ease of use but is not as feature-rich and lacks native coroutine support. Amp provides a powerful asynchronous framework with good coroutine support but has a steeper learning curve and less community support. Swoole distinguishes itself due to its built-in coroutine support, high performance, and extensive feature set, including an integrated HTTP server, WebSocket server, task workers, and timer support. Its robust ecosystem and active community make it a superior choice for building scalable, high-performance PHP applications. This multi-faceted approach to data management underscores the FLoRA engine's adaptability and capacity to handle extensive data loads with efficiency.

As illustrated in Figure~\ref{fig:flora_software_infrastructure}, the operational flow within the FLoRA engine initiates with the aggregation of user-generated trace data at the front-end interface. In a strategic move to circumvent the inefficiencies associated with direct database interactions, all collected data is initially routed to a Redis server, utilising a queue data structure. This server temporarily stores the data in memory, thus optimising data handling processes. Subsequently, the Swoole server is employed to execute asynchronous tasks, which include retrieving data from the Redis server and conducting batch insertions into the MySQL database. The integration of this middleware architecture significantly enhances the overall system performance, ensuring both speed and efficiency in data processing. Furthermore, the FLoRA engine, in conjunction with the Learning Management System (LMS), boasts a versatile deployment capability. It can be easily deployed into various server environments, including but not limited to, Amazon Web Services (AWS), Alibaba Cloud, and Microsoft Azure. This flexibility in deployment options not only extends the system's accessibility to a broader range of educational institutions but also allows for customisation based on specific infrastructural and resource requirements. The FLoRA engine's ability to adapt to different server environments underscores its suitability for a wide array of educational contexts and its potential for global application.

\subsection{Case Studies}

The FLoRA engine was validated in laboratory settings and implemented as supplementary material in various authentic courses at universities in Germany, the Netherlands, and China. For example, in several studies, the task involved a 45-minute learning session during which students were required to study three topics: (1) artificial intelligence, focusing on its basic concepts; (2) differentiation in the classroom, which addresses how teachers can manage differences among learners and the concept of adaptive learning; and (3) scaffolding, an essential method for supporting learners during the learning process. The learning content comprised 5,237 words and six figures. A text readability analysis was conducted to ensure the material was appropriately challenging for students~\citep{lim2021temporal}. The learning task required students to integrate the three topics into an essay of 300-400 words, describing learning in schools in 2035. The participants were undergraduate university students. In these sessions, trace data (lab and authentic classroom settings) and think-aloud data (lab settings only) were collected. Several key research findings have emerged from these studies: (1) instrumentation tools are effective in capturing SRL, as validated by think-aloud data~\citep{van2021instrumentation};  (2) a novel approach that combines theory-driven and data-driven perspectives to validate that the SRL processes extracted from trace data are promising and effective~\citep{fan2022towards}; (3) guidelines for designing and evaluating personalised scaffolds for SRL was developed~\citep{van2023design}; (4) students often diverge from the plans they make at the start of their learning sessions and this inconsistency can be attributed to influences on regulatory processes shaped not only by pre-existing learning conditions (e.g., learning resources) but also by dynamic contextual changes as learners proceed with the task~\citep{li2024analytics}; and (5) while the personalised scaffolds induced SRL activities, they had no significant effect on learning performance~\citep{lim2023effects}. Further research related to the FLoRA studies is ongoing, with more papers expected to be published in the future.

Additional studies have been conducted, modifying the reading materials to include content provided by course coordinators and tutors. This content extends beyond the scope of the original tasks that were used during the project in which the FLoRA engine was developed. These studies aimed to investigate the applicability of FLoRA in diverse learning contexts. For example, reading materials are changed to “Academic English Writing” for a postgraduate course in China and the task length was changed to 2 hours, while in Australia, Finland, Germany and the Netherlands, reading materials are included “AI in education”, “AI in medicine” and “Biology” and the task time was limited to 60min for secondary school students (12-15 year olds) who used FLoRA in their classrooms. These studies also enabled us to compare SRL processes between secondary and higher education students, which is an overlooked topic in the literature~\citep{cheng2025self}. Educators and researchers have provided valuable feedback on the use of the FLoRA engine, emphasising its capacity to unobtrusively and automatically measure and diagnose competency levels in large-scale student cohorts. This capability is primarily attributed to the integration of the FLoRA engine with Moodle. Courses and studies that frequently utilise FLoRA often incorporate supplementary quizzes or questionnaires designed to gather learners’ pre- and post-study knowledge and collect learners’ feedback. These quizzes or questionnaires were set up in the Moodle environment. The FLoRA engine can access the Moodle database in the backend to retrieve the results of these quizzes or questionnaires, enabling the use of data collected by Moodle LMS. This functionality of FLoRA has the potential to facilitate personalised scaffolding by incorporating data-driven insights and that students follow the guidance provided by such personalised scaffolding~\citep{lim2024students}. Ongoing research is focused on examining how FLoRA influences learning outcomes across different learning contexts and whether the SRL processes vary during learning~\citep{li2024analytics,lim2024students}.

\section{Key Comparisons and Limitations}
The FLoRA engine propels the advancement of research in SRL through its specialised design, which enables the detection, measurement, and facilitation of students’ SRL processes across a wide range of learning activities. This paper provides a detailed illustration of the design logic and theoretical foundation underlying all the instrumentation tools within FLoRA. Previous sections include examples demonstrating how the FLoRA engine has been utilised in various studies, with research outcomes confirming the utility of these instrumentation tools in capturing SRL activities. For instance, the timer was used for monitoring, the highlighter and note-taker were associated with both low and high cognitive activities, and the search and planner tools were primarily used for planning and monitoring, aligning well with their intended purposes~\citep{van2021instrumentation}.

\subsection{Comparison Analysis}


For the analogous SRL support systems/engines, including nStudy~\citep{winne2019nstudy}, MetaTutor~\citep{azevedo2022lessons}, and Betty's Brain~\citep{munshi2023analysing}, a comparative analysis is conducted to evaluate them across various dimensions such as program portability and reusability, the theoretical model/framework employed, and the instrumentation tools supported.

Program portability is a critical aspect where FLoRA and nStudy demonstrate superior adaptability. FLoRA’s design, as a suite of JavaScript components, allows for seamless integration with other LMSs or learning environments. Similarly, nStudy~\citep{winne2019nstudy}, functioning as a Google Chrome plugin, is versatile and can operate across various websites. In contrast, MetaTutor~\citep{azevedo2022lessons} and Betty’s Brain~\citep{munshi2023analysing} are comprehensive, self-contained systems tailored for specific tasks, with integrated instrumentation tools. This design necessitates the use of the entire system, which may present a learning curve for new users. However, with FLoRA and nStudy, learners can continue working within familiar environments, easing the transition and enhancing user experience.

Regarding reusability, FLoRA stands out due to its high customisability. It employs a configuration file allowing modifications to tools’ content, scaffolding content, scaffolding triggering timing, and task duration. Different studies can utilise the FLoRA engine by incorporating simple configuration files, obviating the need for program modification. Similarly, nStudy~\citep{winne2019nstudy} also offers tool content customisation. MetaTutor~\citep{azevedo2022lessons} and Betty’s Brain~\citep{munshi2023analysing} can also be employed in various learning scenarios; however, they do not offer a straightforward method for configuring all support instrumentation tools for reuse (such as modifying tool content, enabling or disabling tools, altering scaffolding content, adjusting the timing for scaffolding triggers, and changing task duration) and may require additional development. All systems support textual and image resources, but only FLoRA and nStudy~\citep{winne2019nstudy} extend support to video resources.

Theoretical underpinnings are essential for SRL engines, with FLoRA informed by~\citet{bannert2007metakognition} framework, while the others align with Winne’s and Hadwin’s COPES model~\citep{winne1998studying}. The choice of theoretical underpinning  influences the development of supported tools, ensuring they capture the corresponding SRL processes effectively. FLoRA, in particular, encompasses a comprehensive set of tools, as detailed in Table \ref{tbl:sys-comparison}, sufficient for capturing a range of SRL processes. However, specific tools like the bookmark tool may become redundant in environments with a navigation area, as in the case of FLoRA. If the learning task requires using a large volume of websites or web pages, the bookmark tool would be useful. Additionally, the concept map tool, designed for specific tasks, is not commonly employed in reading and writing tasks.

In summary, the comparison of these SRL support systems reveals that while each has its strengths and unique features, FLoRA exhibits greater portability and reusability, making it more adaptable to various learning environments and tasks. The strength of other systems will be implemented in future development, such as converting the FLoRA engine to a browser plugin and supporting more tools like the concept map tool.

\begin{table}[]
\centering
\caption{Comparison between different SRL support engine/systems}
\label{tbl:sys-comparison}
\begin{tabular}{|p{1cm}|p{2.55cm}|p{3.1cm}|p{3.1cm}|p{2.5cm}|p{2.5cm}|}
  \hline
  \multicolumn{2}{|c|}{} & \textbf{FLoRA} & \textbf{nStudy} & \textbf{MetaTutor} & \textbf{Betty's Brain}  \\
  \hline
  \multicolumn{2}{|p{3.55cm}|}{\textbf{Portability}}  & Compatible with various LMSs  & Compatible with any websites  & Limited for specific tasks & Limited for specific tasks  \\
  \hline
  \multicolumn{2}{|p{3.55cm}|}{\textbf{Reusability}}  & Configurable for tools content, scaffolding content, scaffolding trigger time and task length & Configurable for tools content & Development required & Development required  \\
  \hline
  \multicolumn{2}{|p{3.55cm}|}{\textbf{Support Resources}}  & Textual, images, video & Textual, images, video & Textual, images & Textual, images  \\
  \hline
  \multicolumn{2}{|p{3.55cm}|}{\textbf{Theoretical Model/Framework}}  & \citet{bannert2007metakognition} & \citet{winne1998studying} & \citet{winne1998studying} & \citet{winne1998studying}   \\
  \hline
  \multirow{9}{1cm}{\textbf{Support Tools}} & Bookmark Tool & no & \textbf{\large{yes}} & no & no  \\
  \cline{2-6}
   & Annotation Tool & \textbf{\large{yes}} & \textbf{\large{yes}} & \textbf{\large{yes}} & \textbf{\large{yes}}  \\
  \cline{2-6}
   & Search Tool & \textbf{\large{yes}} & \textbf{\large{yes}} & no & no  \\
  \cline{2-6}
   & Timer Tool & \textbf{\large{yes}} & no & \textbf{\large{yes}} & no  \\
  \cline{2-6}
   & Planner Tool & \textbf{\large{yes}} & no & no & no  \\
  \cline{2-6}
   & Writing Tool & \textbf{\large{yes}} & \textbf{\large{yes}} & no & no  \\
  \cline{2-6}
   & Scaffolding Tool & \textbf{\large{yes}} & no & \textbf{\large{yes}} & \textbf{\large{yes}}  \\
  \cline{2-6}
   & Discussion Tool & \textbf{\large{yes}} & \textbf{\large{yes}} & no & no  \\
  \cline{2-6}
   & Concept Map Tool & no & no & no & \textbf{\large{yes}}  \\
  \hline
\end{tabular}
\end{table}

\subsection{FLoRA Engine Limitations}
The current limitations of the FLoRA engine have been identified through our previous research outcomes, primarily stemming from design constraints that will be addressed in future development. For example, the search function and planner were underutilised, possibly due to the structured nature of the learning materials and the relatively short duration of the tasks. Besides, the FLoRA engine did not have functionality to analyse the text selected for highlights or notes, which could provide deeper insights into processes and constructs of relevance to SRL. Furthermore, FLoRA currently lacks the capability to assess learners’ academic backgrounds and prior knowledge.

The trace parser in the FLoRA engine employs Bannert's~\citeyearpar{bannert2007metakognition} framework to categorise SRL processes. Future research could benefit from incorporating additional models, such as COPEs~\citep{winne1998studying}, to analyse the same trace data. This comparative approach would facilitate a more comprehensive evaluation of the effectiveness of different models in recognising SRL processes, potentially leading to more nuanced understandings of SRL dynamics and enable theory development.

The current scaffolding support in FLoRA has limitations in terms of generalisability. The existing design could not easily be applied to other studies without integrating the specific learning context into the scaffolding framework. The scaffolding design is evidence-based and must be tailored to meet the unique needs of different learning contexts. Although the FLoRA engine can be conveniently configured for tasks of varying duration and scaffold content, the distinctive characteristics of each learning condition must be considered. This includes determining the appropriate time intervals for scaffold triggers and the specific content to be displayed to learners.

In summary, while the FLoRA engine has the potential to significantly promote SRL, its current limitations necessitate further attention. Addressing these limitations through future research and development will be essential for enhancing the engine’s effectiveness and applicability across diverse learning contexts.

\section{Conclusion}

The FLoRA engine offers useful tools for researchers, enabling the acquisition of comprehensive data on students learning behaviours. By examining this data, researchers can link learning stages to metacognitive and cognitive actions, providing critical insights into student development in SRL. Furthermore, FLoRA offers customised and diminishing scaffolding to support the evolving competence of learners, promoting autonomy. To enhance collaborative and open resource sharing in the academic community, FLoRA is available under an open-source license, allowing for extensive customisation across various educational settings and tasks.

Looking ahead, the development agenda for the FLoRA engine is poised to undergo significant expansion, particularly in its data collection capabilities. This expansion is envisioned to be realised through the integration of additional, sophisticated tools, such as video recording tool, academic writing advisor tool, grammar check tool, and collaborative writing tool, enhancing the engine's ability to gather and analyse educational data. The video recording tool will capture students' facial expressions, allowing for the use of computer vision techniques to determine whether students are focused on their learning tasks and to identify signs of confusion. This capability will provide valuable insights into students' engagement and cognitive states, which are critical factors in understanding their SRL processes. The academic writing advisor tool will offer detailed academic writing suggestions, helping students to improve their writing skills. By providing real-time feedback and tailored advice, this tool aims to enhance students' ability to produce well-structured and coherent academic texts. The grammar check tool will automatically identify and suggest corrections for grammatical errors in students' writing. This tool will support students in refining their writing, ensuring clarity, and maintaining high linguistic standards. The collaborative writing tool will facilitate writing tasks in collaborative learning environments. By enabling multiple students to work together on writing assignments, this tool will support the development of collaborative writing skills and foster a deeper understanding of tasks through peer interaction. The primary objective of integrating these tools is to enhance the detection and analysis of students' SRL processes, as well as to monitor their development across various task scenarios. By leveraging these sophisticated tools, the FLoRA engine aims to provide educators and researchers with deeper insights into how students manage and regulate their learning strategies over time. This integration not only enhances the ability to assess the effectiveness of different pedagogical approaches but also enables the customisation of learning experiences to better meet individual student needs. 

In addition to these developments, a significant focus of future development will be the incorporation of chatbots (e.g., ChatGPT) based on large language models into the FLoRA engine. This integration represents a pioneering step in investigating the intricate relationship between a human's SRL and artificial intelligence~\citep{gavsevic2023empowering}. The primary objectives of this initiative include exploring whether AI-powered agents can match or surpass human teachers in effectiveness, understanding the nuances of human-AI interaction for SRL, and evaluating the potential of ChatGPT in enhancing SRL~\citep{jarvela2023human}.

\phantomsection

\section*{Declaration of Conflicting Interest} 

\addcontentsline{toc}{section}{Declaration of Conflicting Interest} 

The author(s) declared no potential conflicts of interest with respect to the research, authorship, and/or publication of this article.

\section*{Funding} 


FLoRA is funded by DFG, ESRC and NWO as part of the Open Research Area (Call 5) under grant number BA 2044/10-1 | GA 2739/1-1 | MO 2698/1-1. This research is also funded partially by the Australian Government through the Australian Research Council (project number DP240100069 and DP220101209) and the Jacobs Foundation (CELLA 2 CERES).
\phantomsection
\bibliography{main}

\begin{thebibliography}{}

\bibitem [\protect \citeauthoryear {%
Ahmad~Uzir%
\ \protect \BOthers {.}}{%
Ahmad~Uzir%
\ \protect \BOthers {.}}{%
{\protect \APACyear {2019}}%
}]{%
ahmad2019discovering}
\APACinsertmetastar {%
ahmad2019discovering}%
\begin{APACrefauthors}%
Ahmad~Uzir, N.%
, Ga{\v{s}}evi{\'c}, D.%
, Matcha, W.%
, Jovanovi{\'c}, J.%
, Pardo, A.%
, Lim, L\BHBI A.%
\BCBL {}\ \BBA {} Gentili, S.%
\end{APACrefauthors}%
\unskip\
\newblock
\APACrefYearMonthDay{2019}{}{}.
\newblock
{\BBOQ}\APACrefatitle {Discovering time management strategies in learning processes using process mining techniques} {Discovering time management strategies in learning processes using process mining techniques}.{\BBCQ}
\newblock
\BIn{} \APACrefbtitle {European Conference on Technology Enhanced Learning} {European conference on technology enhanced learning}\ (\BPGS\ 555--569).
\newblock
\APAChowpublished {\href{https://doi.org/10.1007/978-3-030-29736-7_41}{\color{blue} https://doi.org/10.1007/978-3-030-29736-7\_41}}.
\PrintBackRefs{\CurrentBib}

\bibitem [\protect \citeauthoryear {%
Alexiou%
\ \BBA {} Paraskeva%
}{%
Alexiou%
\ \BBA {} Paraskeva%
}{%
{\protect \APACyear {2015}}%
}]{%
alexiou2015managing}
\APACinsertmetastar {%
alexiou2015managing}%
\begin{APACrefauthors}%
Alexiou, A.%
\BCBT {}\ \BBA {} Paraskeva, F.%
\end{APACrefauthors}%
\unskip\
\newblock
\APACrefYearMonthDay{2015}{}{}.
\newblock
{\BBOQ}\APACrefatitle {Managing time through a self-regulated oriented ePortfolio for undergraduate students} {Managing time through a self-regulated oriented eportfolio for undergraduate students}.{\BBCQ}
\newblock
\BIn{} \APACrefbtitle {Design for Teaching and Learning in a Networked World: 10th European Conference on Technology Enhanced Learning, EC-TEL 2015, Toledo, Spain, September 15-18, 2015, Proceedings 10} {Design for teaching and learning in a networked world: 10th european conference on technology enhanced learning, ec-tel 2015, toledo, spain, september 15-18, 2015, proceedings 10}\ (\BPGS\ 547--550).
\newblock
\APAChowpublished {\href{https://doi.org/10.1007/978-3-319-24258-3_56}{\color{blue} https://doi.org/10.1007/978-3-319-24258-3\_56}}.
\PrintBackRefs{\CurrentBib}

\bibitem [\protect \citeauthoryear {%
Alvarez%
, Jivet%
, P{\'e}rez-Sanagustin%
, Scheffel%
\BCBL {}\ \BBA {} Verbert%
}{%
Alvarez%
\ \protect \BOthers {.}}{%
{\protect \APACyear {2022}}%
}]{%
alvarez2022tools}
\APACinsertmetastar {%
alvarez2022tools}%
\begin{APACrefauthors}%
Alvarez, R\BPBI P.%
, Jivet, I.%
, P{\'e}rez-Sanagustin, M.%
, Scheffel, M.%
\BCBL {}\ \BBA {} Verbert, K.%
\end{APACrefauthors}%
\unskip\
\newblock
\APACrefYearMonthDay{2022}{}{}.
\newblock
{\BBOQ}\APACrefatitle {Tools designed to support self-regulated learning in online learning environments: A systematic review} {Tools designed to support self-regulated learning in online learning environments: A systematic review}.{\BBCQ}
\newblock
\APACjournalVolNumPages{IEEE Transactions on Learning Technologies}{15}{4}{508--522}.
\newblock
\APAChowpublished {\href{https://doi.org/10.1109/TLT.2022.3193271}{\color{blue} https://doi.org/10.1109/TLT.2022.3193271}}.
\PrintBackRefs{\CurrentBib}

\bibitem [\protect \citeauthoryear {%
Azevedo%
\ \protect \BOthers {.}}{%
Azevedo%
\ \protect \BOthers {.}}{%
{\protect \APACyear {2022}}%
}]{%
azevedo2022lessons}
\APACinsertmetastar {%
azevedo2022lessons}%
\begin{APACrefauthors}%
Azevedo, R.%
, Bouchet, F.%
, Duffy, M.%
, Harley, J.%
, Taub, M.%
, Trevors, G.%
\BDBL {}others%
\end{APACrefauthors}%
\unskip\
\newblock
\APACrefYearMonthDay{2022}{}{}.
\newblock
{\BBOQ}\APACrefatitle {Lessons learned and future directions of metatutor: leveraging multichannel data to scaffold self-regulated learning with an intelligent tutoring system} {Lessons learned and future directions of metatutor: leveraging multichannel data to scaffold self-regulated learning with an intelligent tutoring system}.{\BBCQ}
\newblock
\APACjournalVolNumPages{Frontiers in Psychology}{13}{}{}.
\newblock
\APAChowpublished {\href{https://doi.org/10.3389/fpsyg.2022.813632}{\color{blue} https://doi.org/10.3389/fpsyg.2022.813632}}.
\PrintBackRefs{\CurrentBib}

\bibitem [\protect \citeauthoryear {%
Azevedo%
\ \protect \BOthers {.}}{%
Azevedo%
\ \protect \BOthers {.}}{%
{\protect \APACyear {2011}}%
}]{%
azevedo2011metatutor}
\APACinsertmetastar {%
azevedo2011metatutor}%
\begin{APACrefauthors}%
Azevedo, R.%
, Bouchet, F.%
, Harley, J\BPBI M.%
, Feyzi-Behnagh, R.%
, Trevors, G.%
, Duffy, M.%
\BDBL {}others%
\end{APACrefauthors}%
\unskip\
\newblock
\APACrefYearMonthDay{2011}{}{}.
\newblock
{\BBOQ}\APACrefatitle {MetaTutor: An Intelligent Multi-Agent Tutoring System Designed to Detect, Track, Model Foster Self-Regulated Learning} {Metatutor: An intelligent multi-agent tutoring system designed to detect, track, model foster self-regulated learning}.{\BBCQ}
\newblock
\BIn{} \APACrefbtitle {Proceedings of the Fourth Workshop on Self-Regulated Learning in Educational Technologies.} {Proceedings of the fourth workshop on self-regulated learning in educational technologies.}
\newblock
\APAChowpublished {\href{https://doi.org/10.13140/RG.2.1.1334.6640}{\color{blue} https://doi.org/10.13140/RG.2.1.1334.6640}}.
\PrintBackRefs{\CurrentBib}

\bibitem [\protect \citeauthoryear {%
Azevedo%
\ \BBA {} Cromley%
}{%
Azevedo%
\ \BBA {} Cromley%
}{%
{\protect \APACyear {2004}}%
}]{%
azevedo2004does}
\APACinsertmetastar {%
azevedo2004does}%
\begin{APACrefauthors}%
Azevedo, R.%
\BCBT {}\ \BBA {} Cromley, J\BPBI G.%
\end{APACrefauthors}%
\unskip\
\newblock
\APACrefYearMonthDay{2004}{}{}.
\newblock
{\BBOQ}\APACrefatitle {Does training on self-regulated learning facilitate students' learning with hypermedia?} {Does training on self-regulated learning facilitate students' learning with hypermedia?}{\BBCQ}
\newblock
\APACjournalVolNumPages{Journal of educational psychology}{96}{3}{523}.
\newblock
\APAChowpublished {\href{https://doi.org/10.1037/0022-0663.96.3.523}{\color{blue} https://doi.org/10.1037/0022-0663.96.3.523}}.
\PrintBackRefs{\CurrentBib}

\bibitem [\protect \citeauthoryear {%
Azevedo%
\ \BBA {} Ga{\v{s}}evi{\'c}%
}{%
Azevedo%
\ \BBA {} Ga{\v{s}}evi{\'c}%
}{%
{\protect \APACyear {2019}}%
}]{%
azevedo2019analyzing}
\APACinsertmetastar {%
azevedo2019analyzing}%
\begin{APACrefauthors}%
Azevedo, R.%
\BCBT {}\ \BBA {} Ga{\v{s}}evi{\'c}, D.%
\end{APACrefauthors}%
\unskip\
\newblock
\APACrefYearMonthDay{2019}{}{}.
\newblock
\APACrefbtitle {Analyzing multimodal multichannel data about self-regulated learning with advanced learning technologies: Issues and challenges} {Analyzing multimodal multichannel data about self-regulated learning with advanced learning technologies: Issues and challenges}\ (\BVOL~96).
\newblock
\APAChowpublished {\href{https://doi.org/10.1016/j.chb.2019.03.025}{\color{blue} https://doi.org/10.1016/j.chb.2019.03.025}}.
\PrintBackRefs{\CurrentBib}

\bibitem [\protect \citeauthoryear {%
Azevedo%
, Moos%
, Greene%
, Winters%
\BCBL {}\ \BBA {} Cromley%
}{%
Azevedo%
\ \protect \BOthers {.}}{%
{\protect \APACyear {2008}}%
}]{%
azevedo2008externally}
\APACinsertmetastar {%
azevedo2008externally}%
\begin{APACrefauthors}%
Azevedo, R.%
, Moos, D\BPBI C.%
, Greene, J\BPBI A.%
, Winters, F\BPBI I.%
\BCBL {}\ \BBA {} Cromley, J\BPBI G.%
\end{APACrefauthors}%
\unskip\
\newblock
\APACrefYearMonthDay{2008}{}{}.
\newblock
{\BBOQ}\APACrefatitle {Why is externally-facilitated regulated learning more effective than self-regulated learning with hypermedia?} {Why is externally-facilitated regulated learning more effective than self-regulated learning with hypermedia?}{\BBCQ}
\newblock
\APACjournalVolNumPages{Educational Technology Research and Development}{56}{1}{45--72}.
\newblock
\APAChowpublished {\href{https://doi.org/10.1007/s11423-007-9067-0}{\color{blue} https://doi.org/10.1007/s11423-007-9067-0}}.
\PrintBackRefs{\CurrentBib}

\bibitem [\protect \citeauthoryear {%
Bannert%
}{%
Bannert%
}{%
{\protect \APACyear {2007}}%
}]{%
bannert2007metakognition}
\APACinsertmetastar {%
bannert2007metakognition}%
\begin{APACrefauthors}%
Bannert, M.%
\end{APACrefauthors}%
\unskip\
\newblock
\APACrefYearMonthDay{2007}{}{}.
\newblock
{\BBOQ}\APACrefatitle {Metakognition beim Lernen mit Hypermedia} {Metakognition beim lernen mit hypermedia}.{\BBCQ}
\newblock
\APACjournalVolNumPages{Erfassung, Beschreibung und Vermittlung wirksamer metakognitiver Lernstrategien und Regulationsaktivit{\"a}ten. M{\"u}nster: Waxmann}{}{}{}.
\newblock
\APAChowpublished {\href{https://www.waxmann.com/?id=20&cHash=1&buchnr=1872}{\color{blue} https://www.waxmann.com/?id=20\&cHash=1\&buchnr=1872}}.
\PrintBackRefs{\CurrentBib}

\bibitem [\protect \citeauthoryear {%
Bannert%
}{%
Bannert%
}{%
{\protect \APACyear {2009}}%
}]{%
bannert2009promoting}
\APACinsertmetastar {%
bannert2009promoting}%
\begin{APACrefauthors}%
Bannert, M.%
\end{APACrefauthors}%
\unskip\
\newblock
\APACrefYearMonthDay{2009}{}{}.
\newblock
{\BBOQ}\APACrefatitle {Promoting self-regulated learning through prompts} {Promoting self-regulated learning through prompts}.{\BBCQ}
\newblock
\APACjournalVolNumPages{Zeitschrift f{\"u}r P{\"a}dagogische Psychologie}{23}{2}{139--145}.
\newblock
\APAChowpublished {\href{https://doi.org/10.1024/1010-0652.23.2.139}{\color{blue} https://doi.org/10.1024/1010-0652.23.2.139}}.
\PrintBackRefs{\CurrentBib}

\bibitem [\protect \citeauthoryear {%
Bannert%
\ \BBA {} Mengelkamp%
}{%
Bannert%
\ \BBA {} Mengelkamp%
}{%
{\protect \APACyear {2013}}%
}]{%
bannert2013scaffolding}
\APACinsertmetastar {%
bannert2013scaffolding}%
\begin{APACrefauthors}%
Bannert, M.%
\BCBT {}\ \BBA {} Mengelkamp, C.%
\end{APACrefauthors}%
\unskip\
\newblock
\APACrefYearMonthDay{2013}{}{}.
\newblock
{\BBOQ}\APACrefatitle {Scaffolding hypermedia learning through metacognitive prompts} {Scaffolding hypermedia learning through metacognitive prompts}.{\BBCQ}
\newblock
\BIn{} \APACrefbtitle {International handbook of metacognition and learning technologies} {International handbook of metacognition and learning technologies}\ (\BPGS\ 171--186).
\newblock
\APACaddressPublisher{}{Springer}.
\newblock
\APAChowpublished {\href{https://doi.org/10.1007/978-1-4419-5546-3_12}{\color{blue} https://doi.org/10.1007/978-1-4419-5546-3\_12}}.
\PrintBackRefs{\CurrentBib}

\bibitem [\protect \citeauthoryear {%
Bannert%
, Reimann%
\BCBL {}\ \BBA {} Sonnenberg%
}{%
Bannert%
\ \protect \BOthers {.}}{%
{\protect \APACyear {2014}}%
}]{%
bannert2014process}
\APACinsertmetastar {%
bannert2014process}%
\begin{APACrefauthors}%
Bannert, M.%
, Reimann, P.%
\BCBL {}\ \BBA {} Sonnenberg, C.%
\end{APACrefauthors}%
\unskip\
\newblock
\APACrefYearMonthDay{2014}{}{}.
\newblock
{\BBOQ}\APACrefatitle {Process mining techniques for analysing patterns and strategies in students’ self-regulated learning} {Process mining techniques for analysing patterns and strategies in students’ self-regulated learning}.{\BBCQ}
\newblock
\APACjournalVolNumPages{Metacognition and learning}{9}{2}{161--185}.
\newblock
\APAChowpublished {\href{https://doi.org/10.1007/s11409-013-9107-6}{\color{blue} https://doi.org/10.1007/s11409-013-9107-6}}.
\PrintBackRefs{\CurrentBib}

\bibitem [\protect \citeauthoryear {%
Biswas%
, Segedy%
\BCBL {}\ \BBA {} Bunchongchit%
}{%
Biswas%
\ \protect \BOthers {.}}{%
{\protect \APACyear {2016}}%
}]{%
biswas2016design}
\APACinsertmetastar {%
biswas2016design}%
\begin{APACrefauthors}%
Biswas, G.%
, Segedy, J\BPBI R.%
\BCBL {}\ \BBA {} Bunchongchit, K.%
\end{APACrefauthors}%
\unskip\
\newblock
\APACrefYearMonthDay{2016}{}{}.
\newblock
{\BBOQ}\APACrefatitle {From design to implementation to practice a learning by teaching system: Betty’s Brain} {From design to implementation to practice a learning by teaching system: Betty’s brain}.{\BBCQ}
\newblock
\APACjournalVolNumPages{International Journal of Artificial Intelligence in Education}{26}{1}{350--364}.
\newblock
\APAChowpublished {\href{https://doi.org/10.1007/s40593-015-0057-9}{\color{blue} https://doi.org/10.1007/s40593-015-0057-9}}.
\PrintBackRefs{\CurrentBib}

\bibitem [\protect \citeauthoryear {%
Bjork%
, Dunlosky%
\BCBL {}\ \BBA {} Kornell%
}{%
Bjork%
\ \protect \BOthers {.}}{%
{\protect \APACyear {2013}}%
}]{%
bjork2013self}
\APACinsertmetastar {%
bjork2013self}%
\begin{APACrefauthors}%
Bjork, R\BPBI A.%
, Dunlosky, J.%
\BCBL {}\ \BBA {} Kornell, N.%
\end{APACrefauthors}%
\unskip\
\newblock
\APACrefYearMonthDay{2013}{}{}.
\newblock
{\BBOQ}\APACrefatitle {Self-regulated learning: Beliefs, techniques, and illusions} {Self-regulated learning: Beliefs, techniques, and illusions}.{\BBCQ}
\newblock
\APACjournalVolNumPages{Annual review of psychology}{64}{}{417--444}.
\newblock
\APAChowpublished {\href{https://doi.org/10.1146/annurev-psych-113011-143823}{\color{blue} https://doi.org/10.1146/annurev-psych-113011-143823}}.
\PrintBackRefs{\CurrentBib}

\bibitem [\protect \citeauthoryear {%
Boekaerts%
\ \BBA {} Cascallar%
}{%
Boekaerts%
\ \BBA {} Cascallar%
}{%
{\protect \APACyear {2006}}%
}]{%
boekaerts2006far}
\APACinsertmetastar {%
boekaerts2006far}%
\begin{APACrefauthors}%
Boekaerts, M.%
\BCBT {}\ \BBA {} Cascallar, E.%
\end{APACrefauthors}%
\unskip\
\newblock
\APACrefYearMonthDay{2006}{}{}.
\newblock
{\BBOQ}\APACrefatitle {How far have we moved toward the integration of theory and practice in self-regulation?} {How far have we moved toward the integration of theory and practice in self-regulation?}{\BBCQ}
\newblock
\APACjournalVolNumPages{Educational psychology review}{18}{}{199--210}.
\newblock
\APAChowpublished {\href{https://doi.org/10.1007/s10648-006-9013-4}{\color{blue} https://doi.org/10.1007/s10648-006-9013-4}}.
\PrintBackRefs{\CurrentBib}

\bibitem [\protect \citeauthoryear {%
Bowman%
, Jang%
, Kivlighan%
, Schneider%
\BCBL {}\ \BBA {} Ye%
}{%
Bowman%
\ \protect \BOthers {.}}{%
{\protect \APACyear {2020}}%
}]{%
bowman2020impact}
\APACinsertmetastar {%
bowman2020impact}%
\begin{APACrefauthors}%
Bowman, N\BPBI A.%
, Jang, N.%
, Kivlighan, D\BPBI M.%
, Schneider, N.%
\BCBL {}\ \BBA {} Ye, X.%
\end{APACrefauthors}%
\unskip\
\newblock
\APACrefYearMonthDay{2020}{}{}.
\newblock
{\BBOQ}\APACrefatitle {The impact of a goal-setting intervention for engineering students on academic probation} {The impact of a goal-setting intervention for engineering students on academic probation}.{\BBCQ}
\newblock
\APACjournalVolNumPages{Research in Higher Education}{61}{}{142--166}.
\newblock
\APAChowpublished {\href{https://doi.org/10.1007/s11162-019-09555-x}{\color{blue} https://doi.org/10.1007/s11162-019-09555-x}}.
\PrintBackRefs{\CurrentBib}

\bibitem [\protect \citeauthoryear {%
Cerezo%
, Bogar{\'\i}n%
, Esteban%
\BCBL {}\ \BBA {} Romero%
}{%
Cerezo%
\ \protect \BOthers {.}}{%
{\protect \APACyear {2020}}%
}]{%
cerezo2020process}
\APACinsertmetastar {%
cerezo2020process}%
\begin{APACrefauthors}%
Cerezo, R.%
, Bogar{\'\i}n, A.%
, Esteban, M.%
\BCBL {}\ \BBA {} Romero, C.%
\end{APACrefauthors}%
\unskip\
\newblock
\APACrefYearMonthDay{2020}{}{}.
\newblock
{\BBOQ}\APACrefatitle {Process mining for self-regulated learning assessment in e-learning} {Process mining for self-regulated learning assessment in e-learning}.{\BBCQ}
\newblock
\APACjournalVolNumPages{Journal of Computing in Higher Education}{32}{1}{74--88}.
\newblock
\APAChowpublished {\href{https://doi.org/10.1007/s12528-019-09225-y}{\color{blue} https://doi.org/10.1007/s12528-019-09225-y}}.
\PrintBackRefs{\CurrentBib}

\bibitem [\protect \citeauthoryear {%
Cheng%
\ \protect \BOthers {.}}{%
Cheng%
\ \protect \BOthers {.}}{%
{\protect \APACyear {2025}}%
}]{%
cheng2025self}
\APACinsertmetastar {%
cheng2025self}%
\begin{APACrefauthors}%
Cheng, Y.%
, Guan, R.%
, Li, T.%
, Raković, M.%
, Li, X.%
, Fan, Y.%
\BDBL {}Swiecki, Z.%
\end{APACrefauthors}%
\unskip\
\newblock
\APACrefYearMonthDay{2025}{}{}.
\newblock
{\BBOQ}\APACrefatitle {Self-regulated Learning Processes in Secondary Education: A Network Analysis of Trace-based Measures} {Self-regulated learning processes in secondary education: A network analysis of trace-based measures}.{\BBCQ}
\newblock
\APACjournalVolNumPages{}{}{}{392--403}.
\PrintBackRefs{\CurrentBib}

\bibitem [\protect \citeauthoryear {%
Chi%
}{%
Chi%
}{%
{\protect \APACyear {2021}}%
}]{%
chi2021translating}
\APACinsertmetastar {%
chi2021translating}%
\begin{APACrefauthors}%
Chi, M\BPBI T.%
\end{APACrefauthors}%
\unskip\
\newblock
\APACrefYearMonthDay{2021}{}{}.
\newblock
{\BBOQ}\APACrefatitle {Translating a Theory of Active Learning: An Attempt to Close the Research-Practice Gap in Education} {Translating a theory of active learning: An attempt to close the research-practice gap in education}.{\BBCQ}
\newblock
\APACjournalVolNumPages{Topics in Cognitive Science}{13}{3}{441--463}.
\newblock
\APAChowpublished {\href{https://doi.org/10.1111/tops.12539}{\color{blue} https://doi.org/10.1111/tops.12539}}.
\PrintBackRefs{\CurrentBib}

\bibitem [\protect \citeauthoryear {%
Clarebout%
, Elen%
, Collazo%
, Lust%
\BCBL {}\ \BBA {} Jiang%
}{%
Clarebout%
\ \protect \BOthers {.}}{%
{\protect \APACyear {2013}}%
}]{%
clarebout2013metacognition}
\APACinsertmetastar {%
clarebout2013metacognition}%
\begin{APACrefauthors}%
Clarebout, G.%
, Elen, J.%
, Collazo, N\BPBI A\BPBI J.%
, Lust, G.%
\BCBL {}\ \BBA {} Jiang, L.%
\end{APACrefauthors}%
\unskip\
\newblock
\APACrefYearMonthDay{2013}{}{}.
\newblock
{\BBOQ}\APACrefatitle {Metacognition and the use of tools} {Metacognition and the use of tools}.{\BBCQ}
\newblock
\APACjournalVolNumPages{International handbook of metacognition and learning technologies}{}{}{187--195}.
\newblock
\APAChowpublished {\href{https://doi.org/10.1007/978-1-4419-5546-3_13}{\color{blue} https://doi.org/10.1007/978-1-4419-5546-3\_13}}.
\PrintBackRefs{\CurrentBib}

\bibitem [\protect \citeauthoryear {%
Clark%
\ \BBA {} Mayer%
}{%
Clark%
\ \BBA {} Mayer%
}{%
{\protect \APACyear {2023}}%
}]{%
clark2023learning}
\APACinsertmetastar {%
clark2023learning}%
\begin{APACrefauthors}%
Clark, R\BPBI C.%
\BCBT {}\ \BBA {} Mayer, R\BPBI E.%
\end{APACrefauthors}%
\unskip\
\newblock
\APACrefYear{2023}.
\newblock
\APACrefbtitle {E-learning and the science of instruction: Proven guidelines for consumers and designers of multimedia learning} {E-learning and the science of instruction: Proven guidelines for consumers and designers of multimedia learning}.
\newblock
\APACaddressPublisher{}{john Wiley \& sons}.
\newblock
\APAChowpublished {\href{https://doi.org/10.1002/9781119239086}{\color{blue} https://doi.org/10.1002/9781119239086}}.
\PrintBackRefs{\CurrentBib}

\bibitem [\protect \citeauthoryear {%
Cohen%
}{%
Cohen%
}{%
{\protect \APACyear {1960}}%
}]{%
cohen1960coefficient}
\APACinsertmetastar {%
cohen1960coefficient}%
\begin{APACrefauthors}%
Cohen, J.%
\end{APACrefauthors}%
\unskip\
\newblock
\APACrefYearMonthDay{1960}{}{}.
\newblock
{\BBOQ}\APACrefatitle {A coefficient of agreement for nominal scales} {A coefficient of agreement for nominal scales}.{\BBCQ}
\newblock
\APACjournalVolNumPages{Educational and psychological measurement}{20}{1}{37--46}.
\newblock
\APAChowpublished {\href{https://doi.org/10.1177/001316446002000104}{\color{blue} https://doi.org/10.1177/001316446002000104}}.
\PrintBackRefs{\CurrentBib}

\bibitem [\protect \citeauthoryear {%
Davis%
, Triglianos%
, Hauff%
\BCBL {}\ \BBA {} Houben%
}{%
Davis%
\ \protect \BOthers {.}}{%
{\protect \APACyear {2018}}%
}]{%
davis2018srlx}
\APACinsertmetastar {%
davis2018srlx}%
\begin{APACrefauthors}%
Davis, D.%
, Triglianos, V.%
, Hauff, C.%
\BCBL {}\ \BBA {} Houben, G\BHBI J.%
\end{APACrefauthors}%
\unskip\
\newblock
\APACrefYearMonthDay{2018}{}{}.
\newblock
{\BBOQ}\APACrefatitle {SRLx: A personalized learner interface for MOOCs} {Srlx: A personalized learner interface for moocs}.{\BBCQ}
\newblock
\BIn{} \APACrefbtitle {Lifelong Technology-Enhanced Learning: 13th European Conference on Technology Enhanced Learning, EC-TEL 2018, Leeds, UK, September 3-5, 2018, Proceedings 13} {Lifelong technology-enhanced learning: 13th european conference on technology enhanced learning, ec-tel 2018, leeds, uk, september 3-5, 2018, proceedings 13}\ (\BPGS\ 122--135).
\newblock
\APAChowpublished {\href{https://doi.org/10.1007/978-3-319-98572-5_10}{\color{blue} https://doi.org/10.1007/978-3-319-98572-5\_10}}.
\PrintBackRefs{\CurrentBib}

\bibitem [\protect \citeauthoryear {%
Devolder%
, van Braak%
\BCBL {}\ \BBA {} Tondeur%
}{%
Devolder%
\ \protect \BOthers {.}}{%
{\protect \APACyear {2012}}%
}]{%
devolder2012supporting}
\APACinsertmetastar {%
devolder2012supporting}%
\begin{APACrefauthors}%
Devolder, A.%
, van Braak, J.%
\BCBL {}\ \BBA {} Tondeur, J.%
\end{APACrefauthors}%
\unskip\
\newblock
\APACrefYearMonthDay{2012}{}{}.
\newblock
{\BBOQ}\APACrefatitle {Supporting self-regulated learning in computer-based learning environments: systematic review of effects of scaffolding in the domain of science education} {Supporting self-regulated learning in computer-based learning environments: systematic review of effects of scaffolding in the domain of science education}.{\BBCQ}
\newblock
\APACjournalVolNumPages{Journal of Computer Assisted Learning}{28}{6}{557--573}.
\newblock
\APAChowpublished {\href{http://dx.doi.org/10.1111/j.1365-2729.2011.00476.x}{\color{blue} http://dx.doi.org/10.1111/j.1365-2729.2011.00476.x}}.
\PrintBackRefs{\CurrentBib}

\bibitem [\protect \citeauthoryear {%
Di~Vesta%
\ \BBA {} Gray%
}{%
Di~Vesta%
\ \BBA {} Gray%
}{%
{\protect \APACyear {1973}}%
}]{%
di1973listening}
\APACinsertmetastar {%
di1973listening}%
\begin{APACrefauthors}%
Di~Vesta, F\BPBI J.%
\BCBT {}\ \BBA {} Gray, G\BPBI S.%
\end{APACrefauthors}%
\unskip\
\newblock
\APACrefYearMonthDay{1973}{}{}.
\newblock
{\BBOQ}\APACrefatitle {Listening and note taking: II. Immediate and delayed recall as functions of variations in thematic continuity, note taking, and length of listening-review intervals.} {Listening and note taking: Ii. immediate and delayed recall as functions of variations in thematic continuity, note taking, and length of listening-review intervals.}{\BBCQ}
\newblock
\APACjournalVolNumPages{Journal of educational psychology}{64}{3}{278}.
\newblock
\APAChowpublished {\href{https://doi.org/10.1037/h0034589}{\color{blue} https://doi.org/10.1037/h0034589}}.
\PrintBackRefs{\CurrentBib}

\bibitem [\protect \citeauthoryear {%
Fan%
, Lim%
\BCBL {}\ \protect \BOthers {.}}{%
Fan%
, Lim%
\BCBL {}\ \protect \BOthers {.}}{%
{\protect \APACyear {2022}}%
}]{%
fan2022improving}
\APACinsertmetastar {%
fan2022improving}%
\begin{APACrefauthors}%
Fan, Y.%
, Lim, L.%
, van~der Graaf, J.%
, Kilgour, J.%
, Rakovi{\'c}, M.%
, Moore, J.%
\BDBL {}Ga{\v{s}}evi{\'c}, D.%
\end{APACrefauthors}%
\unskip\
\newblock
\APACrefYearMonthDay{2022}{}{}.
\newblock
{\BBOQ}\APACrefatitle {Improving the measurement of self-regulated learning using multi-channel data} {Improving the measurement of self-regulated learning using multi-channel data}.{\BBCQ}
\newblock
\APACjournalVolNumPages{Metacognition and Learning}{}{}{1--31}.
\newblock
\APAChowpublished {\href{https://doi.org/10.1007/s11409-022-09304-z}{\color{blue} https://doi.org/10.1007/s11409-022-09304-z}}.
\PrintBackRefs{\CurrentBib}

\bibitem [\protect \citeauthoryear {%
Fan%
, Matcha%
, Uzir%
, Wang%
\BCBL {}\ \BBA {} Ga{\v{s}}evi{\'c}%
}{%
Fan%
\ \protect \BOthers {.}}{%
{\protect \APACyear {2021}}%
}]{%
fan2021learning}
\APACinsertmetastar {%
fan2021learning}%
\begin{APACrefauthors}%
Fan, Y.%
, Matcha, W.%
, Uzir, N\BPBI A.%
, Wang, Q.%
\BCBL {}\ \BBA {} Ga{\v{s}}evi{\'c}, D.%
\end{APACrefauthors}%
\unskip\
\newblock
\APACrefYearMonthDay{2021}{}{}.
\newblock
{\BBOQ}\APACrefatitle {Learning analytics to reveal links between learning design and self-regulated learning} {Learning analytics to reveal links between learning design and self-regulated learning}.{\BBCQ}
\newblock
\APACjournalVolNumPages{International Journal of Artificial Intelligence in Education}{31}{4}{980--1021}.
\newblock
\APAChowpublished {\href{https://doi.org/10.1007/s40593-021-00249-z}{\color{blue} https://doi.org/10.1007/s40593-021-00249-z}}.
\PrintBackRefs{\CurrentBib}

\bibitem [\protect \citeauthoryear {%
Fan%
\ \protect \BOthers {.}}{%
Fan%
\ \protect \BOthers {.}}{%
{\protect \APACyear {2023}}%
}]{%
fan2023towards}
\APACinsertmetastar {%
fan2023towards}%
\begin{APACrefauthors}%
Fan, Y.%
, Rakovic, M.%
, van Der~Graaf, J.%
, Lim, L.%
, Singh, S.%
, Moore, J.%
\BDBL {}Ga{\v{s}}evi{\'c}, D.%
\end{APACrefauthors}%
\unskip\
\newblock
\APACrefYearMonthDay{2023}{}{}.
\newblock
{\BBOQ}\APACrefatitle {Towards a fuller picture: Triangulation and integration of the measurement of self-regulated learning based on trace and think aloud data} {Towards a fuller picture: Triangulation and integration of the measurement of self-regulated learning based on trace and think aloud data}.{\BBCQ}
\newblock
\APACjournalVolNumPages{Journal of Computer Assisted Learning}{}{}{}.
\newblock
\APAChowpublished {\href{https://doi.org/10.1111/jcal.12801}{\color{blue} https://doi.org/10.1111/jcal.12801}}.
\PrintBackRefs{\CurrentBib}

\bibitem [\protect \citeauthoryear {%
Fan%
, van~der Graaf%
\BCBL {}\ \protect \BOthers {.}}{%
Fan%
, van~der Graaf%
\BCBL {}\ \protect \BOthers {.}}{%
{\protect \APACyear {2022}}%
}]{%
fan2022towards}
\APACinsertmetastar {%
fan2022towards}%
\begin{APACrefauthors}%
Fan, Y.%
, van~der Graaf, J.%
, Lim, L.%
, Rakovi{\'c}, M.%
, Singh, S.%
, Kilgour, J.%
\BDBL {}Ga{\v{s}}evi{\'c}, D.%
\end{APACrefauthors}%
\unskip\
\newblock
\APACrefYearMonthDay{2022}{}{}.
\newblock
{\BBOQ}\APACrefatitle {Towards investigating the validity of measurement of self-regulated learning based on trace data} {Towards investigating the validity of measurement of self-regulated learning based on trace data}.{\BBCQ}
\newblock
\APACjournalVolNumPages{Metacognition and Learning}{}{}{1--39}.
\newblock
\APAChowpublished {\href{https://doi.org/10.1007/s11409-022-09291-1}{\color{blue} https://doi.org/10.1007/s11409-022-09291-1}}.
\PrintBackRefs{\CurrentBib}

\bibitem [\protect \citeauthoryear {%
Fincham%
, Ga{\v{s}}evi{\'c}%
, Jovanovi{\'c}%
\BCBL {}\ \BBA {} Pardo%
}{%
Fincham%
\ \protect \BOthers {.}}{%
{\protect \APACyear {2018}}%
}]{%
fincham2018study}
\APACinsertmetastar {%
fincham2018study}%
\begin{APACrefauthors}%
Fincham, E.%
, Ga{\v{s}}evi{\'c}, D.%
, Jovanovi{\'c}, J.%
\BCBL {}\ \BBA {} Pardo, A.%
\end{APACrefauthors}%
\unskip\
\newblock
\APACrefYearMonthDay{2018}{}{}.
\newblock
{\BBOQ}\APACrefatitle {From study tactics to learning strategies: An analytical method for extracting interpretable representations} {From study tactics to learning strategies: An analytical method for extracting interpretable representations}.{\BBCQ}
\newblock
\APACjournalVolNumPages{IEEE Transactions on Learning Technologies}{12}{1}{59--72}.
\newblock
\APAChowpublished {\href{https://doi.org/10.1109/TLT.2018.2823317}{\color{blue} https://doi.org/10.1109/TLT.2018.2823317}}.
\PrintBackRefs{\CurrentBib}

\bibitem [\protect \citeauthoryear {%
Ga{\v{s}}evi{\'c}%
, Dawson%
\BCBL {}\ \BBA {} Siemens%
}{%
Ga{\v{s}}evi{\'c}%
\ \protect \BOthers {.}}{%
{\protect \APACyear {2015}}%
}]{%
gavsevic2015let}
\APACinsertmetastar {%
gavsevic2015let}%
\begin{APACrefauthors}%
Ga{\v{s}}evi{\'c}, D.%
, Dawson, S.%
\BCBL {}\ \BBA {} Siemens, G.%
\end{APACrefauthors}%
\unskip\
\newblock
\APACrefYearMonthDay{2015}{}{}.
\newblock
{\BBOQ}\APACrefatitle {Let’s not forget: Learning analytics are about learning} {Let’s not forget: Learning analytics are about learning}.{\BBCQ}
\newblock
\APACjournalVolNumPages{TechTrends}{59}{1}{64--71}.
\newblock
\APAChowpublished {\href{https://doi.org/10.1007/s11528-014-0822-x}{\color{blue} https://doi.org/10.1007/s11528-014-0822-x}}.
\PrintBackRefs{\CurrentBib}

\bibitem [\protect \citeauthoryear {%
Ga{\v{s}}evi{\'c}%
, Siemens%
\BCBL {}\ \BBA {} Sadiq%
}{%
Ga{\v{s}}evi{\'c}%
\ \protect \BOthers {.}}{%
{\protect \APACyear {2023}}%
}]{%
gavsevic2023empowering}
\APACinsertmetastar {%
gavsevic2023empowering}%
\begin{APACrefauthors}%
Ga{\v{s}}evi{\'c}, D.%
, Siemens, G.%
\BCBL {}\ \BBA {} Sadiq, S.%
\end{APACrefauthors}%
\unskip\
\newblock
\APACrefYearMonthDay{2023}{}{}.
\newblock
{\BBOQ}\APACrefatitle {Empowering learners for the age of artificial intelligence} {Empowering learners for the age of artificial intelligence}.{\BBCQ}
\newblock
\APACjournalVolNumPages{Computers and Education: Artificial Intelligence}{4}{}{100130}.
\newblock
\APAChowpublished {\href{https://doi.org/10.1016/j.caeai.2023.100130}{\color{blue} https://doi.org/10.1016/j.caeai.2023.100130}}.
\PrintBackRefs{\CurrentBib}

\bibitem [\protect \citeauthoryear {%
Gormley%
\ \BBA {} Tong%
}{%
Gormley%
\ \BBA {} Tong%
}{%
{\protect \APACyear {2015}}%
}]{%
gormley2015elasticsearch}
\APACinsertmetastar {%
gormley2015elasticsearch}%
\begin{APACrefauthors}%
Gormley, C.%
\BCBT {}\ \BBA {} Tong, Z.%
\end{APACrefauthors}%
\unskip\
\newblock
\APACrefYear{2015}.
\newblock
\APACrefbtitle {Elasticsearch: the definitive guide: a distributed real-time search and analytics engine} {Elasticsearch: the definitive guide: a distributed real-time search and analytics engine}.
\newblock
\APACaddressPublisher{}{" O'Reilly Media, Inc."}.
\newblock
\APAChowpublished {\href{https://doi.org/10.5555/2904394}{\color{blue} https://doi.org/10.5555/2904394}}.
\PrintBackRefs{\CurrentBib}

\bibitem [\protect \citeauthoryear {%
Greene%
\ \BBA {} Azevedo%
}{%
Greene%
\ \BBA {} Azevedo%
}{%
{\protect \APACyear {2009}}%
}]{%
greene2009macro}
\APACinsertmetastar {%
greene2009macro}%
\begin{APACrefauthors}%
Greene, J\BPBI A.%
\BCBT {}\ \BBA {} Azevedo, R.%
\end{APACrefauthors}%
\unskip\
\newblock
\APACrefYearMonthDay{2009}{}{}.
\newblock
{\BBOQ}\APACrefatitle {A macro-level analysis of SRL processes and their relations to the acquisition of a sophisticated mental model of a complex system} {A macro-level analysis of srl processes and their relations to the acquisition of a sophisticated mental model of a complex system}.{\BBCQ}
\newblock
\APACjournalVolNumPages{Contemporary educational psychology}{34}{1}{18--29}.
\newblock
\APAChowpublished {\href{https://doi.org/10.1016/j.cedpsych.2008.05.006}{\color{blue} https://doi.org/10.1016/j.cedpsych.2008.05.006}}.
\PrintBackRefs{\CurrentBib}

\bibitem [\protect \citeauthoryear {%
Guo%
}{%
Guo%
}{%
{\protect \APACyear {2022}}%
}]{%
guo2022using}
\APACinsertmetastar {%
guo2022using}%
\begin{APACrefauthors}%
Guo, L.%
\end{APACrefauthors}%
\unskip\
\newblock
\APACrefYearMonthDay{2022}{}{}.
\newblock
{\BBOQ}\APACrefatitle {Using metacognitive prompts to enhance self-regulated learning and learning outcomes: A meta-analysis of experimental studies in computer-based learning environments} {Using metacognitive prompts to enhance self-regulated learning and learning outcomes: A meta-analysis of experimental studies in computer-based learning environments}.{\BBCQ}
\newblock
\APACjournalVolNumPages{Journal of Computer Assisted Learning}{38}{3}{811--832}.
\newblock
\APAChowpublished {\href{https://doi.org/10.1111/jcal.12650}{\color{blue} https://doi.org/10.1111/jcal.12650}}.
\PrintBackRefs{\CurrentBib}

\bibitem [\protect \citeauthoryear {%
Hmelo-Silver%
\ \BBA {} Azevedo%
}{%
Hmelo-Silver%
\ \BBA {} Azevedo%
}{%
{\protect \APACyear {2006}}%
}]{%
hmelo2006understanding}
\APACinsertmetastar {%
hmelo2006understanding}%
\begin{APACrefauthors}%
Hmelo-Silver, C\BPBI E.%
\BCBT {}\ \BBA {} Azevedo, R.%
\end{APACrefauthors}%
\unskip\
\newblock
\APACrefYearMonthDay{2006}{}{}.
\newblock
{\BBOQ}\APACrefatitle {Understanding complex systems: Some core challenges} {Understanding complex systems: Some core challenges}.{\BBCQ}
\newblock
\APACjournalVolNumPages{The Journal of the learning sciences}{15}{1}{53--61}.
\newblock
\APAChowpublished {\href{https://doi.org/10.1207/s15327809jls1501_7}{\color{blue} https://doi.org/10.1207/s15327809jls1501\_7}}.
\PrintBackRefs{\CurrentBib}

\bibitem [\protect \citeauthoryear {%
Huang%
\ \BBA {} Lajoie%
}{%
Huang%
\ \BBA {} Lajoie%
}{%
{\protect \APACyear {2021}}%
}]{%
huang2021process}
\APACinsertmetastar {%
huang2021process}%
\begin{APACrefauthors}%
Huang, L.%
\BCBT {}\ \BBA {} Lajoie, S\BPBI P.%
\end{APACrefauthors}%
\unskip\
\newblock
\APACrefYearMonthDay{2021}{}{}.
\newblock
{\BBOQ}\APACrefatitle {Process analysis of teachers’ self-regulated learning patterns in technological pedagogical content knowledge development} {Process analysis of teachers’ self-regulated learning patterns in technological pedagogical content knowledge development}.{\BBCQ}
\newblock
\APACjournalVolNumPages{Computers \& Education}{166}{}{104169}.
\newblock
\APAChowpublished {\href{https://doi.org/10.1016/j.compedu.2021.104169}{\color{blue} https://doi.org/10.1016/j.compedu.2021.104169}}.
\PrintBackRefs{\CurrentBib}

\bibitem [\protect \citeauthoryear {%
J{\"a}rvel{\"a}%
, Nguyen%
\BCBL {}\ \BBA {} Hadwin%
}{%
J{\"a}rvel{\"a}%
\ \protect \BOthers {.}}{%
{\protect \APACyear {2023}}%
}]{%
jarvela2023human}
\APACinsertmetastar {%
jarvela2023human}%
\begin{APACrefauthors}%
J{\"a}rvel{\"a}, S.%
, Nguyen, A.%
\BCBL {}\ \BBA {} Hadwin, A.%
\end{APACrefauthors}%
\unskip\
\newblock
\APACrefYearMonthDay{2023}{}{}.
\newblock
{\BBOQ}\APACrefatitle {Human and artificial intelligence collaboration for socially shared regulation in learning} {Human and artificial intelligence collaboration for socially shared regulation in learning}.{\BBCQ}
\newblock
\APACjournalVolNumPages{British Journal of Educational Technology}{}{}{}.
\newblock
\APAChowpublished {\href{https://doi.org/10.1111/bjet.13325}{\color{blue} https://doi.org/10.1111/bjet.13325}}.
\PrintBackRefs{\CurrentBib}

\bibitem [\protect \citeauthoryear {%
Lajoie%
}{%
Lajoie%
}{%
{\protect \APACyear {2021}}%
}]{%
lajoie2021student}
\APACinsertmetastar {%
lajoie2021student}%
\begin{APACrefauthors}%
Lajoie, S\BPBI P.%
\end{APACrefauthors}%
\unskip\
\newblock
\APACrefYearMonthDay{2021}{}{}.
\newblock
{\BBOQ}\APACrefatitle {Student modeling for individuals and groups: The BioWorld and HOWARD platforms} {Student modeling for individuals and groups: The bioworld and howard platforms}.{\BBCQ}
\newblock
\APACjournalVolNumPages{International Journal of Artificial Intelligence in Education}{31}{3}{460--475}.
\newblock
\APAChowpublished {\href{https://doi.org/10.1007/s40593-020-00219-x}{\color{blue} https://doi.org/10.1007/s40593-020-00219-x}}.
\PrintBackRefs{\CurrentBib}

\bibitem [\protect \citeauthoryear {%
Lee%
, Watson%
\BCBL {}\ \BBA {} Watson%
}{%
Lee%
\ \protect \BOthers {.}}{%
{\protect \APACyear {2019}}%
}]{%
lee2019systematic}
\APACinsertmetastar {%
lee2019systematic}%
\begin{APACrefauthors}%
Lee, D.%
, Watson, S\BPBI L.%
\BCBL {}\ \BBA {} Watson, W\BPBI R.%
\end{APACrefauthors}%
\unskip\
\newblock
\APACrefYearMonthDay{2019}{}{}.
\newblock
{\BBOQ}\APACrefatitle {Systematic literature review on self-regulated learning in massive open online courses} {Systematic literature review on self-regulated learning in massive open online courses}.{\BBCQ}
\newblock
\APACjournalVolNumPages{Australasian Journal of Educational Technology}{35}{1}{}.
\newblock
\APAChowpublished {\href{https://doi.org/10.14742/ajet.3749}{\color{blue} https://doi.org/10.14742/ajet.3749}}.
\PrintBackRefs{\CurrentBib}

\bibitem [\protect \citeauthoryear {%
Li%
\ \protect \BOthers {.}}{%
Li%
\ \protect \BOthers {.}}{%
{\protect \APACyear {2024}}%
}]{%
li2024analytics}
\APACinsertmetastar {%
li2024analytics}%
\begin{APACrefauthors}%
Li, T.%
, Fan, Y.%
, Srivastava, N.%
, Zeng, Z.%
, Li, X.%
, Khosravi, H.%
\BDBL {}Ga{\v{s}}evi{\'c}, D.%
\end{APACrefauthors}%
\unskip\
\newblock
\APACrefYearMonthDay{2024}{}{}.
\newblock
{\BBOQ}\APACrefatitle {Analytics of Planning Behaviours in Self-Regulated Learning: Links with Strategy Use and Prior Knowledge} {Analytics of planning behaviours in self-regulated learning: Links with strategy use and prior knowledge}.{\BBCQ}
\newblock
\BIn{} \APACrefbtitle {Proceedings of the 14th Learning Analytics and Knowledge Conference} {Proceedings of the 14th learning analytics and knowledge conference}\ (\BPGS\ 438--449).
\newblock
\APAChowpublished {\href{https://doi.org/10.1145/3636555.3636900}{\color{blue} https://doi.org/10.1145/3636555.3636900}}.
\PrintBackRefs{\CurrentBib}

\bibitem [\protect \citeauthoryear {%
Li%
\ \protect \BOthers {.}}{%
Li%
\ \protect \BOthers {.}}{%
{\protect \APACyear {2023}}%
}]{%
li2023learners}
\APACinsertmetastar {%
li2023learners}%
\begin{APACrefauthors}%
Li, T.%
, Lin, J.%
, Iqbal, S.%
, Swiecki, Z.%
, Tsai, Y\BHBI S.%
, Fan, Y.%
\BCBL {}\ \BBA {} Ga{\v{s}}evi{\'c}, D.%
\end{APACrefauthors}%
\unskip\
\newblock
\APACrefYearMonthDay{2023}{}{}.
\newblock
{\BBOQ}\APACrefatitle {Do Learners Appreciate Adaptivity? An Epistemic Network Analysis of How Learners Perceive Adaptive Scaffolding} {Do learners appreciate adaptivity? an epistemic network analysis of how learners perceive adaptive scaffolding}.{\BBCQ}
\newblock
\BIn{} \APACrefbtitle {International Conference on Quantitative Ethnography} {International conference on quantitative ethnography}\ (\BPGS\ 3--17).
\newblock
\APAChowpublished {\href{https://doi.org/10.1007/978-3-031-47014-1_1}{\color{blue} https://doi.org/10.1007/978-3-031-47014-1\_1}}.
\PrintBackRefs{\CurrentBib}

\bibitem [\protect \citeauthoryear {%
Lim%
\ \protect \BOthers {.}}{%
Lim%
\ \protect \BOthers {.}}{%
{\protect \APACyear {2024}}%
}]{%
lim2024students}
\APACinsertmetastar {%
lim2024students}%
\begin{APACrefauthors}%
Lim, L.%
, Bannert, M.%
, van~der Graaf, J.%
, Fan, Y.%
, Rakovic, M.%
, Singh, S.%
\BDBL {}Ga{\v{s}}evi{\'c}, D.%
\end{APACrefauthors}%
\unskip\
\newblock
\APACrefYearMonthDay{2024}{}{}.
\newblock
{\BBOQ}\APACrefatitle {How do students learn with real-time personalized scaffolds?} {How do students learn with real-time personalized scaffolds?}{\BBCQ}
\newblock
\APACjournalVolNumPages{British Journal of Educational Technology}{55}{4}{1309--1327}.
\newblock
\APAChowpublished {\href{https://doi.org/10.1111/bjet.13414}{\color{blue} https://doi.org/10.1111/bjet.13414}}.
\PrintBackRefs{\CurrentBib}

\bibitem [\protect \citeauthoryear {%
Lim%
\ \protect \BOthers {.}}{%
Lim%
\ \protect \BOthers {.}}{%
{\protect \APACyear {2021}}%
}]{%
lim2021temporal}
\APACinsertmetastar {%
lim2021temporal}%
\begin{APACrefauthors}%
Lim, L.%
, Bannert, M.%
, van~der Graaf, J.%
, Molenaar, I.%
, Fan, Y.%
, Kilgour, J.%
\BDBL {}Ga{\v{s}}evi{\'c}, D.%
\end{APACrefauthors}%
\unskip\
\newblock
\APACrefYearMonthDay{2021}{}{}.
\newblock
{\BBOQ}\APACrefatitle {Temporal assessment of self-regulated learning by mining students’ think-aloud protocols} {Temporal assessment of self-regulated learning by mining students’ think-aloud protocols}.{\BBCQ}
\newblock
\APACjournalVolNumPages{Frontiers in Psychology}{12}{}{749749}.
\newblock
\APAChowpublished {\href{https://doi.org/10.3389/fpsyg.2021.749749}{\color{blue} https://doi.org/10.3389/fpsyg.2021.749749}}.
\PrintBackRefs{\CurrentBib}

\bibitem [\protect \citeauthoryear {%
Lim%
\ \protect \BOthers {.}}{%
Lim%
\ \protect \BOthers {.}}{%
{\protect \APACyear {2023}}%
}]{%
lim2023effects}
\APACinsertmetastar {%
lim2023effects}%
\begin{APACrefauthors}%
Lim, L.%
, Bannert, M.%
, van~der Graaf, J.%
, Singh, S.%
, Fan, Y.%
, Surendrannair, S.%
\BDBL {}Ga{\v{s}}evi{\'c}, D.%
\end{APACrefauthors}%
\unskip\
\newblock
\APACrefYearMonthDay{2023}{}{}.
\newblock
{\BBOQ}\APACrefatitle {Effects of real-time analytics-based personalized scaffolds on students’ self-regulated learning} {Effects of real-time analytics-based personalized scaffolds on students’ self-regulated learning}.{\BBCQ}
\newblock
\APACjournalVolNumPages{Computers in Human Behavior}{139}{}{107547}.
\newblock
\APAChowpublished {\href{https://doi.org/10.1016/j.chb.2022.107547}{\color{blue} https://doi.org/10.1016/j.chb.2022.107547}}.
\PrintBackRefs{\CurrentBib}

\bibitem [\protect \citeauthoryear {%
Maier%
\ \BBA {} Klotz%
}{%
Maier%
\ \BBA {} Klotz%
}{%
{\protect \APACyear {2022}}%
}]{%
maier2022personalized}
\APACinsertmetastar {%
maier2022personalized}%
\begin{APACrefauthors}%
Maier, U.%
\BCBT {}\ \BBA {} Klotz, C.%
\end{APACrefauthors}%
\unskip\
\newblock
\APACrefYearMonthDay{2022}{}{}.
\newblock
{\BBOQ}\APACrefatitle {Personalized feedback in digital learning environments: Classification framework and literature review} {Personalized feedback in digital learning environments: Classification framework and literature review}.{\BBCQ}
\newblock
\APACjournalVolNumPages{Computers and Education: Artificial Intelligence}{3}{}{100080}.
\newblock
\APAChowpublished {\href{https://doi.org/10.1016/j.caeai.2022.100080}{\color{blue} https://doi.org/10.1016/j.caeai.2022.100080}}.
\PrintBackRefs{\CurrentBib}

\bibitem [\protect \citeauthoryear {%
Marzouk%
\ \protect \BOthers {.}}{%
Marzouk%
\ \protect \BOthers {.}}{%
{\protect \APACyear {2016}}%
}]{%
marzouk2016if}
\APACinsertmetastar {%
marzouk2016if}%
\begin{APACrefauthors}%
Marzouk, Z.%
, Rakovic, M.%
, Liaqat, A.%
, Vytasek, J.%
, Samadi, D.%
, Stewart-Alonso, J.%
\BDBL {}Nesbit, J\BPBI C.%
\end{APACrefauthors}%
\unskip\
\newblock
\APACrefYearMonthDay{2016}{}{}.
\newblock
{\BBOQ}\APACrefatitle {What if learning analytics were based on learning science?} {What if learning analytics were based on learning science?}{\BBCQ}
\newblock
\APACjournalVolNumPages{Australasian Journal of Educational Technology}{32}{6}{}.
\newblock
\APAChowpublished {\href{https://doi.org/10.14742/ajet.3058}{\color{blue} https://doi.org/10.14742/ajet.3058}}.
\PrintBackRefs{\CurrentBib}

\bibitem [\protect \citeauthoryear {%
Matcha%
\ \protect \BOthers {.}}{%
Matcha%
\ \protect \BOthers {.}}{%
{\protect \APACyear {2019}}%
}]{%
matcha2019detection}
\APACinsertmetastar {%
matcha2019detection}%
\begin{APACrefauthors}%
Matcha, W.%
, Ga{\v{s}}evi{\'c}, D.%
, Ahmad~Uzir, N.%
, Jovanovi{\'c}, J.%
, Pardo, A.%
, Maldonado-Mahauad, J.%
\BCBL {}\ \BBA {} P{\'e}rez-Sanagust{\'\i}n, M.%
\end{APACrefauthors}%
\unskip\
\newblock
\APACrefYearMonthDay{2019}{}{}.
\newblock
{\BBOQ}\APACrefatitle {Detection of learning strategies: A comparison of process, sequence and network analytic approaches} {Detection of learning strategies: A comparison of process, sequence and network analytic approaches}.{\BBCQ}
\newblock
\BIn{} \APACrefbtitle {European conference on technology enhanced learning} {European conference on technology enhanced learning}\ (\BPGS\ 525--540).
\newblock
\APAChowpublished {\href{https://doi.org/10.1007/978-3-030-29736-7_39}{\color{blue} https://doi.org/10.1007/978-3-030-29736-7\_39}}.
\PrintBackRefs{\CurrentBib}

\bibitem [\protect \citeauthoryear {%
Matcha%
, Ga\v{s}evi\'{c}%
\BCBL {}\ \protect \BOthers {.}}{%
Matcha%
, Ga\v{s}evi\'{c}%
\BCBL {}\ \protect \BOthers {.}}{%
{\protect \APACyear {2020}}%
}]{%
matcha2020analytics}
\APACinsertmetastar {%
matcha2020analytics}%
\begin{APACrefauthors}%
Matcha, W.%
, Ga\v{s}evi\'{c}, D.%
, Uzir, N\BPBI A.%
, Jovanovic, J.%
, Pardo, A.%
, Lim, L.%
\BDBL {}Tsai, Y\BHBI S.%
\end{APACrefauthors}%
\unskip\
\newblock
\APACrefYearMonthDay{2020}{}{}.
\newblock
{\BBOQ}\APACrefatitle {Analytics of Learning Strategies: Role of Course Design and Delivery Modality.} {Analytics of learning strategies: Role of course design and delivery modality.}{\BBCQ}
\newblock
\APACjournalVolNumPages{Journal of Learning Analytics}{7}{2}{45--71}.
\newblock
\APAChowpublished {\href{https://doi.org/10.18608/jla.2020.72.3}{\color{blue} https://doi.org/10.18608/jla.2020.72.3}}.
\PrintBackRefs{\CurrentBib}

\bibitem [\protect \citeauthoryear {%
Matcha%
, Uzir%
, Gašević%
\BCBL {}\ \BBA {} Pardo%
}{%
Matcha%
, Uzir%
\BCBL {}\ \protect \BOthers {.}}{%
{\protect \APACyear {2020}}%
}]{%
matcha2019systematic}
\APACinsertmetastar {%
matcha2019systematic}%
\begin{APACrefauthors}%
Matcha, W.%
, Uzir, N\BPBI A.%
, Gašević, D.%
\BCBL {}\ \BBA {} Pardo, A.%
\end{APACrefauthors}%
\unskip\
\newblock
\APACrefYearMonthDay{2020}{}{}.
\newblock
{\BBOQ}\APACrefatitle {A Systematic Review of Empirical Studies on Learning Analytics Dashboards: A Self-Regulated Learning Perspective} {A systematic review of empirical studies on learning analytics dashboards: A self-regulated learning perspective}.{\BBCQ}
\newblock
\APACjournalVolNumPages{IEEE Transactions on Learning Technologies}{13}{2}{226-245}.
\newblock
\APAChowpublished {\href{https://doi.org/10.1109/TLT.2019.2916802}{\color{blue} https://doi.org/10.1109/TLT.2019.2916802}}.
\PrintBackRefs{\CurrentBib}

\bibitem [\protect \citeauthoryear {%
Molenaar%
, Roda%
, van Boxtel%
\BCBL {}\ \BBA {} Sleegers%
}{%
Molenaar%
\ \protect \BOthers {.}}{%
{\protect \APACyear {2012}}%
}]{%
molenaar2012dynamic}
\APACinsertmetastar {%
molenaar2012dynamic}%
\begin{APACrefauthors}%
Molenaar, I.%
, Roda, C.%
, van Boxtel, C.%
\BCBL {}\ \BBA {} Sleegers, P.%
\end{APACrefauthors}%
\unskip\
\newblock
\APACrefYearMonthDay{2012}{}{}.
\newblock
{\BBOQ}\APACrefatitle {Dynamic scaffolding of socially regulated learning in a computer-based learning environment} {Dynamic scaffolding of socially regulated learning in a computer-based learning environment}.{\BBCQ}
\newblock
\APACjournalVolNumPages{Computers \& Education}{59}{2}{515--523}.
\newblock
\APAChowpublished {\href{https://doi.org/10.1016/j.compedu.2011.12.006}{\color{blue} https://doi.org/10.1016/j.compedu.2011.12.006}}.
\PrintBackRefs{\CurrentBib}

\bibitem [\protect \citeauthoryear {%
Munshi%
\ \protect \BOthers {.}}{%
Munshi%
\ \protect \BOthers {.}}{%
{\protect \APACyear {2023}}%
}]{%
munshi2023analysing}
\APACinsertmetastar {%
munshi2023analysing}%
\begin{APACrefauthors}%
Munshi, A.%
, Biswas, G.%
, Baker, R.%
, Ocumpaugh, J.%
, Hutt, S.%
\BCBL {}\ \BBA {} Paquette, L.%
\end{APACrefauthors}%
\unskip\
\newblock
\APACrefYearMonthDay{2023}{}{}.
\newblock
{\BBOQ}\APACrefatitle {Analysing adaptive scaffolds that help students develop self-regulated learning behaviours} {Analysing adaptive scaffolds that help students develop self-regulated learning behaviours}.{\BBCQ}
\newblock
\APACjournalVolNumPages{Journal of Computer Assisted Learning}{39}{2}{351--368}.
\newblock
\APAChowpublished {\href{https://doi.org/10.1111/jcal.12761}{\color{blue} https://doi.org/10.1111/jcal.12761}}.
\PrintBackRefs{\CurrentBib}

\bibitem [\protect \citeauthoryear {%
Nelson%
\ \BBA {} Narens%
}{%
Nelson%
\ \BBA {} Narens%
}{%
{\protect \APACyear {1994}}%
}]{%
nelson1994investigate}
\APACinsertmetastar {%
nelson1994investigate}%
\begin{APACrefauthors}%
Nelson, T.%
\BCBT {}\ \BBA {} Narens, L.%
\end{APACrefauthors}%
\unskip\
\newblock
\APACrefYearMonthDay{1994}{}{}.
\newblock
\APACrefbtitle {Why investigate metacognition. Metacognition: Knowing about knowing, 13, 1-25.} {Why investigate metacognition. metacognition: Knowing about knowing, 13, 1-25.}
\newblock
\APAChowpublished {\href{https://doi.org/10.7551/mitpress/4561.003.0003}{\color{blue} https://doi.org/10.7551/mitpress/4561.003.0003}}.
\PrintBackRefs{\CurrentBib}

\bibitem [\protect \citeauthoryear {%
Nussbaumer%
\ \protect \BOthers {.}}{%
Nussbaumer%
\ \protect \BOthers {.}}{%
{\protect \APACyear {2014}}%
}]{%
nussbaumer2014framework}
\APACinsertmetastar {%
nussbaumer2014framework}%
\begin{APACrefauthors}%
Nussbaumer, A.%
, Kravcik, M.%
, Renzel, D.%
, Klamma, R.%
, Berthold, M.%
\BCBL {}\ \BBA {} Albert, D.%
\end{APACrefauthors}%
\unskip\
\newblock
\APACrefYearMonthDay{2014}{}{}.
\newblock
{\BBOQ}\APACrefatitle {A framework for facilitating self-regulation in responsive open learning environments} {A framework for facilitating self-regulation in responsive open learning environments}.{\BBCQ}
\newblock
\APACjournalVolNumPages{arXiv preprint arXiv:1407.5891}{}{}{}.
\newblock
\APAChowpublished {\href{https://doi.org/10.48550/arXiv.1407.5891}{\color{blue} https://doi.org/10.48550/arXiv.1407.5891}}.
\PrintBackRefs{\CurrentBib}

\bibitem [\protect \citeauthoryear {%
Oluwatosin%
}{%
Oluwatosin%
}{%
{\protect \APACyear {2014}}%
}]{%
oluwatosin2014client}
\APACinsertmetastar {%
oluwatosin2014client}%
\begin{APACrefauthors}%
Oluwatosin, H\BPBI S.%
\end{APACrefauthors}%
\unskip\
\newblock
\APACrefYearMonthDay{2014}{}{}.
\newblock
{\BBOQ}\APACrefatitle {Client-server model} {Client-server model}.{\BBCQ}
\newblock
\APACjournalVolNumPages{IOSR Journal of Computer Engineering}{16}{1}{67--71}.
\newblock
\APAChowpublished {\href{http://dx.doi.org/10.9790/0661-16195771}{\color{blue} http://dx.doi.org/10.9790/0661-16195771}}.
\PrintBackRefs{\CurrentBib}

\bibitem [\protect \citeauthoryear {%
Paans%
, Molenaar%
, Segers%
\BCBL {}\ \BBA {} Verhoeven%
}{%
Paans%
\ \protect \BOthers {.}}{%
{\protect \APACyear {2019}}%
}]{%
paans2019temporal}
\APACinsertmetastar {%
paans2019temporal}%
\begin{APACrefauthors}%
Paans, C.%
, Molenaar, I.%
, Segers, E.%
\BCBL {}\ \BBA {} Verhoeven, L.%
\end{APACrefauthors}%
\unskip\
\newblock
\APACrefYearMonthDay{2019}{}{}.
\newblock
{\BBOQ}\APACrefatitle {Temporal variation in children's self-regulated hypermedia learning} {Temporal variation in children's self-regulated hypermedia learning}.{\BBCQ}
\newblock
\APACjournalVolNumPages{Computers in Human Behavior}{96}{}{246--258}.
\newblock
\APAChowpublished {\href{https://doi.org/10.1016/j.chb.2018.04.002}{\color{blue} https://doi.org/10.1016/j.chb.2018.04.002}}.
\PrintBackRefs{\CurrentBib}

\bibitem [\protect \citeauthoryear {%
Panadero%
}{%
Panadero%
}{%
{\protect \APACyear {2017}}%
}]{%
panadero2017review}
\APACinsertmetastar {%
panadero2017review}%
\begin{APACrefauthors}%
Panadero, E.%
\end{APACrefauthors}%
\unskip\
\newblock
\APACrefYearMonthDay{2017}{}{}.
\newblock
{\BBOQ}\APACrefatitle {A review of self-regulated learning: Six models and four directions for research} {A review of self-regulated learning: Six models and four directions for research}.{\BBCQ}
\newblock
\APACjournalVolNumPages{Frontiers in psychology}{}{}{422}.
\newblock
\APAChowpublished {\href{https://doi.org/10.3389/fpsyg.2017.00422}{\color{blue} https://doi.org/10.3389/fpsyg.2017.00422}}.
\PrintBackRefs{\CurrentBib}

\bibitem [\protect \citeauthoryear {%
Papamitsiou%
\ \BBA {} Economides%
}{%
Papamitsiou%
\ \BBA {} Economides%
}{%
{\protect \APACyear {2014}}%
}]{%
papamitsiou2014learning}
\APACinsertmetastar {%
papamitsiou2014learning}%
\begin{APACrefauthors}%
Papamitsiou, Z.%
\BCBT {}\ \BBA {} Economides, A\BPBI A.%
\end{APACrefauthors}%
\unskip\
\newblock
\APACrefYearMonthDay{2014}{}{}.
\newblock
{\BBOQ}\APACrefatitle {Learning analytics and educational data mining in practice: A systematic literature review of empirical evidence} {Learning analytics and educational data mining in practice: A systematic literature review of empirical evidence}.{\BBCQ}
\newblock
\APACjournalVolNumPages{Journal of Educational Technology \& Society}{17}{4}{49--64}.
\newblock
\APAChowpublished {\href{https://www.jstor.org/stable/jeductechsoci.17.4.49}{\color{blue} https://www.jstor.org/stable/jeductechsoci.17.4.49}}.
\PrintBackRefs{\CurrentBib}

\bibitem [\protect \citeauthoryear {%
Reimann%
}{%
Reimann%
}{%
{\protect \APACyear {2009}}%
}]{%
reimann2009time}
\APACinsertmetastar {%
reimann2009time}%
\begin{APACrefauthors}%
Reimann, P.%
\end{APACrefauthors}%
\unskip\
\newblock
\APACrefYearMonthDay{2009}{}{}.
\newblock
{\BBOQ}\APACrefatitle {Time is precious: Variable-and event-centred approaches to process analysis in CSCL research} {Time is precious: Variable-and event-centred approaches to process analysis in cscl research}.{\BBCQ}
\newblock
\APACjournalVolNumPages{International Journal of Computer-Supported Collaborative Learning}{4}{}{239--257}.
\newblock
\APAChowpublished {\href{https://doi.org/10.1007/s11412-009-9070-z}{\color{blue} https://doi.org/10.1007/s11412-009-9070-z}}.
\PrintBackRefs{\CurrentBib}

\bibitem [\protect \citeauthoryear {%
Saint%
, Fan%
, Ga{\v{s}}evi{\'c}%
\BCBL {}\ \BBA {} Pardo%
}{%
Saint%
\ \protect \BOthers {.}}{%
{\protect \APACyear {2022}}%
}]{%
saint2022temporally}
\APACinsertmetastar {%
saint2022temporally}%
\begin{APACrefauthors}%
Saint, J.%
, Fan, Y.%
, Ga{\v{s}}evi{\'c}, D.%
\BCBL {}\ \BBA {} Pardo, A.%
\end{APACrefauthors}%
\unskip\
\newblock
\APACrefYearMonthDay{2022}{}{}.
\newblock
{\BBOQ}\APACrefatitle {Temporally-focused analytics of self-regulated learning: A systematic review of literature} {Temporally-focused analytics of self-regulated learning: A systematic review of literature}.{\BBCQ}
\newblock
\APACjournalVolNumPages{Computers and Education: Artificial Intelligence}{}{}{100060}.
\newblock
\APAChowpublished {\href{https://doi.org/10.1016/j.caeai.2022.100060}{\color{blue} https://doi.org/10.1016/j.caeai.2022.100060}}.
\PrintBackRefs{\CurrentBib}

\bibitem [\protect \citeauthoryear {%
Saint%
, Ga{\v{s}}evi{\'c}%
, Matcha%
, Uzir%
\BCBL {}\ \BBA {} Pardo%
}{%
Saint%
, Ga{\v{s}}evi{\'c}%
\BCBL {}\ \protect \BOthers {.}}{%
{\protect \APACyear {2020}}%
}]{%
saint2020combining}
\APACinsertmetastar {%
saint2020combining}%
\begin{APACrefauthors}%
Saint, J.%
, Ga{\v{s}}evi{\'c}, D.%
, Matcha, W.%
, Uzir, N\BPBI A.%
\BCBL {}\ \BBA {} Pardo, A.%
\end{APACrefauthors}%
\unskip\
\newblock
\APACrefYearMonthDay{2020}{}{}.
\newblock
{\BBOQ}\APACrefatitle {Combining analytic methods to unlock sequential and temporal patterns of self-regulated learning} {Combining analytic methods to unlock sequential and temporal patterns of self-regulated learning}.{\BBCQ}
\newblock
\BIn{} \APACrefbtitle {Proceedings of the tenth international conference on learning analytics \& knowledge} {Proceedings of the tenth international conference on learning analytics \& knowledge}\ (\BPGS\ 402--411).
\newblock
\APAChowpublished {\href{https://doi.org/10.1145/3375462.3375487}{\color{blue} https://doi.org/10.1145/3375462.3375487}}.
\PrintBackRefs{\CurrentBib}

\bibitem [\protect \citeauthoryear {%
Saint%
, Whitelock-Wainwright%
, Ga{\v{s}}evi{\'c}%
\BCBL {}\ \BBA {} Pardo%
}{%
Saint%
, Whitelock-Wainwright%
\BCBL {}\ \protect \BOthers {.}}{%
{\protect \APACyear {2020}}%
}]{%
saint2020trace}
\APACinsertmetastar {%
saint2020trace}%
\begin{APACrefauthors}%
Saint, J.%
, Whitelock-Wainwright, A.%
, Ga{\v{s}}evi{\'c}, D.%
\BCBL {}\ \BBA {} Pardo, A.%
\end{APACrefauthors}%
\unskip\
\newblock
\APACrefYearMonthDay{2020}{}{}.
\newblock
{\BBOQ}\APACrefatitle {Trace-SRL: a framework for analysis of microlevel processes of self-regulated learning from trace data} {Trace-srl: a framework for analysis of microlevel processes of self-regulated learning from trace data}.{\BBCQ}
\newblock
\APACjournalVolNumPages{IEEE Transactions on Learning Technologies}{13}{4}{861--877}.
\newblock
\APAChowpublished {\href{http://dx.doi.org/10.1109/TLT.2020.3027496}{\color{blue} http://dx.doi.org/10.1109/TLT.2020.3027496}}.
\PrintBackRefs{\CurrentBib}

\bibitem [\protect \citeauthoryear {%
Shaffer%
, Collier%
\BCBL {}\ \BBA {} Ruis%
}{%
Shaffer%
\ \protect \BOthers {.}}{%
{\protect \APACyear {2016}}%
}]{%
shaffer2016tutorial}
\APACinsertmetastar {%
shaffer2016tutorial}%
\begin{APACrefauthors}%
Shaffer, D\BPBI W.%
, Collier, W.%
\BCBL {}\ \BBA {} Ruis, A\BPBI R.%
\end{APACrefauthors}%
\unskip\
\newblock
\APACrefYearMonthDay{2016}{}{}.
\newblock
{\BBOQ}\APACrefatitle {A tutorial on epistemic network analysis: Analyzing the structure of connections in cognitive, social, and interaction data} {A tutorial on epistemic network analysis: Analyzing the structure of connections in cognitive, social, and interaction data}.{\BBCQ}
\newblock
\APACjournalVolNumPages{Journal of Learning Analytics}{3}{3}{9--45}.
\newblock
\APAChowpublished {\href{https://doi.org/10.18608/jla.2016.33.3}{\color{blue} https://doi.org/10.18608/jla.2016.33.3}}.
\PrintBackRefs{\CurrentBib}

\bibitem [\protect \citeauthoryear {%
Sharma%
\ \BBA {} Hannafin%
}{%
Sharma%
\ \BBA {} Hannafin%
}{%
{\protect \APACyear {2007}}%
}]{%
sharma2007scaffolding}
\APACinsertmetastar {%
sharma2007scaffolding}%
\begin{APACrefauthors}%
Sharma, P.%
\BCBT {}\ \BBA {} Hannafin, M\BPBI J.%
\end{APACrefauthors}%
\unskip\
\newblock
\APACrefYearMonthDay{2007}{}{}.
\newblock
{\BBOQ}\APACrefatitle {Scaffolding in technology-enhanced learning environments} {Scaffolding in technology-enhanced learning environments}.{\BBCQ}
\newblock
\APACjournalVolNumPages{Interactive learning environments}{15}{1}{27--46}.
\newblock
\APAChowpublished {\href{https://doi.org/10.1080/10494820600996972}{\color{blue} https://doi.org/10.1080/10494820600996972}}.
\PrintBackRefs{\CurrentBib}

\bibitem [\protect \citeauthoryear {%
Siadaty%
, Ga\v{s}evi\'{c}%
\BCBL {}\ \BBA {} Hatala%
}{%
Siadaty%
\ \protect \BOthers {.}}{%
{\protect \APACyear {2016}}%
}]{%
siadaty2016trace}
\APACinsertmetastar {%
siadaty2016trace}%
\begin{APACrefauthors}%
Siadaty, M.%
, Ga\v{s}evi\'{c}, D.%
\BCBL {}\ \BBA {} Hatala, M.%
\end{APACrefauthors}%
\unskip\
\newblock
\APACrefYearMonthDay{2016}{}{}.
\newblock
{\BBOQ}\APACrefatitle {Trace-based micro-analytic measurement of self-regulated learning processes} {Trace-based micro-analytic measurement of self-regulated learning processes}.{\BBCQ}
\newblock
\APACjournalVolNumPages{Journal of Learning Analytics}{3}{1}{183--214}.
\newblock
\APAChowpublished {\href{https://doi.org/10.18608/jla.2016.31.11}{\color{blue} https://doi.org/10.18608/jla.2016.31.11}}.
\PrintBackRefs{\CurrentBib}

\bibitem [\protect \citeauthoryear {%
Srivastava%
\ \protect \BOthers {.}}{%
Srivastava%
\ \protect \BOthers {.}}{%
{\protect \APACyear {2022}}%
}]{%
srivastava2022effects}
\APACinsertmetastar {%
srivastava2022effects}%
\begin{APACrefauthors}%
Srivastava, N.%
, Fan, Y.%
, Rakovic, M.%
, Singh, S.%
, Jovanovic, J.%
, Van Der~Graaf, J.%
\BDBL {}others%
\end{APACrefauthors}%
\unskip\
\newblock
\APACrefYearMonthDay{2022}{}{}.
\newblock
{\BBOQ}\APACrefatitle {Effects of Internal and External Conditions on Strategies of Self-regulated Learning: A Learning Analytics Study} {Effects of internal and external conditions on strategies of self-regulated learning: A learning analytics study}.{\BBCQ}
\newblock
\BIn{} \APACrefbtitle {LAK22: 12th International Learning Analytics and Knowledge Conference} {Lak22: 12th international learning analytics and knowledge conference}\ (\BPGS\ 392--403).
\newblock
\APAChowpublished {\href{https://doi.org/10.1145/3506860.3506972}{\color{blue} https://doi.org/10.1145/3506860.3506972}}.
\PrintBackRefs{\CurrentBib}

\bibitem [\protect \citeauthoryear {%
Uzir%
\ \protect \BOthers {.}}{%
Uzir%
\ \protect \BOthers {.}}{%
{\protect \APACyear {2020}}%
}]{%
uzir2020analytics}
\APACinsertmetastar {%
uzir2020analytics}%
\begin{APACrefauthors}%
Uzir, N\BPBI A.%
, Ga{\v{s}}evi{\'c}, D.%
, Jovanovi{\'c}, J.%
, Matcha, W.%
, Lim, L\BHBI A.%
\BCBL {}\ \BBA {} Fudge, A.%
\end{APACrefauthors}%
\unskip\
\newblock
\APACrefYearMonthDay{2020}{}{}.
\newblock
{\BBOQ}\APACrefatitle {Analytics of time management and learning strategies for effective online learning in blended environments} {Analytics of time management and learning strategies for effective online learning in blended environments}.{\BBCQ}
\newblock
\BIn{} \APACrefbtitle {Proceedings of the tenth international conference on learning analytics \& knowledge} {Proceedings of the tenth international conference on learning analytics \& knowledge}\ (\BPGS\ 392--401).
\newblock
\APAChowpublished {\href{http://dx.doi.org/10.1145/3375462.3375493}{\color{blue} http://dx.doi.org/10.1145/3375462.3375493}}.
\PrintBackRefs{\CurrentBib}

\bibitem [\protect \citeauthoryear {%
van~der Graaf%
\ \protect \BOthers {.}}{%
van~der Graaf%
\ \protect \BOthers {.}}{%
{\protect \APACyear {2021}}%
}]{%
van2021instrumentation}
\APACinsertmetastar {%
van2021instrumentation}%
\begin{APACrefauthors}%
van~der Graaf, J.%
, Lim, L.%
, Fan, Y.%
, Kilgour, J.%
, Moore, J.%
, Bannert, M.%
\BDBL {}Molenaar, I.%
\end{APACrefauthors}%
\unskip\
\newblock
\APACrefYearMonthDay{2021}{}{}.
\newblock
{\BBOQ}\APACrefatitle {Do instrumentation tools capture self-regulated learning?} {Do instrumentation tools capture self-regulated learning?}{\BBCQ}
\newblock
\BIn{} \APACrefbtitle {LAK21: 11th international learning analytics and knowledge conference} {Lak21: 11th international learning analytics and knowledge conference}\ (\BPGS\ 438--448).
\newblock
\APAChowpublished {\href{https://doi.org/10.1145/3448139.3448181}{\color{blue} https://doi.org/10.1145/3448139.3448181}}.
\PrintBackRefs{\CurrentBib}

\bibitem [\protect \citeauthoryear {%
van~der Graaf%
\ \protect \BOthers {.}}{%
van~der Graaf%
\ \protect \BOthers {.}}{%
{\protect \APACyear {2022}}%
}]{%
van2022dynamics}
\APACinsertmetastar {%
van2022dynamics}%
\begin{APACrefauthors}%
van~der Graaf, J.%
, Lim, L.%
, Fan, Y.%
, Kilgour, J.%
, Moore, J.%
, Ga{\v{s}}evi{\'c}, D.%
\BDBL {}Molenaar, I.%
\end{APACrefauthors}%
\unskip\
\newblock
\APACrefYearMonthDay{2022}{}{}.
\newblock
{\BBOQ}\APACrefatitle {The dynamics between self-regulated learning and learning outcomes: An exploratory approach and implications} {The dynamics between self-regulated learning and learning outcomes: An exploratory approach and implications}.{\BBCQ}
\newblock
\APACjournalVolNumPages{Metacognition and Learning}{17}{3}{745--771}.
\newblock
\APAChowpublished {\href{https://doi.org/10.1007/s11409-022-09308-9}{\color{blue} https://doi.org/10.1007/s11409-022-09308-9}}.
\PrintBackRefs{\CurrentBib}

\bibitem [\protect \citeauthoryear {%
van~der Graaf%
\ \protect \BOthers {.}}{%
van~der Graaf%
\ \protect \BOthers {.}}{%
{\protect \APACyear {2023}}%
}]{%
van2023design}
\APACinsertmetastar {%
van2023design}%
\begin{APACrefauthors}%
van~der Graaf, J.%
, Rakovi{\'c}, M.%
, Fan, Y.%
, Lim, L.%
, Singh, S.%
, Bannert, M.%
\BDBL {}Molenaar, I.%
\end{APACrefauthors}%
\unskip\
\newblock
\APACrefYearMonthDay{2023}{}{}.
\newblock
{\BBOQ}\APACrefatitle {How to design and evaluate personalized scaffolds for self-regulated learning} {How to design and evaluate personalized scaffolds for self-regulated learning}.{\BBCQ}
\newblock
\APACjournalVolNumPages{Metacognition and Learning}{}{}{1--28}.
\newblock
\APAChowpublished {\href{https://doi.org/10.1007/s11409-023-09361-y}{\color{blue} https://doi.org/10.1007/s11409-023-09361-y}}.
\PrintBackRefs{\CurrentBib}

\bibitem [\protect \citeauthoryear {%
Veenman%
}{%
Veenman%
}{%
{\protect \APACyear {2007}}%
}]{%
veenman2007assessment}
\APACinsertmetastar {%
veenman2007assessment}%
\begin{APACrefauthors}%
Veenman, M\BPBI V.%
\end{APACrefauthors}%
\unskip\
\newblock
\APACrefYearMonthDay{2007}{}{}.
\newblock
{\BBOQ}\APACrefatitle {The assessment and instruction of self-regulation in computer-based environments: a discussion} {The assessment and instruction of self-regulation in computer-based environments: a discussion}.{\BBCQ}
\newblock
\APACjournalVolNumPages{Metacognition and Learning}{2}{2}{177--183}.
\newblock
\APAChowpublished {\href{https://doi.org/10.1007/s11409-007-9017-6}{\color{blue} https://doi.org/10.1007/s11409-007-9017-6}}.
\PrintBackRefs{\CurrentBib}

\bibitem [\protect \citeauthoryear {%
Winne%
}{%
Winne%
}{%
{\protect \APACyear {2017}}%
}]{%
winne2017learning}
\APACinsertmetastar {%
winne2017learning}%
\begin{APACrefauthors}%
Winne, P\BPBI H.%
\end{APACrefauthors}%
\unskip\
\newblock
\APACrefYearMonthDay{2017}{}{}.
\newblock
{\BBOQ}\APACrefatitle {Learning analytics for self-regulated learning} {Learning analytics for self-regulated learning}.{\BBCQ}
\newblock
\APACjournalVolNumPages{Handbook of learning analytics}{}{}{241--249}.
\newblock
\APAChowpublished {\href{https://doi.org/10.18608/hla17.021}{\color{blue} https://doi.org/10.18608/hla17.021}}.
\PrintBackRefs{\CurrentBib}

\bibitem [\protect \citeauthoryear {%
Winne%
}{%
Winne%
}{%
{\protect \APACyear {2018}}%
}]{%
winne2018theorizing}
\APACinsertmetastar {%
winne2018theorizing}%
\begin{APACrefauthors}%
Winne, P\BPBI H.%
\end{APACrefauthors}%
\unskip\
\newblock
\APACrefYearMonthDay{2018}{}{}.
\newblock
{\BBOQ}\APACrefatitle {Theorizing and researching levels of processing in self-regulated learning} {Theorizing and researching levels of processing in self-regulated learning}.{\BBCQ}
\newblock
\APACjournalVolNumPages{British Journal of Educational Psychology}{88}{1}{9--20}.
\newblock
\APAChowpublished {\href{https://doi.org/10.1111/bjep.12173}{\color{blue} https://doi.org/10.1111/bjep.12173}}.
\PrintBackRefs{\CurrentBib}

\bibitem [\protect \citeauthoryear {%
Winne%
\ \BBA {} Hadwin%
}{%
Winne%
\ \BBA {} Hadwin%
}{%
{\protect \APACyear {1998}}%
}]{%
winne1998studying}
\APACinsertmetastar {%
winne1998studying}%
\begin{APACrefauthors}%
Winne, P\BPBI H.%
\BCBT {}\ \BBA {} Hadwin, A.%
\end{APACrefauthors}%
\unskip\
\newblock
\APACrefYearMonthDay{1998}{}{}.
\newblock
\APACrefbtitle {Studying as self-regulated learning. Metacognition in educational theory and practice, D. HACKER, J. DUNLOSKY, AND A. GRAESSER, Eds.} {Studying as self-regulated learning. metacognition in educational theory and practice, d. hacker, j. dunlosky, and a. graesser, eds.}
\newblock
\APAChowpublished {\href{https://psycnet.apa.org/record/1998-07283-011}{\color{blue} https://psycnet.apa.org/record/1998-07283-011}}.
\newblock
\APACaddressPublisher{}{Mahwah, NJ: Erlbaum}.
\PrintBackRefs{\CurrentBib}

\bibitem [\protect \citeauthoryear {%
Winne%
\ \BBA {} Perry%
}{%
Winne%
\ \BBA {} Perry%
}{%
{\protect \APACyear {2000}}%
}]{%
winne2000measuring}
\APACinsertmetastar {%
winne2000measuring}%
\begin{APACrefauthors}%
Winne, P\BPBI H.%
\BCBT {}\ \BBA {} Perry, N\BPBI E.%
\end{APACrefauthors}%
\unskip\
\newblock
\APACrefYearMonthDay{2000}{}{}.
\newblock
{\BBOQ}\APACrefatitle {Measuring self-regulated learning} {Measuring self-regulated learning}.{\BBCQ}
\newblock
\BIn{} \APACrefbtitle {Handbook of self-regulation} {Handbook of self-regulation}\ (\BPGS\ 531--566).
\newblock
\APACaddressPublisher{}{Elsevier}.
\newblock
\APAChowpublished {\href{https://doi.org/10.1016/B978-012109890-2/50045-7}{\color{blue} https://doi.org/10.1016/B978-012109890-2/50045-7}}.
\PrintBackRefs{\CurrentBib}

\bibitem [\protect \citeauthoryear {%
Winne%
\ \protect \BOthers {.}}{%
Winne%
\ \protect \BOthers {.}}{%
{\protect \APACyear {2019}}%
}]{%
winne2019nstudy}
\APACinsertmetastar {%
winne2019nstudy}%
\begin{APACrefauthors}%
Winne, P\BPBI H.%
, Teng, K.%
, Chang, D.%
, Lin, M\BPBI P\BHBI C.%
, Marzouk, Z.%
, Nesbit, J\BPBI C.%
\BDBL {}Vytasek, J.%
\end{APACrefauthors}%
\unskip\
\newblock
\APACrefYearMonthDay{2019}{}{}.
\newblock
{\BBOQ}\APACrefatitle {nStudy: Software for learning analytics about processes for self-regulated learning} {nstudy: Software for learning analytics about processes for self-regulated learning}.{\BBCQ}
\newblock
\APACjournalVolNumPages{Journal of Learning Analytics}{6}{2}{95--106}.
\newblock
\APAChowpublished {\href{https://doi.org/10.18608/jla.2019.62.7}{\color{blue} https://doi.org/10.18608/jla.2019.62.7}}.
\PrintBackRefs{\CurrentBib}

\bibitem [\protect \citeauthoryear {%
Wisniewski%
, Zierer%
\BCBL {}\ \BBA {} Hattie%
}{%
Wisniewski%
\ \protect \BOthers {.}}{%
{\protect \APACyear {2020}}%
}]{%
wisniewski2020power}
\APACinsertmetastar {%
wisniewski2020power}%
\begin{APACrefauthors}%
Wisniewski, B.%
, Zierer, K.%
\BCBL {}\ \BBA {} Hattie, J.%
\end{APACrefauthors}%
\unskip\
\newblock
\APACrefYearMonthDay{2020}{}{}.
\newblock
{\BBOQ}\APACrefatitle {The power of feedback revisited: A meta-analysis of educational feedback research} {The power of feedback revisited: A meta-analysis of educational feedback research}.{\BBCQ}
\newblock
\APACjournalVolNumPages{Frontiers in Psychology}{10}{}{3087}.
\newblock
\APAChowpublished {\href{https://doi.org/10.3389/fpsyg.2019.03087}{\color{blue} https://doi.org/10.3389/fpsyg.2019.03087}}.
\PrintBackRefs{\CurrentBib}

\bibitem [\protect \citeauthoryear {%
Yau%
\ \BBA {} Joy%
}{%
Yau%
\ \BBA {} Joy%
}{%
{\protect \APACyear {2009}}%
}]{%
yau2009mobile}
\APACinsertmetastar {%
yau2009mobile}%
\begin{APACrefauthors}%
Yau, J\BHBI K.%
\BCBT {}\ \BBA {} Joy, M.%
\end{APACrefauthors}%
\unskip\
\newblock
\APACrefYearMonthDay{2009}{}{}.
\newblock
{\BBOQ}\APACrefatitle {A mobile context-aware framework for supporting self-regulated learners} {A mobile context-aware framework for supporting self-regulated learners}.{\BBCQ}
\newblock
\APAChowpublished {\href{https://wrap.warwick.ac.uk/id/eprint/60327/}{\color{blue} https://wrap.warwick.ac.uk/id/eprint/60327/}}.
\PrintBackRefs{\CurrentBib}

\bibitem [\protect \citeauthoryear {%
Yousef%
, Chatti%
, Danoyan%
, Th{\"u}s%
\BCBL {}\ \BBA {} Schroeder%
}{%
Yousef%
\ \protect \BOthers {.}}{%
{\protect \APACyear {2015}}%
}]{%
yousef2015video}
\APACinsertmetastar {%
yousef2015video}%
\begin{APACrefauthors}%
Yousef, A\BPBI M\BPBI F.%
, Chatti, M\BPBI A.%
, Danoyan, N.%
, Th{\"u}s, H.%
\BCBL {}\ \BBA {} Schroeder, U.%
\end{APACrefauthors}%
\unskip\
\newblock
\APACrefYearMonthDay{2015}{}{}.
\newblock
{\BBOQ}\APACrefatitle {Video-mapper: A video annotation tool to support collaborative learning in moocs} {Video-mapper: A video annotation tool to support collaborative learning in moocs}.{\BBCQ}
\newblock
\APACjournalVolNumPages{Proceedings of the Third European MOOCs Stakeholders Summit EMOOCs}{}{}{131--140}.
\newblock
\APAChowpublished {\href{http://www.emoocs2015.eu/sites/default/files/Papers.pdf}{\color{blue} http://www.emoocs2015.eu/sites/default/files/Papers.pdf}}.
\PrintBackRefs{\CurrentBib}

\bibitem [\protect \citeauthoryear {%
Zimmerman%
}{%
Zimmerman%
}{%
{\protect \APACyear {1986}}%
}]{%
zimmerman1986becoming}
\APACinsertmetastar {%
zimmerman1986becoming}%
\begin{APACrefauthors}%
Zimmerman, B\BPBI J.%
\end{APACrefauthors}%
\unskip\
\newblock
\APACrefYearMonthDay{1986}{}{}.
\newblock
{\BBOQ}\APACrefatitle {Becoming a self-regulated learner: Which are the key subprocesses?} {Becoming a self-regulated learner: Which are the key subprocesses?}{\BBCQ}
\newblock
\APACjournalVolNumPages{Contemporary educational psychology}{11}{4}{307--313}.
\newblock
\APAChowpublished {\href{https://doi.org/10.1016/0361-476X(86)90027-5}{\color{blue} https://doi.org/10.1016/0361-476X(86)90027-5}}.
\PrintBackRefs{\CurrentBib}

\bibitem [\protect \citeauthoryear {%
Zimmerman%
}{%
Zimmerman%
}{%
{\protect \APACyear {2000}}%
}]{%
zimmerman2000attaining}
\APACinsertmetastar {%
zimmerman2000attaining}%
\begin{APACrefauthors}%
Zimmerman, B\BPBI J.%
\end{APACrefauthors}%
\unskip\
\newblock
\APACrefYearMonthDay{2000}{}{}.
\newblock
{\BBOQ}\APACrefatitle {Attaining self-regulation: A social cognitive perspective} {Attaining self-regulation: A social cognitive perspective}.{\BBCQ}
\newblock
\BIn{} \APACrefbtitle {Handbook of self-regulation} {Handbook of self-regulation}\ (\BPGS\ 13--39).
\newblock
\APACaddressPublisher{}{Elsevier}.
\newblock
\APAChowpublished {\href{https://doi.org/10.1016/B978-012109890-2/50031-7}{\color{blue} https://doi.org/10.1016/B978-012109890-2/50031-7}}.
\PrintBackRefs{\CurrentBib}

\bibitem [\protect \citeauthoryear {%
Zimmerman%
}{%
Zimmerman%
}{%
{\protect \APACyear {2013}}%
}]{%
zimmerman2013cognitive}
\APACinsertmetastar {%
zimmerman2013cognitive}%
\begin{APACrefauthors}%
Zimmerman, B\BPBI J.%
\end{APACrefauthors}%
\unskip\
\newblock
\APACrefYearMonthDay{2013}{}{}.
\newblock
{\BBOQ}\APACrefatitle {From cognitive modeling to self-regulation: A social cognitive career path} {From cognitive modeling to self-regulation: A social cognitive career path}.{\BBCQ}
\newblock
\APACjournalVolNumPages{Educational psychologist}{48}{3}{135--147}.
\newblock
\APAChowpublished {\href{http://dx.doi.org/10.1080/00461520.2013.794676}{\color{blue} http://dx.doi.org/10.1080/00461520.2013.794676}}.
\PrintBackRefs{\CurrentBib}

\end{thebibliography}


\begin{appendices}


\section{Appendix A - Library of Action Label and Processing Label}\label{sec:appendixB}

\begin{table}[hb]
\centering
\caption{Action library table}
\label{tab:action_library_table} 
\renewcommand{\arraystretch}{1.5}
\begin{tabular}{p{0.8cm} p{4.5cm} p{10.5cm}}

\hline\noalign{\smallskip}
\# & Actions & Descriptions\\
\noalign{\smallskip}\hline\noalign{\smallskip}
1 & GENERAL\_INSTRUCTION & learners read or re-read the general instructions and learning goals\\
2 & RUBRIC & learners read or re-read the rubric for essay writing\\
3 & RELEVANT\_READING & learners read and learn learning content for the first time\\
4 & RELEVANT\_RE-READING & learners re-read and review for learning content which they have read before\\
5 & IRRELEVANT\_READING & Learners read the pages which are not relevant to the learning goal and learning task\\
6 & IRRELEVANT\_RE-READING & Learners re-read the pages which are not relevant to the learning goal and learning task\\

7 & NAVIGATION & (1)learners quickly navigate through pages by quick reading and clicking (2)learners look/scroll at catalog area or page (3)learners open the table of content page\\
8 & OPEN\_ESSAY & learners: (1)open; (2)read, and (3)think about the essay\\
9 & WRITE\_ESSAY & learners: (1)write; (2)edit or re-write\\
10 & EDIT\_ANNOTATION & On all pages, learners: (1)create;(2)delete; or (3)edit the annotations\\
11 & READ\_ANNOTATION & On all pages, learners: (1)read;(2)re-read; (3)look at annotations\\
12 & LABEL\_ANNOTATION & Learners create tags for annotations\\

13 & SEARCH\_ANNOTATION & On all pages, learners use search function to search notes or highlights\\
14 & TIMER & On all pages, learners check timer during learning\\
15 & SEARCH\_CONTENT & On all pages, learners use search function to search contents\\
16 & PLANNER & On all pages, learners: (1)create planner; (2)edit planner or (3)check planner\\
17 & OFF\_TASK & Learners do not have any action for relatively long time\\
18 & SCAFFOLDING & Learners interact with scaffoldings\\

\noalign{\smallskip}\hline
\end{tabular}
\end{table}

\begin{table}
\centering
\caption{Scaffolding Sub-Action library table}
\label{tab:scaffolding_sub_action_library_table} 
\renewcommand{\arraystretch}{1.5}
\begin{tabular}{p{0.8cm} p{4.5cm} p{5.5cm}}

\hline\noalign{\smallskip}
\# & Actions & Sub-action\\
\noalign{\smallskip}\hline\noalign{\smallskip}
1 & SCAFFOLDING & Message\_Triggered\\
2 & SCAFFOLDING & Message\_Displayed\\
3 & SCAFFOLDING & Notification\_Clicked\\
4 & SCAFFOLDING & Message\_Closed\\
5 & SCAFFOLDING & MessageOption\_Checked\\
6 & SCAFFOLDING & MessageOption\_UnChecked\\

7 & SCAFFOLDING & CreateChecklist\\
8 & SCAFFOLDING & CurrToDoList\_Displayed\\
9 & SCAFFOLDING & PrevToDoList\_Displayed\\
10 & SCAFFOLDING & CurrToDoList\_Edit\\
11 & SCAFFOLDING & PrevToDoList\_Edit\\
12 & SCAFFOLDING & ToDoList\_Closed\\

13 & SCAFFOLDING & CurrToDoListItem\_Checked\\
14 & SCAFFOLDING & CurrToDoListItem\_UnChecked\\
15 & SCAFFOLDING & PrevToDoListItem\_Checked\\
16 & SCAFFOLDING & PrevToDoListItem\_UnChecked\\
17 & SCAFFOLDING & CurrToDoList\_Re-Ordered\\
18 & SCAFFOLDING & PrevToDoList\_Re-Ordered\\

19 & SCAFFOLDING & PrevToDoListItem\_ClickedLink\\
20 & SCAFFOLDING & NextToDoListItem\_ClickedLink\\

\noalign{\smallskip}\hline
\end{tabular}
\end{table}

\begin{figure}[!h]\centering 
\includegraphics[angle=270,scale=0.65]{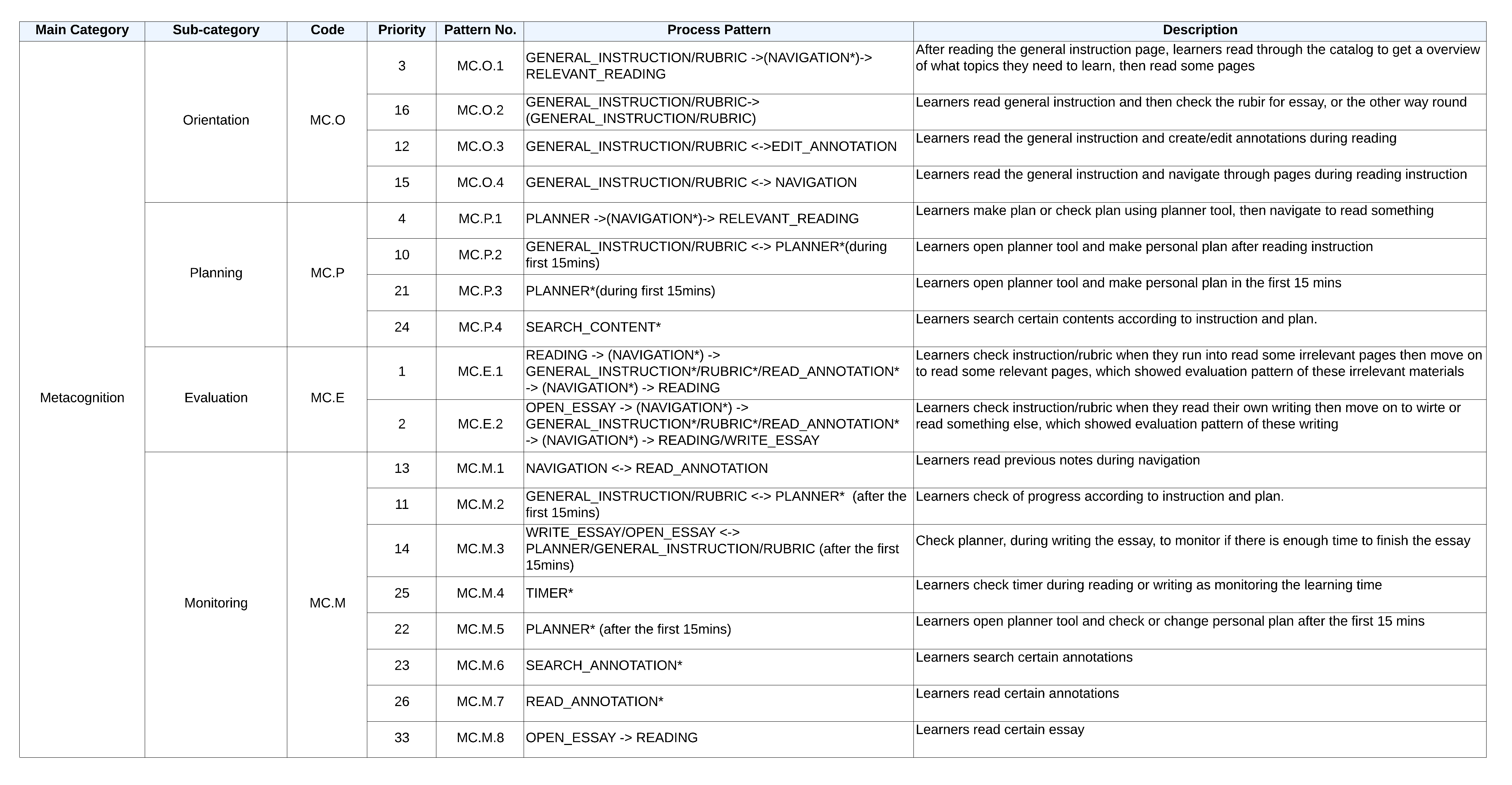}
\caption{Process label library - Metacognition}
\label{fig:process_label_metacognition}
\end{figure}

\begin{figure}[!h]\centering 
\includegraphics[angle=270,scale=0.35]{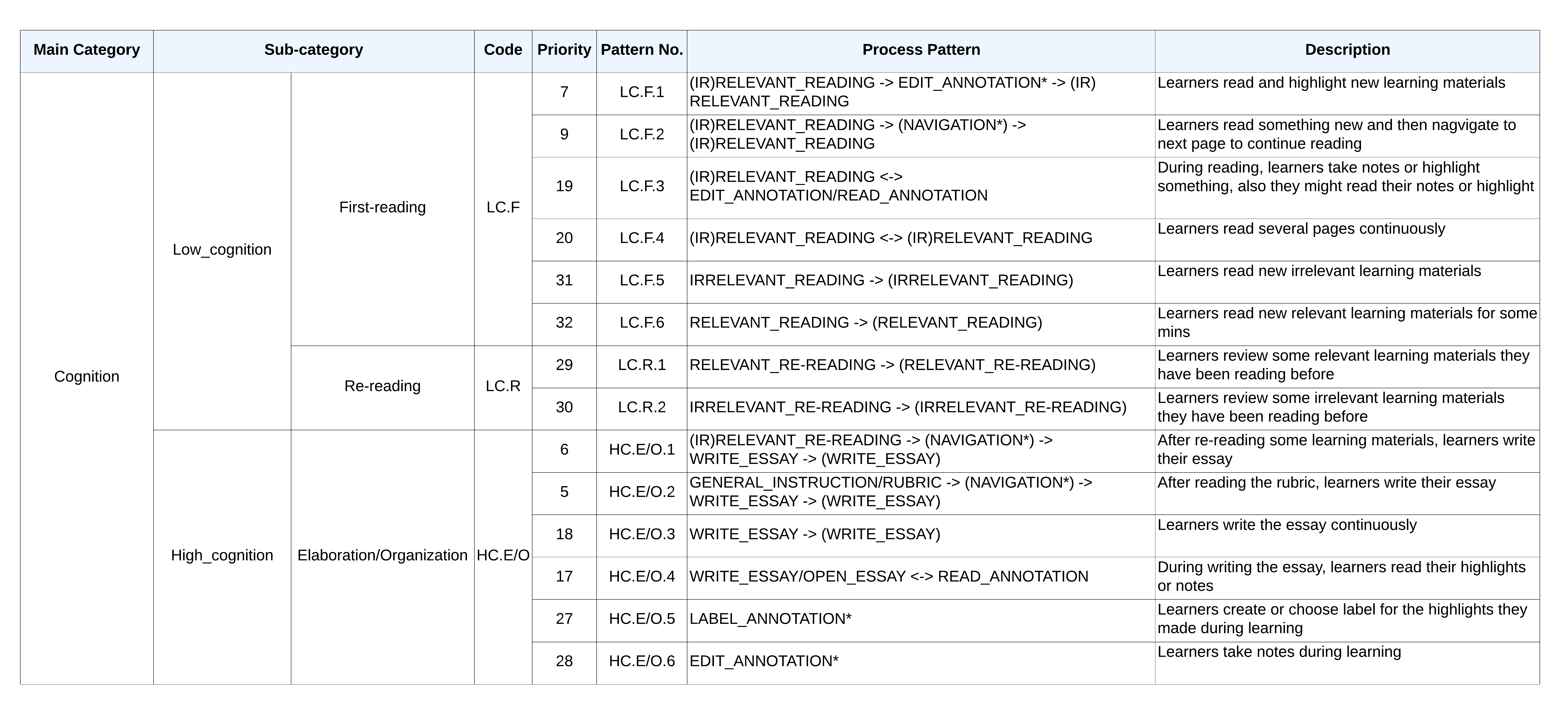}
\caption{Process label library - Cognition}
\label{fig:process_label_cognition}
\end{figure}

\end{appendices}

\end{document}